\documentclass[aps,showpacs,pre,amsmath,amsfonts,amssymb,superscriptaddress,nofootinbib,notitlepage,a4paper]{revtex4}

\pdfoutput=1

\usepackage{cancel}
\usepackage{graphicx}
\usepackage{float}
\usepackage{psfrag}
\usepackage[naturalnames]{hyperref}

\newcommand{\beq}{\begin{equation}} 
\newcommand{\eeq}{\end{equation}}
\newcommand{\bem}{\begin{multline}}
\newcommand{\bes}{\begin{split}} \newcommand{\ees}{\end{split}} 
\newcommand{\bea}{\begin{eqnarray}} \newcommand{\eea}{\end{eqnarray}}

\def\d{{\delta}}
\def\w{{\omega}}
\def\s{{\sigma}}

\def\G{{\Gamma}}
\def\Gr{{\Gamma_{\rm r}}}
\def\qr{q_{\rm r}}

\def\a{{\alpha}}
\def\asat{\alpha_{\rm sat}}
\def\aalgo{\alpha_{\rm alg}}
\def\ac{\alpha_{\rm c}}
\def\ad{\alpha_{\rm d}}
\def\adu{\alpha_{\rm d,u}}
\def\aKS{\alpha_{\rm KS}}
\def\ar{\alpha_{\rm r}}
\def\af{\alpha_{\rm f}}
\def\aru{\alpha_{\rm r,u}}
\def\aopt{\alpha_{\rm opt}}
\def\xd{{x_{\rm d}}}
\def\l{{\lambda}}
\def\e{{\epsilon}}
\def\eopt{{\epsilon_{\rm opt}}}

\def\us{{\underline{\sigma}}}

\def\di{{\partial i}}
\def\da{{\partial a}}

\def\dima{{\partial i \setminus a}}
\def\dami{{\partial a \setminus i}}

\def\heta{{\widehat{\eta}}}

\def\hq{{\widehat{q}}}

\def\hz{{\widehat{z}}}

\def\hQ{{\widehat{Q}}}
\def\hP{{\widehat{P}}}

\def\dd{{\rm d}}

\def\Z{{\cal Z}}

\def\tp{{\widetilde{p}}}

\def\E{{\mathbb{E}}}

\def\I{{\mathbb{I}}}

\def\N{{\cal N}}

\def\ve{\varepsilon}
\def\hve{\widehat{\varepsilon}}

\def\cP{{\cal P}}

\def\hcP{\widehat{\cal P}}

\begin{document}
\bibliographystyle{myunsrt}
\title{Biased landscapes for random Constraint Satisfaction Problems}

\author{Louise Budzynski}
\affiliation{Laboratoire de physique th\'eorique de l'Ecole normale sup\'erieure, PSL Research University, CNRS, Sorbonne Universit\'es, 24 rue Lhomond, 75231 Paris Cedex 05, France} 

\author{Federico Ricci-Tersenghi}
\affiliation{Diparimento di Fisica, Sapienza Universit\`a di Roma,
  and Nanotec-CNR, UOS di Roma, and INFN-Sezione di Roma 1,
  P.le Aldo Moro 5, 00185 Roma Italy}

\author{Guilhem Semerjian}
\affiliation{Laboratoire de physique th\'eorique de l'Ecole normale sup\'erieure, PSL Research University, CNRS, Sorbonne Universit\'es, 24 rue Lhomond, 75231 Paris Cedex 05, France}

\begin{abstract}
The typical complexity of Constraint Satisfaction Problems (CSPs) can be investigated by means of random ensembles of instances. The latter exhibit many threshold phenomena besides their satisfiability phase transition, in particular a clustering or dynamic phase transition (related to the tree reconstruction problem) at which their typical solutions shatter into disconnected components. In this paper we study the evolution of this phenomenon under a bias that breaks the uniformity among solutions of one CSP instance, concentrating on the bicoloring of $k$-uniform random hypergraphs. We show that for small $k$ the clustering transition can be delayed in this way to higher density of constraints, and that this strategy has a positive impact on the performances of Simulated Annealing algorithms. We characterize the modest gain that can be expected in the large $k$ limit from the simple implementation of the biasing idea studied here. This paper contains also a contribution of a more methodological nature, made of a review and extension of the methods to determine numerically the discontinuous dynamic transition threshold.

\end{abstract}

\maketitle

\tableofcontents

\section{Introduction}

In Constraint Satisfaction Problems (CSPs) a set of $N$ discrete-valued variables are subjected to $M$ constraints; the decision version of the problem consists in answering yes or no to the question ``is there an assignment of the variables that satisfies all the constraints simultaneously ?'' Computational complexity theory~\cite{GareyJohnson79,Papadimitriou94} classifies the difficulty of these problems according to the existence of efficient (polynomial time) algorithms able to solve all their possible instances.

Besides this worst-case point of view an important effort has been devoted to the characterization of the ``typical'' difficulty of CSPs, where typical is defined with respect to a random ensemble of instances, a property being considered typical if it occurs with a probability going to one in the thermodynamic (large size) limit. Many random CSPs have been studied, notably $k$-SAT and $q$-COL; in this paper we will focus on the bicoloring of random hypergraphs (related to $k$-NAESAT), in which the $N$ variables can each take two values (colors), each of the $M$ constraints is generated by choosing uniformly at random a $k$-uplet of distinct variables, and impose that both colors appear in the configuration of these $k$ variables (i.e. they forbid monochromatic neighborhoods). The interactions induced by these constraints have thus the structure of an Erd\H os-R\'enyi random hypergraph, and the thermodynamic limit corresponds to $N,M \to \infty$ with fixed ratio $\a = M/N$ and arity parameter $k$. The random bicoloring problem exhibits the same rich phenomenology as $k$-SAT and $q$-COL, while being slightly simpler from a technical point of view.

Random CSPs exhibit threshold phenomena in the large size limit, the probability of some properties jumping abruptly from 1 to 0 as a function of the control parameter $\a$. The most prominent of these phase transitions occur at the satisfiability threshold $\asat(k)$: for $\a < \asat(k)$ typical instances are satisfiable, i.e. admit configurations of variables that satisfy all constraints simultaneously, while for $\a > \asat(k)$ a random instance is typically unsatisfiable.

Random CSPs bear a formal similarity with mean-field spin-glasses, the interactions induced by the constraints being of a frustrating nature while lacking a finite-dimensional structure thanks to the randomness in their construction. This analogy has been exploited in depth through the application of methods first developed in the context of statistical mechanics of disordered sytems, namely the replica and cavity method, to random CSPs~\cite{MonassonZecchina99b,BiroliMonasson00,MezardParisi02,MertensMezard06,krzakala2007gibbs}. This line of study has provided predictions of $\asat(k)$ for many models, but also unveiled many other phase transitions for the structure of the set of solutions in the satisfiable phase, and led to the proposal of new algorithms that exploit this detailed picture of the solution space. Many of these (heuristic) predictions have been confirmed rigorously later on~\cite{AchlioptasRicci06,AchlioptasCoja-Oghlan08,molloy_col_freezing,ding2014proof}.

In this paper we will pay a particular attention to the dynamic phase transition that occurs at some threshold $\ad(k) < \asat(k)$, which is also known as the clustering and reconstruction transition. This transition can indeed be described from various perspectives; looking at the set of solutions of typical instances, $\ad(k)$ separates a regime where this set is rather well-connected, any solution can be reached from any other one via nearby intermediate solutions, while for $\a > \ad(k)$ the solutions break apart into distinct clusters (or pure states) which are internally well-connected, but separated one from the other by regions of the configuration space void from solutions. This transition marks also the birth of a specific type of long-range correlations between variables, known as point-to-set correlations, which implies the solvability of an information-theoretic problem called tree reconstruction~\cite{MoPe03}. These correlations forbid in turn the rapid equilibration of the stochastic processes that respect the detailed balance condition~\cite{MontanariSemerjian06b}, hence the name dynamic given to $\ad(k)$. As a matter of fact the static properties of the model are smooth at $\ad(k)$ and are only sensitive to a further condensation transition $\ac(k)$ that affects the number of dominant clusters~\cite{krzakala2007gibbs}.

Despite the rather detailed picture of the set of solutions of random CSPs sketched above, many questions remain open, in particular concerning the behavior of algorithms that attempts at finding such solutions. These algorithms can be of very different types, proceeding through a local search in the space of configurations~\cite{SelmanKautz94,SemerjianMonasson03,ArdeliusAurell06,AlavaArdelius07}, or assigning variables one by one, according to either simple heuristics~\cite{FrancoPaull83,transition_lb,Achltcs,Co01} or detailed information provided by message passing algorithms (Belief or Survey Propagation) inspired by statistical mechanics considerations~\cite{MezardParisi02,BraunsteinMezard05,MaPaRi15,Allerton,RiSe09}. These dynamics are most of the time ``out-of-equilibrium'' processes, either because their definition breaks explicitly the detailed balance (reversibility) conditions, or because they will not be able to remain equilibrated during their evolution on reasonably accessible time scales. The great diversity of these algorithms and their out-of-equilibrium character makes very challenging the attempts to characterize the putative algorithmic barrier $\aalgo(k)$ above which no algorithm is able to find a solution in polynomial time for a typical random instance (assuming of course P$\neq$NP), and to relate it to the structural phase transitions undergone by the set of solutions. For small values of $k$ numerical experiments suggest that carefully designed local search algorithms~\cite{SelmanKautz94,ArdeliusAurell06,AlavaArdelius07} or Survey Propagation implementations~\cite{MaPaRi15} work (i.e. find solutions in polynomial time) up to densities very close to the satisfiability threshold, thus setting lowerbounds on $\aalgo(k)$ almost coinciding with the upperbound $\asat(k)$. The situation is quite different in the large $k$ limit, that allows for some analytical simplifications. Let us recall that in this limit the satisfiability threshold occurs at $\asat(k) \sim 2^{k-1} \ln 2$, while the asymptotic expansion of the clustering threshold is $\ad(k) \sim 2^{k-1} (\ln k)/k$ (the quantitative statements are made here for the hypergraph bicoloring problem, but the qualitative picture is the same for many random CSPs). Numerical experiments cannot access directly the large $k$ limit, but simple enough algorithms can be studied analytically for all $k$; sequential assignment algorithms that use simple heuristics to guide their choices can be described in terms of differential equations~\cite{FrancoPaull83,transition_lb,Achltcs,Co01} and shown to work in polynomial time up to densities of the order $\asat(k)/k$, with a constant prefactor depending on the heuristic chosen. This scaling was improved in~\cite{Amin_algo}, where an algorithm was shown to work up to densities of constraints coinciding at leading order with $\ad(k)$. This leaves a multiplicative gap of order $k$ (neglecting the sub-dominant logarithmic correction) from the satisfiability transition, hence a wide range of parameters $\a$ for which typical instances are known to have solutions, yet no provably efficient algorithm is known at present to find them. Some negative results have also been obtained, \cite{GaSu14} proved that no ``local'' algorithm (in a precise sense) can find solutions in polynomial time for densities $\a$ larger than $\ad(k) \ln k$ (asymptotically at large $k$), see also~\cite{Allerton,RiSe09,Coja12,Hetterich,CoFr14,CoHaHe17} for other analysis of some specific algorithms.

A further structural property of the set of solutions of a CSP has been studied under the name of ``frozen variables''. Roughly speaking, a frozen variable is a variable that takes the same value in all the solutions of a cluster~\cite{ZdeborovaKrzakala07,MontanariRicci08}, or equivalently a variable whose flip from a solution to another one requires rearranging an extensive number of other variables~\cite{MontanariSemerjian06,Semerjian07}. Two distinct phase transitions can then be defined: the so-called rigidity transition $\ar$ marks the appearance of a positive fraction of frozen variables in typical solutions of a random CSP, while above the freezing transition $\af$ all solutions have this property. In the intermediate regime $[\ar,\af]$ unfrozen solutions still exist but are exponentially less numerous than the typical, frozen ones. The determination of $\ar$ is relatively easy, as it concerns a property of the typical solutions, and in the large $k$ limit the rigidity threshold $\ar$ is very close to the dynamic one $\ad$~\cite{molloy_col_freezing,molloy_csp_freezing,Sly08,MoReTe11_recclus,SlyZhang16}. On the contrary the freezing transition at $\af$ is intrisically a large deviation result, and its determination is quite challenging even for heuristic statistical mechanics methods. It has been predicted in~\cite{BrDaSeZd16} to occur close to the satisfiability threshold for large $k$ (more precisely at $\af \sim \asat/2$), in line with the rigorous result $\af \le (4/5) \asat$ from~\cite{AchlioptasRicci06,achlioptas2009random}. 

Frozen variables induce a very strong form of correlations, and it seems impossible to construct in polynomial time a solution containing an extensive number of frozen variables, that must all be set in a consistent way (except in the very special case of XORSAT due to the linear structure of its constraints)~\cite{ZdeborovaKrzakala07,KrzakalaZdeborova07b}. As a matter of fact the best performing solving algorithms in the range $\a\in[\ar,\af]$ do not find typical solutions with frozen variables, but atypical solution without frozen variables~\cite{ManevaMossel05,BraunsteinZecchina04,DallAstaRamezanpour08,MaPaRi15}. It is thus natural to conjecture that $\af$ is an upperbound on $\aalgo$, but this still leaves a very wide gap between the threshold of the best known algorithms and $\af$.

In this paper we study probability measures over the set of solutions of random CSPs, for which not all solutions are equally probable but have a weight depending on their tendency to form frozen variables. The same perspective has been used in a few recent works~\cite{BrDaSeZd16,BaInLuSaZe15_long,BaBo16,MaSeSeZa18}, namely to consider a biased probability measure over the set of solutions of a random CSP (in~\cite{BaInLuSaZe15_long,BaBo16} the local entropy, or density of solutions in configuration space, is used to weight differently the solutions, in~\cite{BrDaSeZd16} this role is played by the number of frozen variables, while in~\cite{MaSeSeZa18} hard sphere particles are considered as a CSP, with a bias due to an additional pairwise interaction between particles). Indeed the structural phase transitions mentioned above concern the uniform measure over solutions, and it has been demonstrated that the threshold for properties that are typical in the uniform ensemble (in particular the existence of frozen variables) can be significantly moved by an appropriate bias~\cite{BrDaSeZd16} (biased measures with weights depending on the number of satisfying literals in a clause were also studied in~\cite{KrMeZd14} but with the different goal of allowing for a quiet planting of solutions in $k$-SAT instances). This opens some hope to diminish the algorithmic gap, by giving more weight to solutions that are ``easier'' to find, for instance because they contain less frozen variables, and to turn atypical properties of the uniform measure into typical ones of the biased measure.

We will concentrate in particular on the increase of the dynamic threshold $\ad$ that results from a well-chosen bias between solutions. The algorithmic motivation for the study of this threshold comes from the Simulated Annealing~\cite{KirkpatrickGelatt83} procedure: below $\ad$ a Markov Chain reversible with respect to a finite-temperature probability distribution should be able to equilibrate in polynomial time, hence to find solutions (non-uniformly) once the temperature is lowered slowly enough (if there is no reentrance in temperature). We shall implement here a relatively simple version of this idea, introducing soft interactions between the variables inside a constraint of the original CSP. We will demonstrate that for small values of $k$ this allows indeed to increase the dynamic threshold $\ad$, and check that this effect improves the performance of Simulated Annealing. On a more negative side we shall see that this simple implementation of the idea is not powerful enough to modify the leading order of the large $k$ algorithmic gap, but motivates the study of more elaborate biasing strategies.

The rest of the paper is organized as follows. In Section~\ref{sec_def_cavity} we define more precisely the model under study and present the equations that describe its behavior in the framework of the cavity method from statistical mechanics. Section~\ref{sec_methodology} is of a more methodological nature, and contains a review and extension of the numerical procedures to determine accurately the dynamic threshold in models exhibiting a discontinuous 1RSB transition. This Section can be skipped by a reader mostly interested in the results we obtained, which are presented in Sec.~\ref{sec_results_cavity} for what concerns the phase diagrams predicted by the cavity method, in particular the increase of the dynamic threshold of the biased measure with respect to the uniform one, and in Sec.~\ref{sec_results_sa} for numerical simulations on finite size samples via Simulated Annealing. We study in Section~\ref{sec_largek} the limit of large $k$ and derive an asymptotic upperbound on the possible increase of the dynamic threshold, before presenting our conclusions and perspectives in Sec.~\ref{sec_conclu}.


\section{Definition of the model and statistical mechanics formalism}
\label{sec_def_cavity}

\subsection{Definition of the model}

\begin{figure}[hbtp]
\begin{center}
\includegraphics[scale=0.5]{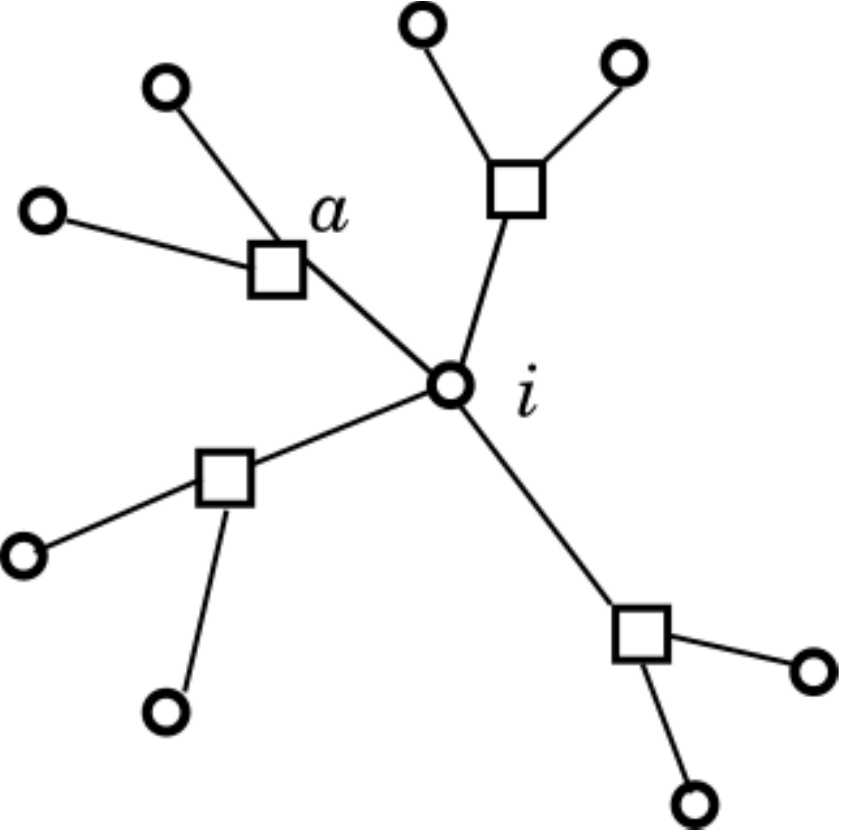}
\caption{Example of a factor graph with $k=3$, $N=8$, $M=4$: vertices are represented by circles, and hyperedges by squares. An edge is drawn between the $a$-th hyperedge and the $i$-th vertex if and only if $i \in \da$.}
\label{factorgraph}
\end{center}
\end{figure}

We will consider in this paper the $k$-uniform hypergraph bicoloring problem (related to the $k$-NAESAT problem)~\cite{CastellaniNapolano03,DallAstaRamezanpour08,CojaZdeb12,BaCoRa14,AchlioptasMoore06,CoPa12,DiSlSu13_naeksat,AchlioptasCoja-Oghlan08,BrDaSeZd16}. An instance of this constraint satisfaction problem (CSP) is defined by an hypergraph $G=(V,E)$ where $V$ is a set of $N$ vertices, and $E$ a set of $M$ hyperedges, each of them containing $k$ vertices (see Fig.~\ref{factorgraph} for a representation of $G$ as a factor graph). We shall denote $\da$ the set of vertices contained in the $a$-th hyperedge, and similarly $\di$ the set of hyperedges adjacent to the $i$-th vertex. The variables of this CSP are $N$ Boolean variables, represented as Ising spins $\s_i \in \{-1,1\}$, living on the vertices of $G$. We will denote $\us=(\s_1,\dots,\s_N)$ the global configuration of the variables, and $\us_S=\{\s_i\}_{i \in S}$ the configuration of the variables in a subset $S$ of the vertices. A constraint (or clause) is associated to each hyperedge $a \in E$; the $a$-th constraint is satisfied by the configuration $\us$ if and only if there is at least one $+1$ and one $-1$ among the $k$ variables of $\us_\da$, in such a way that the edge is not monochromatic (i.e. that not all variables adjacent to it are equal). A configuration $\us$ is called a solution of the CSP if it satisfies the $M$ constraints simultaneously.

A convenient way to study the set of solutions $S(G)$ of a given instance (assuming it is non-empty) is to consider the uniform probability measure over the solutions,
\begin{align}
\label{measure}
\mu(\us) = \frac{1}{Z(G)}\prod_{a=1}^{M}\w(\us_{\da}) \ ,
\end{align}
where the normalization factor $Z(G)$ (also called partition function) counts the number of solutions, and $\w(\s_1,\dots,\s_k)$ is the indicator function of the event ``the $k$ variables $\s_1,\dots,\s_k$ are not all equal''; the $a$-th term in this product is thus equal to $1$ if the $a$-th constraint is satisfied, and to $0$ otherwise.

We shall actually study in this paper a measure of the form (\ref{measure}), but with a more generic form for the function $\w(\s_1,\dots,\s_k)$. We will assume that $\w$ is invariant under all permutation of its $k$ arguments; as the latter are binary variables $\w$ can only depend on the number $p$ of $-1$ among its arguments, and we will denote $\w_p \ge 0$ the value it then assumes. This translates into the formula
\begin{align}
\label{generalConstraint}
\w(\s_1,\dots,\s_k) = \w_p \qquad \text{if} \qquad \sum_{i=1}^k\s_i=k-2p \ .
\end{align}
The uniform measure over the solutions of the bicoloring problem is recovered for the choice $\w_0=\w_k=0$, $\w_1=\dots=\w_{k-1}=1$. If one chooses instead $\w_p$ to depend on $p$ for $p \in \{1,\dots,k-1\}$, while keeping $\w_0=\w_k=0$, one obtains a probability measure $\mu$ that is still supported solely on the proper bicolorings of $G$, but is not uniform anymore. As explained in the introduction our goal in this paper is to explore the properties of $\mu$ that arises from this bias between solutions of the CSP. We will sometimes relax the constraint $\w_0=\w_k=0$, to model the effect of a positive temperature that allows some constraints to be violated. We will in any case always assume that $\w_p = \w_{k-p}$: this ensures that the global spin-flip symmetry $\mu(-\us)=\mu(\us)$ (which is indeed a property of the set of proper bicolorings) is preserved.

We shall characterize the properties of $\mu(\us)$ for ``typical'' hypergraphs, by studying random instances; even if most of the approach can be generalized to more generic random ensembles, for concreteness we will concentrate on Erd\H os-R\'enyi (ER) random hypergraphs. An instance of this ensemble is generated by drawing, independently for each of its hyperedges $a=1,\dots, M$, the adjacent vertices $\da$ as an uniformly random $k$-uplet among the $\binom{N}{k}$ possible ones. We will be interested in the large size (thermodynamic) limit, in which both $N$ and $M$ go to infinity at a fixed ratio $\a=M/N$. We recall that in this limit such random hypergraphs converge locally to hypertrees: the neighborhood within a fixed distance around an uniformly chosen vertex is, with a probability going to 1 in the large size limit, acyclic. More precisely, the local limit tree is a Galton-Watson branching process, in which the law $p_d$ for the degree $|\partial i|$ of an uniformly chosen vertex $i$ is the Poisson distribution of average $\a k$. Thanks to the properties of the Poisson law this is also the probability that an uniformly chosen vertex among an uniformly chosen hyperedge has degree $d+1$ (i.e. the offspring probability in the Galton-Watson tree is also given by $p_d$).

\subsection{BP equations and Bethe free-energy}

In order to determine the typical properties of the measure $\mu(\us)$, and of the free-entropy density $(\ln Z)/N$, we shall exploit the cavity method~\cite{MezardParisi01,MezardParisi03,MezardMontanari07}, a formalism particularly efficient for interacting particle models on sparse random structures. As recalled above these structures are locally tree-like; the first step of the cavity method amounts in consequence to study such models on finite trees. In that case one can exploit the recursive nature of trees to derive an exact description of $\mu(\us)$ in terms of its marginals, from which also $\ln Z$ can be expressed.

More precisely, let us introduce the messages $\eta_{i \to a}$ and $\heta_{a \to i}$ on each edge $(i,a)$ of the factor graph, that are the marginal probability laws of $\s_i$ in amputated graphs where some interactions are discarded; $\eta_{i \to a}$ is the marginal of $\s_i$ in the factor graph where one removes the hyperedge $a$, and $\heta_{a \to i}$ is the marginal of $\s_i$ in the factor graph where one removes all the hyperedges in $\dima$. Removing an interaction in a tree breaks it into independent subtrees, which allows to write recursive equations between these messages:
\begin{align}
\eta_{i\to a}(\s_i) &= \frac{1}{z_0^{ia}}\prod_{b \in \dima} \heta_{b\to i}(\s_i) \ , \label{eq_BP_itoa} \\
\heta_{a\to i}(\s_i) &= \frac{1}{\hz_0^{ai}} \sum_{\us_{\dami}} \w(\us_{\da}) \prod_{j \in \dami} \eta_{j\to a}(\s_j) \ ,
\end{align}
where the constants $z_0^{ia}$ and $\hz_0^{ai}$ are normalizing factors. These equations are valid for any (discrete) domain of the spins $\s_i$; as we are studying the case where $\s_i=\pm 1$, we can parametrize the probability laws $\eta_{i \to a}$ and $\heta_{a \to i}$ by their mean values,  defining 
$\eta_{i \to a}(\s_i)=(1+h_{i \to a} \s_i)/2$ and $\heta_{a \to i}(\s_i)=(1+u_{a \to i}\s_i)/2$, with $h_{i \to a},u_{a \to i} \in [-1,1]$. The recursive equations can be rewritten with this parametrization as
\beq
h_{i\to a} = f(\{u_{b\to i}\}_{b \in \dima}) \ , \qquad
u_{a\to i} = g(\{h_{j\to a}\}_{j \in \dami}) \ ,
\label{eq_BP_hu}
\eeq
where the functions $f$ and $g$ read explicitly
\bea
f(u_1,\dots,u_d) &=& \frac{\underset{i=1}{\overset{d}{\prod}} (1+u_i) - \underset{i=1}{\overset{d}{\prod}} (1-u_i) }{z_0(u_1,\dots,u_d)} \ , \qquad
z_0(u_1,\dots,u_d) = \underset{i=1}{\overset{d}{\prod}} (1+u_i) + \underset{i=1}{\overset{d}{\prod}} (1-u_i) \ ,
\label{eq_f} \\
g(h_1,\dots,h_{k-1}) &=& \frac{\underset{\s_1,\dots,\s_k}{\sum}\w(\s_1,\dots,\s_k) \s_k \underset{i=1}{\overset{k-1}{\prod}} (1+h_i \s_i) }{\hz_0(h_1,\dots,h_{k-1}) } \ , \ \ 
\hz_0(h_1,\dots,h_{k-1}) = \underset{\s_1,\dots,\s_k}{\sum}\w(\s_1,\dots,\s_k) \underset{i=1}{\overset{k-1}{\prod}} (1+h_i \s_i) 
\label{eq_g}
\eea
The function $f$ has been written here for a vertex of degree $d+1$.

On a tree the equations (\ref{eq_BP_hu}) admit a single solution, that can be found by iterating (\ref{eq_BP_hu}) from the leaves towards the interior of the graph. Once this solution is determined one can easily compute the marginal probability of $\s_i$ under $\mu$ (using the formula in (\ref{eq_BP_itoa}) with all messages incoming onto $i$), as well as the partition function $Z(G)$:
\begin{align}
\label{BetheFreeEnergy}
\frac{1}{N}\ln Z(G) = 
\frac{1}{N}\sum_{i=1}^{N} \ln \Z_0^{\rm v} (\{u_{a \to i}\}_{a \in \di})
+ \frac{1}{N}\sum_{a=1}^{M} \ln \Z_0^{\rm c} (\{h_{i \to a}\}_{i \in \da})
- \frac{1}{N}\sum_{(i,a)} \ln \Z_0^{\rm e}(h_{i\to a}, u_{a\to i}) \ ,
\end{align}
where the last sum runs over the edges of the factor graph, and the local partition functions are defined as:
\begin{align}
\Z_0^{\rm v}(u_1,\dots,u_d) &= \sum_{\s} \prod_{i=1}^d \left(\frac{1+\s u_i}{2}\right) \ , \\
\Z_0^{\rm c}(h_1,\dots,h_k) &= \sum_{\s_1,\dots,\s_k} \w(\s_1,\dots,\s_k) \prod_{i=1}^k \left(\frac{1+\s_i h_i}{2}\right) \ , \\
\Z_0^{\rm e}(h,u) &= \sum_{\s} \left(\frac{1+\s h}{2}\right)\left(\frac{1+\s u}{2}\right) \ .
\end{align}

The recursive equations (\ref{eq_BP_hu}) and the associated expression (\ref{BetheFreeEnergy}) of the log partition function (a.k.a. free-entropy density) are exact if the factor graph is a finite tree; they can however be used heuristically on any factor graph, even in the presence of cycles. The resulting message passing iterative algorithm that searches for fixed points of (\ref{eq_BP_hu}) is then known as Belief Propagation (BP)~\cite{Pearl88,KschischangFrey01,YedidiaFreeman03}, and the expression (\ref{BetheFreeEnergy}) is called the Bethe-Peierls approximation for the free-entropy. As typical random graphs are locally tree-like one can reasonably hope that this approach is asymptotically exact in the thermodynamic limit for typical instances. This is indeed the basis of the cavity method, with nevertheless some subtleties in the treatment of the long loops that are present in random graphs.

\subsection{Replica symmetric cavity method}

The aim of the cavity method is to characterize the properties of the measure $\mu$ defined in (\ref{measure}), for typical random graphs in the thermodynamic limit, in particular the value of the quenched free-entropy density
\beq
\Phi(\a,\{\w_p\}) = \lim_{N \to \infty} \frac{1}{N} \E[\ln Z(G)] \ ,
\label{eq_Phi}
\eeq
around which $(\ln Z(G))/N$ concentrates thanks to the self-averaging phenomenon.

There are different versions of the cavity method, that rely on self-consistent hypotheses of various complexity on the effect of the long loops. In the simplest version, called replica symmetric (RS), one assumes a fast decay of the correlations between distant variables in the measure $\mu(\us)$, in such a way that the BP equations converge to a unique fixed point on a typical large instance, and that the measure is well described by the locally tree-like approximation. Consider then an uniformly chosen directed edge $i \to a$ in a random hypergraph, and call $\cP^{RS}$ the probability law of the fixed-point message $h_{i \to a}$ thus obtained. We shall denote similarly $\hcP^{RS}$ the probability of the messages $u_{a \to i}$; within the decorrelation hypothesis of the RS cavity method the incoming messages on a given vertex (resp. hyperedge) are i.i.d. with the law $\hcP^{RS}$ (resp. $\cP^{RS}$). For this to be self-consistent the recursion equations (\ref{eq_BP_hu}) must become equalities in distribution, or in other words the laws $\cP^{RS}$ and $\hcP^{RS}$ must obey the following equations:
\begin{align}
\label{RSeqn}
\cP^{RS}(h) &= \sum_{d=0}^{\infty} p_d\int \left(\prod_{i=1}^{d} \dd u_i \hcP^{RS}(u_i) \right) 
\, \d(h-f(u_1,\dots,u_d)) \ , \\
\nonumber
\hcP^{RS}(u) &= \int \left(\prod_{i=1}^{k-1} \dd h_i \cP^{RS}(h_i) \right) \, \d(u-g(h_1,\dots,h_{k-1})) \ . \\
\end{align}
The RS cavity prediction for the free-entropy (\ref{eq_Phi}) is then obtained by averaging the Bethe expression (\ref{BetheFreeEnergy}) with respect to the message distributions $\cP^{RS}$ and $\hcP^{RS}$, which yields:
\bea
\Phi^{RS} &=& \sum_{d=0}^{\infty} p_d \int \left( \prod_{i=1}^{d} \dd u_i \hcP^{RS}(u_i) \right) \ln \Z_0^{\rm v}(u_1,\dots,u_d) + \a \int \left( \prod_{i=1}^{k} \dd h_i \cP^{RS}(h_i) \right) \ln \Z_0^{\rm c}(h_1,\dots,h_k) \nonumber
\\ && -\a k \int \dd h \dd u \cP^{RS}(h) \hcP^{RS}(u) \ln \Z_0^{\rm e}(h,u) \ .
\label{eq_Phi_RS}
\eea

As we assume that $\w_p=\w_{k-p}$, i.e. that the model is invariant under the global spin-flip symmetry, the RS equations admit as a solution the uniform distributions $\cP^{RS}(h) = \d(h)$, $\hcP^{RS}(u) = \d(u)$. For frustrated models with an antiferromagnetic character this is the relevant solution (a spontaneous breaking of the symmetry between positive and negative spins would induce an alternating order of the N\'eel type, that is admissible on a tree but incompatible with random graphs that contain cycles of odd lengths), hence one obtains explicity the value of the free-entropy by inserting the trivial solution of the RS equations into (\ref{eq_Phi_RS}):
\begin{align}
\label{quenchedRSenergy}
\Phi^{RS}(\a,\{\w_p\}) = \ln 2 + \a \ln \left(\frac{1}{2^k}\sum_{\s_1,\dots,\s_k} \w(\s_1,\dots,\s_k) \right)
= \ln 2 + \a \ln \left(\frac{1}{2^k}\sum_{p=0}^k \binom{k}{p} \w_p \right) \ .
\end{align}
Note that this expression actually coincides with the annealed (first moment) computation $\lim_{N \to \infty} (\ln \E [ Z(G)])/N$.

In the special case $\w_0=\w_k=0$, $\w_1=\dots=\w_{k-1}=1$, for which $\mu(\us)$ corresponds to the uniform measure over proper bicolorings, the partition function $Z(G)$ counts the number of solutions, hence the free-entropy $\ln Z$ is equal to the entropy of the uniform measure. The prediction of the RS cavity method is thus (using a subscript u for uniform):
\begin{align}
s^{RS}_{\rm u}(\a)=\Phi^{RS}_{\rm u}(\a) = \ln 2 + \a \ln \left(1-\frac{1}{2^{k-1}}\right) \ .
\end{align}

For a generic choice of parameters $\{\w_p\}$ the free-entropy $\ln Z$ differs from the (Shannon) entropy of the measure $\mu(\us)$. The latter can be obtained by a Legendre transform with respect to the parameters $\{\w_p\}$; one way to justify this statement is to remember that for a probability measure of the form (\ref{measure}) one has
\beq
S(\mu)= - \sum_\us \mu(\us) \ln \mu(\us) = \ln Z - \sum_{a=1}^M \sum_\us \mu(\us) \ln \w(\us_\da) \ .
\eeq
From the joint marginal of the variables around a constraint in the trivial RS solution, and from the RS free-entropy (\ref{quenchedRSenergy}) one thus obtains the RS prediction for the entropy density
\begin{align}
s^{RS}(\a,\{\w_p\}) = \ln 2 + \a \ln \left( \frac{1}{2^k}\sum_{p=0}^k \binom{k}{p} \w_p\right) - \a \ \frac{\underset{p=0}{\overset{k}{\sum}} \binom{k}{p} \w_p \ln \w_p}{\underset{p=0}{\overset{k}{\sum}} \binom{k}{p} \w_p}
\ .
\label{eq_sRS}
\end{align}
This quantity is a decreasing function of $\a$ and becomes negative for $\a > \a^{s=0}$, with
\beq
\a^{s=0}(\{\w_p\}) = \frac{\ln 2}{
\frac{\underset{p=0}{\overset{k}{\sum}} \binom{k}{p} \w_p \ln \w_p}{\underset{p=0}{\overset{k}{\sum}} \binom{k}{p} \w_p} - \ln \left( \frac{1}{2^k} \underset{p=0}{\overset{k}{\sum}} \binom{k}{p} \w_p\right)
} \ .
\label{eq_s0_generic}
\eeq
The negativity of the entropy for $\a > \a^{s=0}$ is a clear sign of the failure of the RS assumptions, the Shannon entropy of a discrete probability measure being non-negative.

\subsection{1RSB formalism}

\subsubsection{1RSB cavity equations}

The hypothesis underlying the RS cavity method must break down when the density of interactions per variable $\a$ becomes too large; a first hint of this phenomenon, called Replica Symmetry Breaking (RSB), is the negativity of the RS entropy at large enough $\a$, which is impossible for a system with discrete degrees of freedom. As a matter of fact RSB can occur before $\a^{s=0}$; increasing $\a$ above a certain threshold causes the appearance of long-range correlations between distant variables under the measure $\mu$, which contradicts the RS hypothesis. In such a case it becomes necessary to use more refined versions of the cavity method, that are able to deal with this RSB phenomenon~\cite{MezardParisi01}. At its first non-trivial level, called 1RSB for one step of RSB, the cavity method postulates the existence of a partition of the configuration space $\{-1,1\}^N$ into ``pure states'', or clusters, such that the restriction of the measure $\mu$ to a pure state has good decorrelation properties. This restricted measure can then be treated as the full measure in the RS cavity method, i.e. with BP equations to describe its marginal probabilities, and the Bethe free-entropy to compute its partition function.

To be more quantitative let us index with $\gamma$ the partition of the configuration space into clusters, and denote $Z_\gamma$ the contribution to the partition function of the $\gamma$-th cluster, as well as $\{u_{a \to i}^\gamma,h_{i \to a}^\gamma\}$ the solution of the BP equations that describe it. The 1RSB cavity method aims at computing the potential
\beq
\Phi_1(m) = \lim_{N \to \infty} \frac{1}{N} \ln \left(\sum_\gamma (Z_\gamma)^m \right) \ ,
\label{eq_def_Phi1}
\eeq
where the so-called Parisi parameter $m$ allows to weight differently the various pure states, according to their relative weights. This quantity contains precious informations about the pure-state decomposition; suppose indeed that, at the leading exponential order, there are $e^{N \Sigma(\phi)}$ pure states $\gamma$ with $Z_\gamma =e^{N \phi}$ (again neglecting sub-exponential corrections). The so-called complexity $\Sigma(\phi)$ plays thus the role of an entropy density, with pure states replacing usual configurations, and captures the RSB phenomenon quantitatively. The potential $\Phi_1(m)$ and the complexity $\Sigma(\phi)$ are Legendre transforms of each other~\cite{Monasson95}; evaluating (\ref{eq_def_Phi1}) via the Laplace method yields indeed
\beq
\Phi_1(m) = \sup_\phi \, [\Sigma(\phi) + m \phi] \ ,
\eeq
which can be inverted in terms of the conjugated parameter as
\beq
\Sigma(m) = \Phi_1(m)-m\frac{\dd}{\dd m}\Phi_1(m) \ .
\label{eq_Legendre_Sigma}
\eeq

In order to compute $\Phi_1$ one introduces, for a given sample and a given edge $(i,a)$ of the factor graph, two distributions $P_{i \to a}$ and $\hP_{a \to i}$, that encode the laws of $h_{i \to a}^\gamma$ and $u_{a \to i}^\gamma$ when the pure state $\gamma$ is chosen randomly with a probability proportional to $Z_\gamma^m$. These distributions are found to obey self-consistent equations of the form
\beq
P_{i\to a} = F(\{\hP_{b\to i}\}_{b \in \dima}) \ , \qquad
\hP_{a\to i} = G(\{P_{j\to a}\}_{j \in \dami}) \ ,
\eeq
where $P=F(\hP_1,\dots,\hP_d)$ is a shorthand for
\beq
P(h) = \frac{1}{z_1(\hP_1,\dots,\hP_d)} 
\int \left(\prod_{i=1}^d \dd u_i \hP_i(u_i)\right) \, \d(h-f(u_1,\dots,u_d)) \, z_0(u_1,\dots,u_d)^m \ ,
\label{eq_F}
\eeq
and $\hP=G(P_1,\dots,P_{k-1})$ means
\beq
\hP(u) = \frac{1}{\hz_1(P_1,\dots,P_{k-1})} \int \left(\prod_{i=1}^{k-1} \dd h_i P_i(h_i) \right) \,  \d(u-g(h_1,\dots,h_{k-1})) \, \hz_0(h_1,\dots,h_{k-1})^m \ ;
\label{eq_G}
\eeq
the functions $f$, $z_0$, and $g$, $\hz_0$ were defined in Eqs.~(\ref{eq_f}) and (\ref{eq_g}), respectively, and the factors $z_1$ and $\hz_1$ ensure the normalization of the distributions $P(h)$ and $\hP(u)$.

In order to deal with random hypergraphs one introduces the probability distributions over the 1RSB messages $\cP^{1RSB}(P)$ and $\hcP^{1RSB}(\hP)$ that obeys the consistency relations similar to (\ref{RSeqn}),
\begin{align}
\cP^{1RSB}(P) &= \sum_{d=0}^{\infty} p_d \int \left(\prod_{i=1}^d \dd \hP_i \, \hcP^{1RSB}(\hP_i) \right) \, \d[P-F(\hP_1,\dots,\hP_d)] \ , \label{eq_1RSB_cavity}
\\
\nonumber
\hcP^{1RSB}(\hP) &= \int \left( \prod_{i=1}^{k-1} \dd P_i \, \cP^{1RSB}(P_i) \right) \, \d[\hP-G(P_1,\dots,P_{k-1})] \ .
\end{align}
The 1RSB potential for typical random hypergraphs can then be computed from the solution of these equations as
\bea
\Phi_1(m) &=& \sum_{d=0}^{\infty} p_d \int \left(\prod_{i=1}^d \dd\hP_i \hcP^{1RSB}(\hP_i) \right) \ln \Z_1^{\rm v}(\hP_1,\dots,\hP_d) 
+ \a \int \left(\prod_{i=1}^k \dd P_i \cP^{1RSB}(P_i)\right) \ln \Z_1^{\rm c}(P_1,\dots,P_k) \nonumber \\
&& -\a k \int \dd P \dd \hP \cP^{1RSB}(P)\hcP^{1RSB}(\hP) \ln \Z_1^{\rm e}(P,\hP) \ ,
\label{free_energy1RSB}
\eea
with:
\bea
\Z_1^{\rm v}(\hP_1,\dots,\hP_d)  &=& \int \left(\prod_{i=1}^d \dd u_i \hP_i(u_i)\right) \,  (\Z_0^{\rm v}(u_1,\dots,u_d))^m \ , \\
\Z_1^{\rm c}(P_1,\dots,P_k) &=& \int \left(\prod_{i=1}^k \dd h_i P_i(h_i)\right) (\Z_0^{\rm c}(h_1,\dots,h_k))^m \ , \\
\Z_1^{\rm e}(P,\hP) &=& \int \dd h \dd u P(h) \hP(u) (\Z_0^{\rm e}(h,u))^m \ .
\eea
Finally the 1RSB prediction for the free-entropy is
\beq
\Phi^{1RSB} = \inf_{m \in [0,1]} \frac{\Phi_1(m)}{m} \ .
\label{eq_Phi1RSB}
\eeq

Note that the 1RSB equations always admit the RS solution as a special case, when the distributions $P$ in the support of $\cP^{1RSB}$ are Dirac measures. In most models this trivial solution of the 1RSB equations is the only one at small values of $\a$; then $\Phi_1(m)=m\Phi^{RS}$, and the thermodynamic prediction of the RS and 1RSB versions of the cavity method coincides. Increasing the number of constraints of the system non-trivial solutions of the 1RSB equations can appear; the dynamic threshold $\ad$ is defined as the smallest value of $\a$ for which the 1RSB equations with $m=1$ admit a solution distinct from the RS one. A further distinction has then to be made: if the associated complexity $\Sigma(m=1)$ is positive the extremum in (\ref{eq_Phi1RSB}) is reached for $m=1$ and $\Phi^{1RSB}=\Phi^{RS}$. In such a ``dynamic 1RSB'' situation the typical configurations of the Gibbs measure are supported on an exponentially large number of pure states, in such a way that the total free-entropy (or any correlation functions between a finite number of variables) is unable to detect the difference with a RS situation. On the contrary when $\Sigma(m=1)<0$ the extremum in (\ref{eq_Phi1RSB}) selects a non-trivial value $m_{\rm s}<1$ of the Parisi parameter, the Gibbs measure condenses on a sub-exponential number of clusters, and correlations between finite sets of variables unveil the RSB phenomenon. One calls condensation threhold $\ac$ the smallest value of $\a$ for which a solution of the 1RSB equations with $\Sigma(m=1)<0$ exists, which corresponds to a point of non-analyticity of the free-entropy density.

\subsubsection{Simplifications for $m=1$}
\label{sec_simplifications_m1}

The complete 1RSB equations have a rather intricate structure, as they are self-consistent equations for probability distributions over probability distributions, which make in particular their numerical resolution rather cumbersome. Fortunately for special values of the parameter $m$ (i.e. $m=0$ and $m=1$) they can be largely simplified. We shall sketch here this simplification procedure for $m=1$, which as explained above is the important one for the determination of the dynamic and condensation phase transitions; for further details the reader is referred to~\cite{MontanariRicci08} where the simplifications are explained in more details and in a general setting, and to~\cite{MezardMontanari06} where their connections with the tree reconstruction problem are discussed.

The crucial technical property of the equations that opens the door to simplifications at $m=1$ is the fact that, for this value, the normalization constant $z_1$ in (\ref{eq_F}) does not depend on the whole distributions $\hP_1$, \dots, $\hP_d$, but only on their average values $\int \dd u_i \hP_i(u_i) u_i$ (a similar statement holds for $\hz_1$ in (\ref{eq_G})). Conditional on this average values $F$ is thus a multilinear function of its arguments, the equations (\ref{eq_1RSB_cavity}) can then be averaged and closed on the mean distributions $Q$ and $\hQ$ defined as:
\begin{align}
Q &= \int \dd P \, \cP^{1RSB}(P) \, P \ , \qquad 
\hQ = \int \dd \hP \, \hcP^{1RSB}(\hP)\, \hP \ ,
\end{align}
which must be symmetric probability laws (i.e. $Q(h)=Q(-h)$ and $\hQ(u)=\hQ(-u)$) for the global spin-flip symmetry to be preserved. These two quantities are solutions of
\bea
Q(h) &=& \sum_{d=0}^{\infty} p_d \int \left( \prod_{i=1}^{d} \dd u_i \hQ(u_i) \right) \, \d(h-f(u_1,\dots,u_d)) \frac{z_0(u_1,\dots,u_d)}{z_0(0,\dots,0)} \ , \\
\hQ(u) &=& \int \left(\prod_{i=1}^{k-1} \dd h_i Q(h_i) \right) 
\d(u-g( h_1,\dots, h_{k-1})) \frac{\hz_0(h_1,\dots, h_{k-1})}{\hz_0(0,\dots,0)}
\ .
\eea
These equations are definitely simpler than the full 1RSB equations, as they bear on probability distributions instead of distributions of distributions; they have however one inconvenient feature, in particular for their numerical resolution, namely the reweighting terms $z_0$ and $\hz_0$ which prevents their direct interpretation as recursive distributional equations. To get around this difficulty we shall define, for $\s=\pm 1$, the distributions $Q_{\s}(h)=(1+h\s)Q(h)$ and $\hQ_{\s}(u)=(1+u\s)\hQ(u)$. Thanks to the symmetry of $Q$ and $\hQ$ these are well-normalized, and are related to the original distributions by $Q(h)=(Q_+(h)+Q_-(h))/2$. One can then show that they obey the following equations,
\begin{align}
Q_{\s}(h) &= \sum_{d=0}^{\infty} p_d \int \left( \prod_{i=1}^{d} \dd u_i \hQ_{\s}(u_i) \right) \, \d(h-f(u_1,\dots,u_d)) \ , \\
\nonumber
\hQ_{\s}(u) &= \sum_{\s_1,...,\s_{k-1}} \tp(\s_1,...,\s_{k-1}|\s) 
\int \left(\prod_{i=1}^{k-1} \dd h_i Q_{\s_i}(h_i) \right) 
\d(u-g( h_1,\dots, h_{k-1})) \ , 
\end{align}
with the conditional probability distribution:
\begin{align}
\tp(\s_1, \dots ,\s_{k-1}|\s) = \frac{\w(\s_1, \dots ,\s_{k-1},\s)}{\underset{\s'_1,\dots,\s'_{k-1}}{\sum} \w(\s'_1, \dots ,\s'_{k-1},\s) } \ .
\end{align}
Noting finally that the global flip-spin symmetry implies $Q_-(h)=Q_+(-h)$ and $\hQ_-(u)=\hQ_+(-u)$, one can write closed equations, without reweighting terms, on $Q_+$ and $\hQ_+$ solely:
\begin{align}
\label{1RSBeqnSimplifs}
Q_+^{(t+1)}(h) &= \sum_{d=0}^{\infty} p_d \int \left( \prod_{i=1}^{d} \dd u_i \hQ_+^{(t)}(u_i) \right) \, \d(h-f(u_1,\dots,u_d)) \ , \\
\nonumber
\hQ_+^{(t)}(u) &= \sum_{\s_1,...,\s_{k-1}} \tp(\s_1,...,\s_{k-1}|+) 
\int \left(\prod_{i=1}^{k-1} \dd h_i Q_+^{(t)}(h_i) \right) 
\d(u-g( \s_1 h_1,\dots, \s_{k-1} h_{k-1})) \ ,
\end{align}
where for future use we introduced discrete time indices $t$ on these distributions, and where
\begin{align}
\label{tildeP}
\tp(\s_1, \dots ,\s_{k-1}|+) = \frac{\w(\s_1, \dots ,\s_{k-1},+)}{\underset{\s'_1,\dots,\s'_{k-1}}{\sum} \w(\s'_1, \dots ,\s'_{k-1},+) } =
\frac{\overset{k-1}{\underset{p=0}{\sum}} \w_p \,  
\I\left[ \overset{k-1}{\underset{i=1}{\sum}} \s_i=k-1-2p\right]}{\overset{k-1}{\underset{p=0}{\sum}} \binom{k-1}{p} \w_p } \ .
\end{align}
Let us now turn to the thermodynamic computations; at $m=1$ one can check that $\Phi_1(m=1) = \Phi^{RS}$ as given in (\ref{quenchedRSenergy}). To compute the complexity at $m=1$ from (\ref{eq_Legendre_Sigma}) we need to take the derivative with respect to $m$ of $\Phi_1$ from (\ref{free_energy1RSB}). Because of its variational character ((\ref{free_energy1RSB}) is stationary with respect to variations of $\cP^{1RSB}$ and $\hcP^{1RSB}$ as long as the 1RSB cavity equations (\ref{eq_1RSB_cavity}) are fulfilled) only the explicit dependency on $m$ has to be differentiated. Doing the simplification at $m=1$ yields then an expression in terms of $Q_+$ and $\hQ_+$:
 \bea
\left. \frac{\dd}{\dd m}\Phi_1(m)\right|_{m=1} &=& 
\sum_{d=0}^{\infty} p_d \int \left(\prod_{i=1}^d \dd u_i \hQ_+(u_i)\right) 
\, \ln \Z_0^{\rm v}(u_1,\dots,u_d) \\
&& + \a \sum_{\s_1,\dots,\s_k} p(\s_1,\dots,\s_k) 
\int \left(\prod_{i=1}^k \dd h_i Q_+(h_i)\right) \,  \ln \Z_0^{\rm c}(\s_1 h_1,\dots, \s_k h_k) 
\\ \nonumber
&& - \a k \int \dd h \dd u \, Q_+(h) \hQ_+(u) \, \ln \Z_0^{\rm e}(h,u) \ ,
\eea
with
\beq
p(\s_1,\dots,\s_k) = \frac{\w (\s_1,\dots,\s_k)}{
\underset{\s'_1,\dots,\s'_k}{\sum}\w(\s'_1,\dots,\s'_k)} \ .
\eeq 

The form (\ref{1RSBeqnSimplifs}) of the 1RSB equations at $m=1$ is particularly convenient for an approximate numerical resolution with a procedure known as population dynamics~\cite{MezardParisi01}. Suppose indeed that $\hQ_+^{(t)}(u)$ can be approximated by the empirical distribution over a large sample of representative elements:
\beq
\hQ_+^{(t)}(u) \approx \frac{1}{{\cal N}} \sum_{i=1}^{{\cal N}} \delta(u - u_i^{(t)}) \ .
\eeq
Inserting this form in the r.h.s. of the first line of (\ref{1RSBeqnSimplifs}) yields an approximation for $Q_+^{(t)}(h)$ as
\beq
Q_+^{(t+1)}(h) \approx \frac{1}{{\cal N}} \sum_{i=1}^{{\cal N}} \delta(h - h_i^{(t+1)}) \ ,
\label{eq_def_popu_Qp}
\eeq
where each of the representants $h_i^{(t+1)}$ is constructed independently by drawing an integer $d$ from the law $p_d$, then $d$ indices $i_1,\dots, i_d$ uniformly at random in $\{1,\dots,{\cal N}\}$ and setting $h_i^{(t+1)}=g(u_{i_1}^{(t)},\dots,u_{i_d}^{(t)})$. The second line of (\ref{1RSBeqnSimplifs}) can similarly be translated into a rule for generating a population of fields $u^{(t)}$ from the population of fields $h^{(t)}$. The size ${\cal N}$ of the population used controls the computational cost of the procedure, and the numerical accuracy (in the limit ${\cal N}\to \infty$ empirical distributions converge to the exact ones). Iterating these two steps many times one converges to a fixed point solution of (\ref{1RSBeqnSimplifs}), which can either be the trivial one $Q_+(h)=\delta(h)$, $\hQ_+(u)=\delta(u)$, or a non-trivial solution. The dynamic transition $\ad$ is precisely the threshold that separates these two behaviors. It has been shown in~\cite{MezardMontanari06} that the equations (\ref{1RSBeqnSimplifs}) can also be interpreted in terms of an information-theoretic problem called tree reconstruction~\cite{MoPe03,Mossel01,JansonMossel04}. In the latter one considers a rooted Galton-Watson random tree and use it as an information channel, with spin variables located on the vertices. The value $\s$ of the spin at the root is broadcasted along the hyperedges, according to the free-boundary Gibbs measure with local interaction $\omega$. The question in this context is whether the observation of the variables at distance $t$ from the root contains a non-vanishing information on $\s$, in the limit $t\to\infty$; in which case one says that the problem is reconstructible. As $Q_+^{(t)}(h)$, after $t$ iterations of (\ref{1RSBeqnSimplifs}), is the distribution of the posterior magnetization of the root conditional on the observation of the variables at distance $t$, in the broadcast process with $\s=+1$, the reconstructibility of the tree problem is equivalent to the existence of a non-trivial solution of the 1RSB equations with $m=1$, and the dynamic threshold $\ad$ coincides with the reconstruction transition of the associated tree problem. Moreover this connection unveils a natural initial condition for the iterative numerical resolution of (\ref{1RSBeqnSimplifs}),
\beq
Q_+^{(t=0)}(h)=\delta(h-1) \ ,
\label{eq_initialcondition}
\eeq
corresponding to the perfect observation of the variables at distance $t$ from the root.

\subsubsection{The local instability of the RS solution (Kesten-Stigum bound)}
\label{sec_KS}

The properties of the measure $\mu(\us)$ change drastically when, upon increasing $\a$, one moves from the RS phase to the 1RSB phase. The transition between the two situations, at the critical (dynamic) density $\ad$, shows up as the appearance of a non-trivial solution of the 1RSB equations at $m=1$. Depending on the models this bifurcation, on which we shall come back in more details in Sec.~\ref{sec_methodology}, can occur in a continuous or a discontinuous way. We shall discuss here a bound on $\ad$, known as the Kesten-Stigum~\cite{KestenStigum66,MoPe03} transition in the context of the tree reconstruction problem, or as the de Almeida-Thouless~\cite{AlmeidaThouless78} transition for mean-field spin-glasses, that is tight for continuous bifurcations and that in any case provide an easy to compute analytical upper bound on $\ad$ (besides the bound $\ad < \a^{s=0}$ we already discussed).

Let us recall that the 1RSB equations (\ref{eq_1RSB_cavity}) always admit as a solution the RS distribution, with $\cP^{1RSB}(P) = \delta[P-P_{\rm triv}]$, $P_{\rm triv}(h)=\delta(h)$ (and similarly for $\hcP^{1RSB}(\hP)$). One way to test the existence of a non-trivial solution of the 1RSB equations is to investigate the local stability of the RS solution. Suppose indeed that the distributions in the support of $\cP^{1RSB}$ are close to $P_{\rm triv}$, i.e. that they are supported on small values of $h$. One can then expand (\ref{eq_F},\ref{eq_G}) and study the evolution of their average moments under the iterations of (\ref{eq_1RSB_cavity}); the global spin-flip symmetry imposes that the average mean remains zero. The first non-trivial moment is thus the variance, which is found after a short computation (see for instance~\cite{RiSeZd18} or App. B in~\cite{gabrie2017phase} for more details) to grow if and only if $\a k (k-1) \theta^2 > 1$, where $\theta$ is the derivative of $g(u_1,\dots,u_{k-1})$ with respect to one of its arguments, evaluated on the trivial fixed-point. From the expression (\ref{eq_g}) we thus obtain the Kesten-Stigum threshold $\aKS$ above which the trival solution of the 1RSB equations is unstable (and there must then exist a non-trivial solution) as
\beq
\aKS = \frac{1}{k(k-1)\theta^2}  \ , \qquad
\theta = \frac{\underset{\s_1,\dots,\s_k}{\sum}\w(\s_1,\dots,\s_k) \s_1 \s_2}{\underset{\s_1,\dots,\s_k}{\sum}\w(\s_1,\dots,\s_k)} = 
\frac{\underset{p=0}{\overset{k-2}{\sum}}\binom{k-2}{p} (\w_p - 2 \w_{p+1}+\w_{p+2}) }{\underset{p=0}{\overset{k}{\sum}}\binom{k}{p} \w_p  } \ .
\label{eq_KS_generic}
\eeq

\subsubsection{The presence of hard-fields in the 1RSB solution with $m=1$ (rigidity threshold)}
\label{sec_1RSB_hardfields}

For the special value $m=1$ of the Parisi parameter we have obtained in (\ref{1RSBeqnSimplifs}) a simplified form of the 1RSB equations; even if much simpler than the general formalism, these equations bear on probability distributions ($Q_+(h)$ and $\hQ_+(u)$) and cannot be solved analytically in general. One can however make some more explicit statements when $\w_0=\w_k=0$ and $\w_p>0$ for $p\in [1,k-1]$, i.e. when the constraints forbid monochromatic hyperedges, but allow all bichromatic configurations (even if they can give different weights to these configurations). In this case the distribution $Q_+(h)$ can contain a Dirac peak in $h=1$, corresponding to ``hard fields'' that impose strictly the value of some variables (that are called frozen, or rigid) inside one pure state. Let us denote $q$ (resp. $\hq$) the weight of $h=1$ (resp. $u=1$) under $Q_+$ (resp. $\hQ_+$). One can then obtain from (\ref{1RSBeqnSimplifs}) closed equations on $q$ and $\hq$:
\bea
q&=&1-\sum_{d=0}^\infty p_d (1-\hq)^d = 1-e^{-\a k \hq} \ , \\
\hq &=& \tp(-,\dots,-|+) \, q^{k-1} \ ;
\eea
indeed the expression of $f$ in (\ref{eq_f}) reveals that $h=1$ as soon as one of the neighboring constraint sends the hard field $u_i=1$, while (\ref{eq_g}) shows that a variable is forced to a certain value $\s$ by a constraint only when the $k-1$ other variables are simultaneously forced to $-\s$. Eliminating $\hq$ one sees that $q$ is solution of
\beq
q=1-\exp(-\G q^{k-1}) \ , \qquad \text{with} \ \ \G = \a k \tp(-,\dots,-|+) \ .
\eeq
For $k \ge 3$ a non-trivial solution to this equation appears discontinuously, when $\G$ exceeds a critical value $\Gr$. The value of $\Gr$, and the associated solution $\qr$, are the solutions of
\beq
\begin{cases}
\qr=1-\exp(-\Gr \qr^{k-1}) \\
1=(k-1)\Gr \qr^{k-2} \exp(-\Gr \qr^{k-1}) 
\end{cases} \ ;
\label{eq_qrGr}
\eeq
see Sec.~\ref{sec_methodology} for more explanations on the origin of these equations. One can close the equation on $\qr$, that obeys $1=(k-1)\ln(1-\qr)(1-1/\qr)$, from which $\Gr$ is obtained as $\Gr = - \frac{\ln(1-\qr)}{\qr^{k-1}}$. Note that $\qr$ and $\Gr$ depend solely on $k$. 

Translating back to the parameters $\a$, $\{\w_p\}$, one sees that for any choice of $\{\w_p\}$ such that $\w_0=\w_k=0$ and $\w_p>0$ for $p\in [1,k-1]$ there exists a ``rigidity threshold'' $\ar(k,\{\w_p\})$ above which the equation on the probability of hard-fields admits a non-trivial solution, with (recalling the expression of $\tp$ from (\ref{tildeP}))
\beq
\ar(k,\{\w_p\}) =\frac{1}{k} \Gr(k) \frac{\overset{k-1}{\underset{p=1}{\sum}} \binom{k-1}{p} \w_p}{\w_1} 
= \frac{1}{k} \Gr(k) \frac{\overset{k-1}{\underset{p=1}{\sum}} \binom{k}{p} \w_p}{2 \w_1} 
\ ,
\label{eq_rigidity_generic}
\eeq
where in the last step we exploited the symmetry $\w_p=\w_{k-p}$. We shall denote $\aru$ the value of this threshold for the uniform case $\w_1=\dots=\w_{k-1}=1$, i.e. when all proper bicolorings are weighted equally, in such a way that
\beq
\aru(k) = \frac{1}{k} \Gr(k) (2^{k-1}-1) \ .
\eeq

This rigidity threshold is an upperbound on the dynamic transition: if there exists a solution to the 1RSB equations at $m=1$ containing hard-fields, this is certainly a non-trivial solution of the 1RSB equations. The inequality $\ad \le \ar$ is in general strict, i.e. there can be non-trivial solution of the 1RSB equations at $m=1$ that do not contain any hard-field; this has been seen numerically in many problems, and proven rigorously for the graph $q$-coloring problem in the large $q$ limit in~\cite{Sly08,SlyZhang16}. This rigidity threshold corresponds actually to a transition for a strong form of reconstructibility in the tree reconstruction interpretation: when $\a > \ar$, with positive probability the observation of far away variables allows to infer the value of the root without possibility of error. Let us also underline that among all the parameters $\{\omega_p\}$ that define the bias among proper bicolorings $\omega_1=\omega_{k-1}$ plays a special role in the expression (\ref{eq_rigidity_generic}) of the rigidity threshold. Indeed hard fields are propagated along constraints that are ``almost violated'', in the sense that they contain a single variable of a given color. Penalizing such ``almost monochromatic'' hyperedges tends thus to avoid the percolation of frozen variables.

\section{On the numerical determination of the dynamic transition}
\label{sec_methodology}

The dynamic threshold $\ad$ is the smallest value of $\a$ such that the 1RSB equations at $m=1$ admit a non-trivial solution (besides the RS trivial one in which all fields are equal to zero). Depending on the models the appearance of a non-trivial solution can occur either in a continuous or a discontinuous way. In the former case one has $\ad=\aKS$, the bifurcation occurs via the local instability of the trivial fixed point studied in Sec.~\ref{sec_KS}, and $\ad$ is thus known analytically. In the latter case $\ad < \aKS$, the birth of the non-trivial solution occurs non-perturbatively and cannot be detected from the properties of the trivial fixed point. The accurate numerical determination of $\ad$ when the transition is discontinuous is a rather difficult task. It corresponds to study a bifurcation for a fixed-point equation of the form $Q=F(Q,\a)$, where $Q$ is a probability distribution and $F$ a functional on this space, depending on the parameter $\a$. We shall discuss later on the different numerical strategies that can be followed to determine $\ad$, in particular one that, to the best of our knowledge, is new in this context. To explain these different methods it is instructive to study first a much simpler case, in which the unknown $Q$ is replaced by a real number, that we shall instead denote $x$ for clarity.

\subsection{Scalar bifurcations}

Let us consider a function $f(x,\a)$, smooth in its two real arguments, and the associated discrete dynamical system $x^{(t+1)} = f(x^{(t)},\a)$, parametrized by $\a$. We recall some basic facts in this setting: the stationary configurations of the dynamical system are the solutions $x_*(\a)$ of the fixed point equation $x=f(x,\a)$. Their (linear) stability is determined by the coefficient $\lambda(\a) =(\partial_x f)(x_*(\a),\a)$ (here and below we denote $(\partial_xf)$, $(\partial_\a f)$ and so on the partial derivatives of the function $f$); a fixed point $x_*(\a)$ is indeed stable under iterations if $|\lambda(\a)|<1$, and unstable if $|\lambda(\a)|>1$. We also recall the implicit function theorem: if $(x_0,\a_0)$ is a solution of $f(x_0,\a_0)=x_0$, and if $(\partial_x f) (x_0,\a_0) \neq 1$, then there is a unique smooth function $x_*(\a)$ with $f(x_*(\a),\a)=x_*(\a)$ in a neighborhood of $\a_0$, with $x_*(\a_0)=x_0$. Hence the bifurcations of the fixed point equation, i.e. the modifications in the number of solutions, or the singularities of these solutions, are associated to points where $(\partial_x f) (x(\a_0),\a_0) = 1$, in order for the implicit function theorem to be unapplicable. At these points the stability parameter $\lambda(\a)$ reaches its critical value $1$.

\begin{figure}[hbtp]
\begin{center}
\includegraphics[scale=0.6]{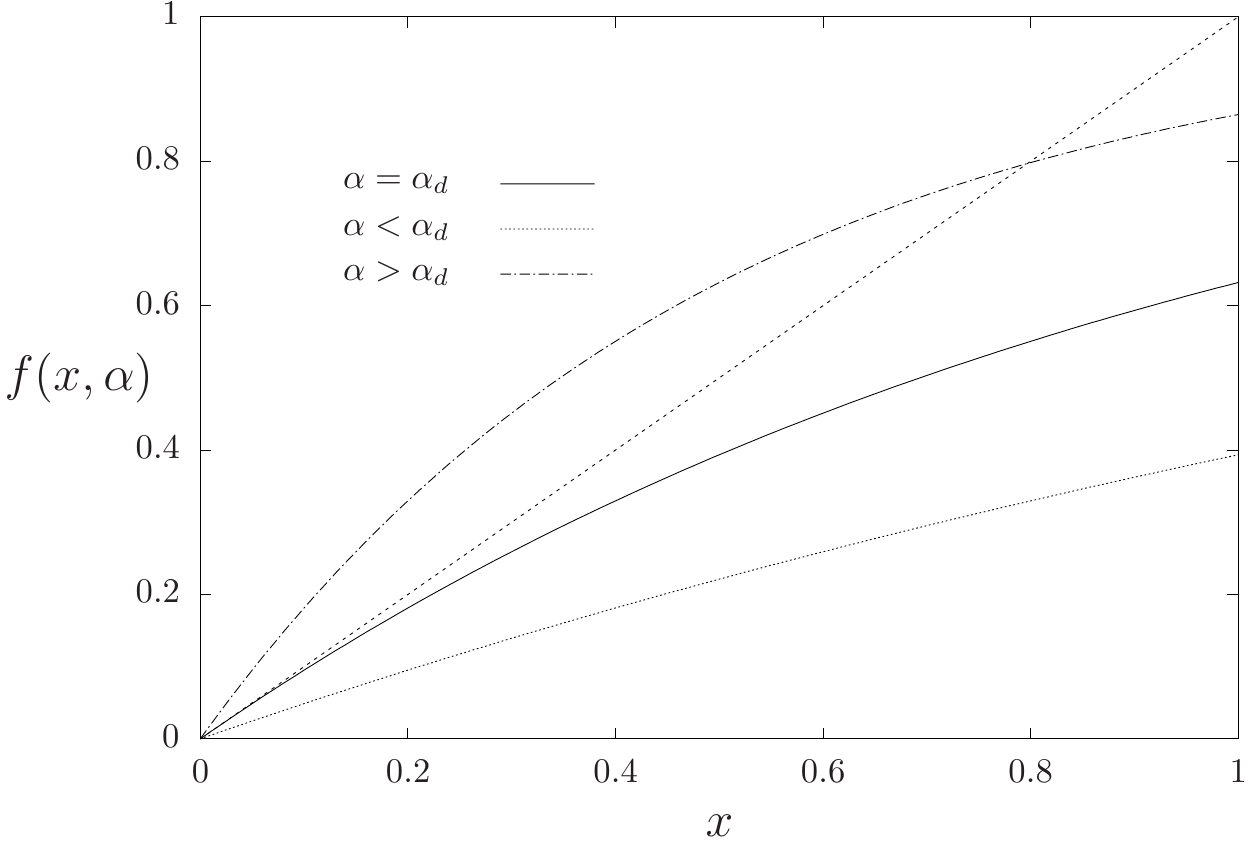}
\hspace{1cm}
\includegraphics[scale=0.6]{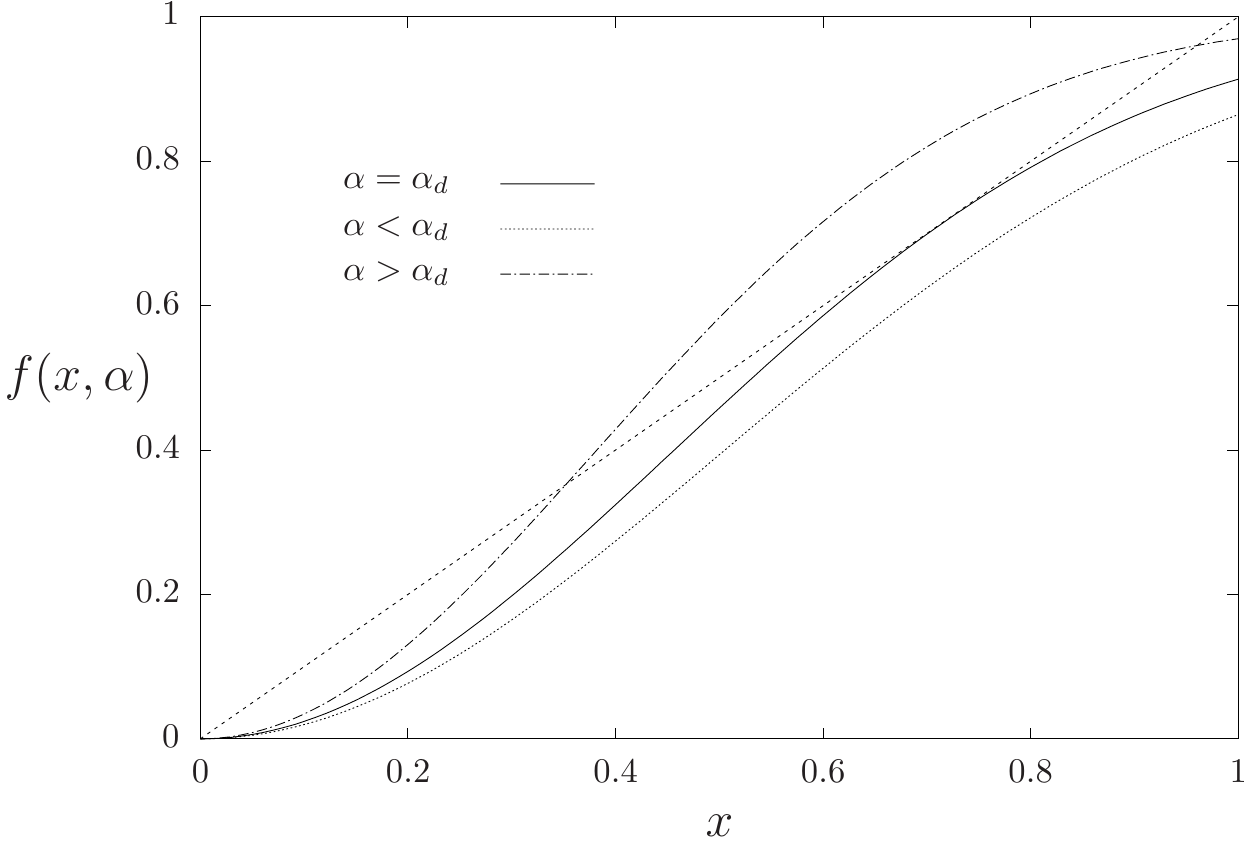}
\caption{Left panel: example of a continuous bifurcation with $f(x,\a)=1-e^{-\a x}$, for which $\ad=1$. Right panel: example of a discontinuous bifurcation with $f(x,\a)=1-e^{-\a x^2}$, for which $\ad=2.45541$.}
\label{fig_bifurcation_scalaire}
\end{center}
\end{figure}

To be more concrete we shall make the additional hypotheses that the order parameter $x$ is restricted to non-negative values ($x\ge 0$), and that $f(0,\a)=0$ for all $\a$. Let us assume that this trivial fixed point, that exists for all $\a$, is the unique solution for small enough values of $\a$, and becomes non-unique when $\a$ exceeds a threshold $\ad$. The two simplest ways to implement these hypotheses are sketched in Fig.~\ref{fig_bifurcation_scalaire}, corresponding to a continuous bifurcation on the left panel, a discontinuous one on the right. Let us state a series of simple facts on these two types of phase transitions, that will be enlightening when we turn to the functional case.

Consider first the continuous case illustrated on the left panel of Fig.~\ref{fig_bifurcation_scalaire}. The bifurcation occurs at the critical parameter value $\ad$ defined by $(\partial_x f) (0,\ad)=1$, the trivial fixed point being stable (resp. unstable) for $\a < \ad$ (resp. $\a > \ad$). For $\a > \ad$ there exists a non-trivial branch of stable fixed points $x_*(\a)>0$; in the neighborhood of the bifurcation the latter behaves as
\beq
x_*(\a) = K (\a-\ad) + o((\a-\ad)) \qquad
\text{when} \ \ \a \to \ad^+ \ ,
\eeq
with $K=-2 (\partial_{x \a}f)/(\partial_{x x}f)$, the derivatives being computed in $(0,\ad)$ (here and in the following the expressions of the various constants $K$ can be obtained by a Taylor expansion of the equation $x=f(x,\a)$ around the bifurcation point, at the lowest non-trivial order). The stability parameter of the non-trivial solution, $\lambda(\a)=(\partial_x f)(x_*(\a),\a)$, reaches its marginal value 1 at the bifurcation as
\beq
\lambda(\a) = 1 - K' (\a-\ad) + o(\a-\ad) \qquad
\text{when} \ \ \a \to \ad^+ \ ,
\label{eq_lambda_scalaire_continue}
\eeq
with $K'=(\partial_{x \a}f)$.

\begin{figure}[hbtp]
\begin{center}
\includegraphics[scale=0.6]{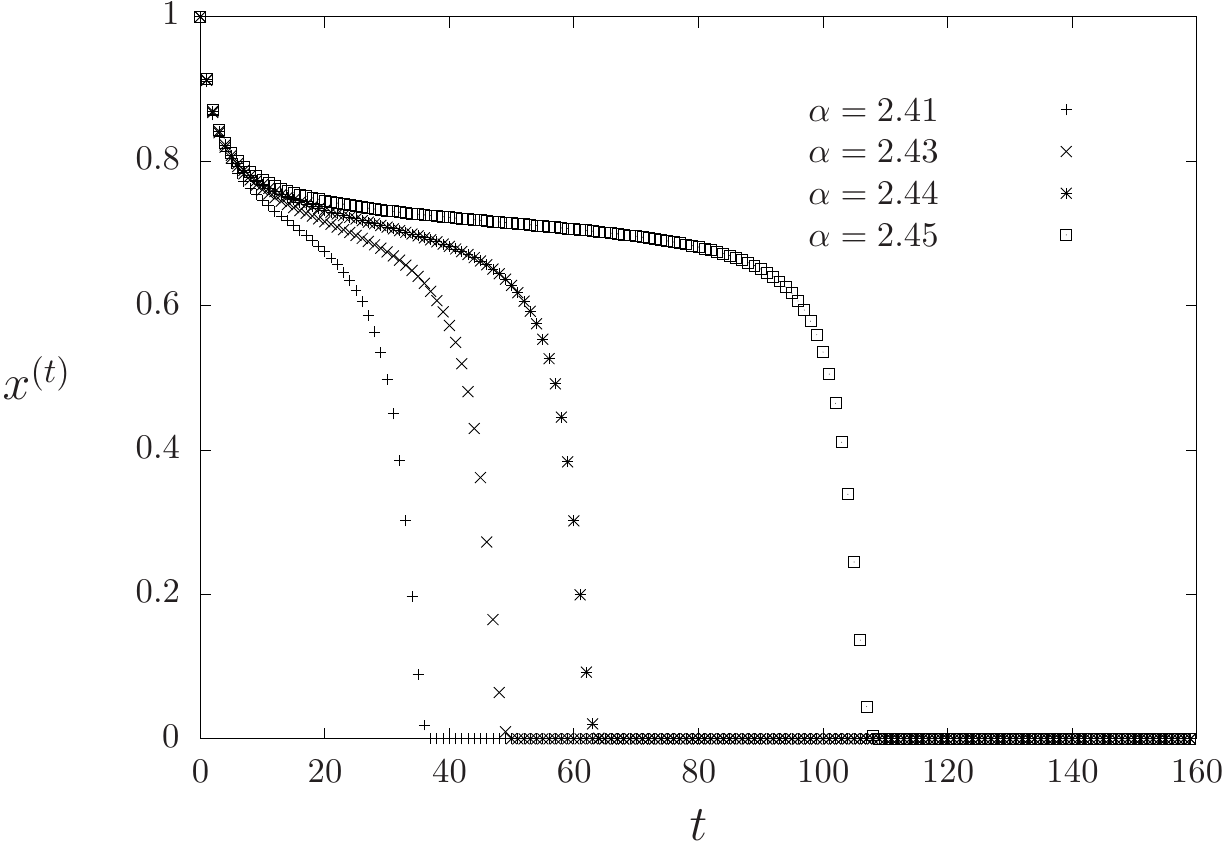}
\hspace{1cm}
\includegraphics[scale=0.6 ]{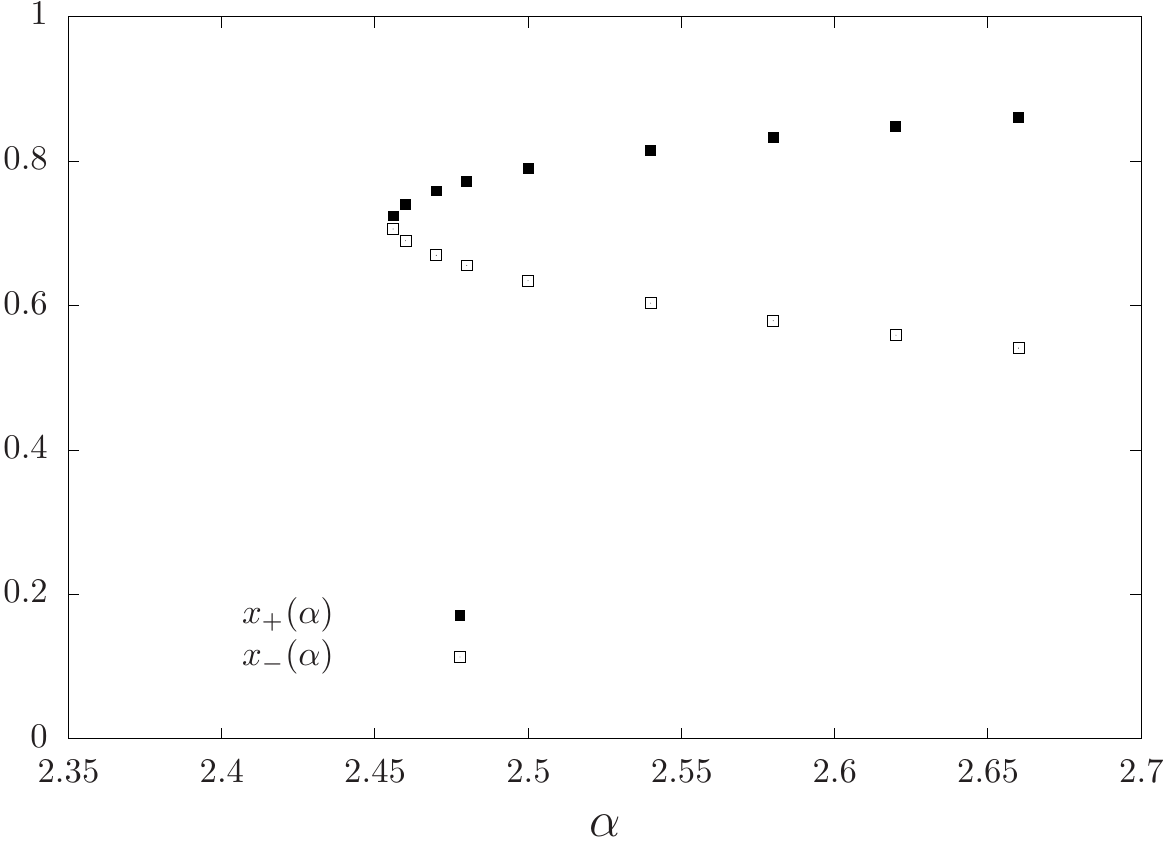}
\caption{Properties of the iterates and fixed points for the discontinuous bifurcation undergone by the function $f(x,\a)=1-e^{-\a x^2}$. Left panel: $x^{(t)}$ as a function of $t$ for a few values of $\a$ close but strictly less than $\ad$. Right panel: the non-trivial solutions $x_\pm(\a)$ for $\a\ge \ad$.}
\label{fig_bifurcation_scalaire_discontinue}
\end{center}
\end{figure}

Let us now turn to the discontinuous case (cf. the right panel of Fig.~\ref{fig_bifurcation_scalaire}), and emphasize the main properties of the critical behavior of the bifurcation. The trivial fixed point is stable for all values of the parameter $\a$; the bifurcation occurs at $\ad$ with the abrupt appearance of a solution $\xd>0$. These two quantities can be determined by solving the system of equations
\beq
\begin{cases}
\xd = f(\xd,\ad) \ , \\
1 = (\partial_x f)(\xd,\ad) \ .
\end{cases}
\eeq
For $\a > \ad$ there are two branches of non-trivial solutions $x_-(\a) < \xd < x_+(\a)$ that emerge from $\xd$ (see the right panel of Fig.~\ref{fig_bifurcation_scalaire_discontinue}); in the neighborhood of $\ad$ they behave as
\beq
x_\pm(\a) = \xd \pm K \sqrt{\a-\ad} + o(\sqrt{\a-\ad}) \qquad
\text{when} \ \ \a \to \ad^+ \ ,
\label{x2_behavior}
\eeq
where the coefficient $K$ can be computed from the expansion of $f$ around the bifurcation point (explicitly, $K=\sqrt{-2 (\partial_\a f) / (\partial_{xx}f)}$, the derivatives being computed in $(\xd,\ad)$). For $\a>\ad$ $x_+(\a)$ (resp. $x_-(\a)$) is linearly stable (resp. unstable); the stability parameter $\lambda(\a)=(\partial_x f)(x_+(\a),\a)$ of the stable non-trivial branch reaches its critical value 1 at the bifurcation, with a critical exponent 1/2:
\beq
\lambda(\a) = 1 - K' \sqrt{\a-\ad} + o(\sqrt{\a-\ad}) \qquad
\text{when} \ \ \a \to \ad^+ \ ,
\label{eq_scalaire_lambda}
\eeq
with $K'=\sqrt{-2 (\partial_\a f)  (\partial_{xx}f)}$. We present in the left panel of Fig.~\ref{fig_bifurcation_scalaire_discontinue} the iterates $x^{(t+1)}=f(x^{(t)},\a)$, for a few values of $\a < \ad$, starting from an initial condition $x^{(0)} > \xd$. Their long time limit is of course $0$, the only fixed point in this phase, but when $\a \to \ad^-$ the decay is slower and slower, with a large number of iterations spent around a plateau value at $\xd$. More quantitatively one can define $t_*(\a)$ as the minimal $t$ such that $x^{(t)} \le \xd - \epsilon$, and obtain that
\beq
t_*(\a) \sim K'' (\ad-\a)^{-1/2} \qquad \text{when} \ \ \a \to \ad^- \ ,
\label{eq_scalaire_divergencet}
\eeq
with $K''=2\pi/K'$, independently of $x^{(0)}$ and $\epsilon$ (as long as $0<\epsilon<\xd$). Actually a whole scaling function describing the evolution of $x^{(t)}$ around the plateau can be derived, see~\cite{MontanariSemerjian06} for more details.

\subsection{Discontinuous functional bifurcations}

Let us now come back to our original goal, namely the determination of the dynamic threshold $\ad$ above which appears a non-trivial solution of the 1RSB equations at $m=1$. As in the scalar case this transition can occur either in a continuous or in a discontinuous way; the former case was analytically dealt with in Sec.~\ref{sec_KS}, we shall hence concentrate now on the discontinuous transitions.

\begin{figure}[hbtp]
   \begin{tabular}{cc}
      \includegraphics[scale=0.6]{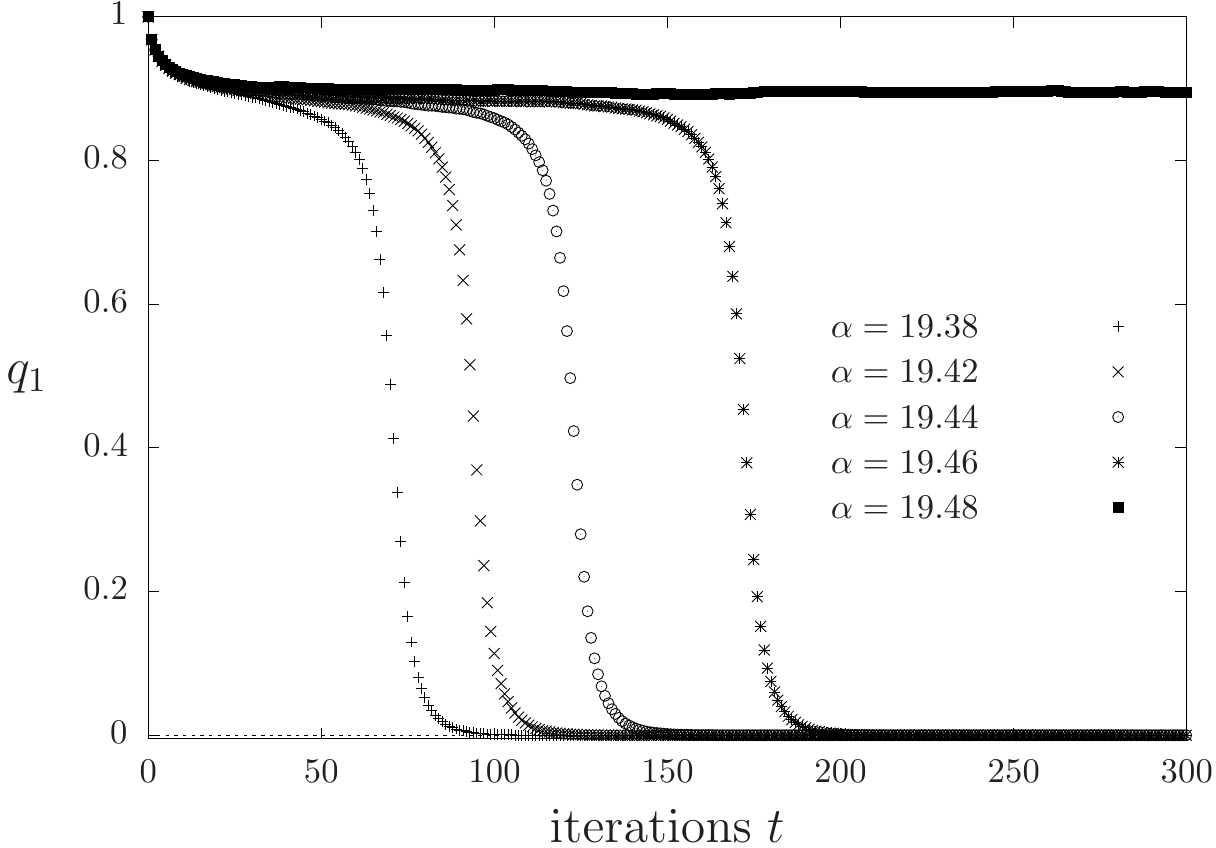}
\hspace{1cm}
      \includegraphics[scale=0.6]{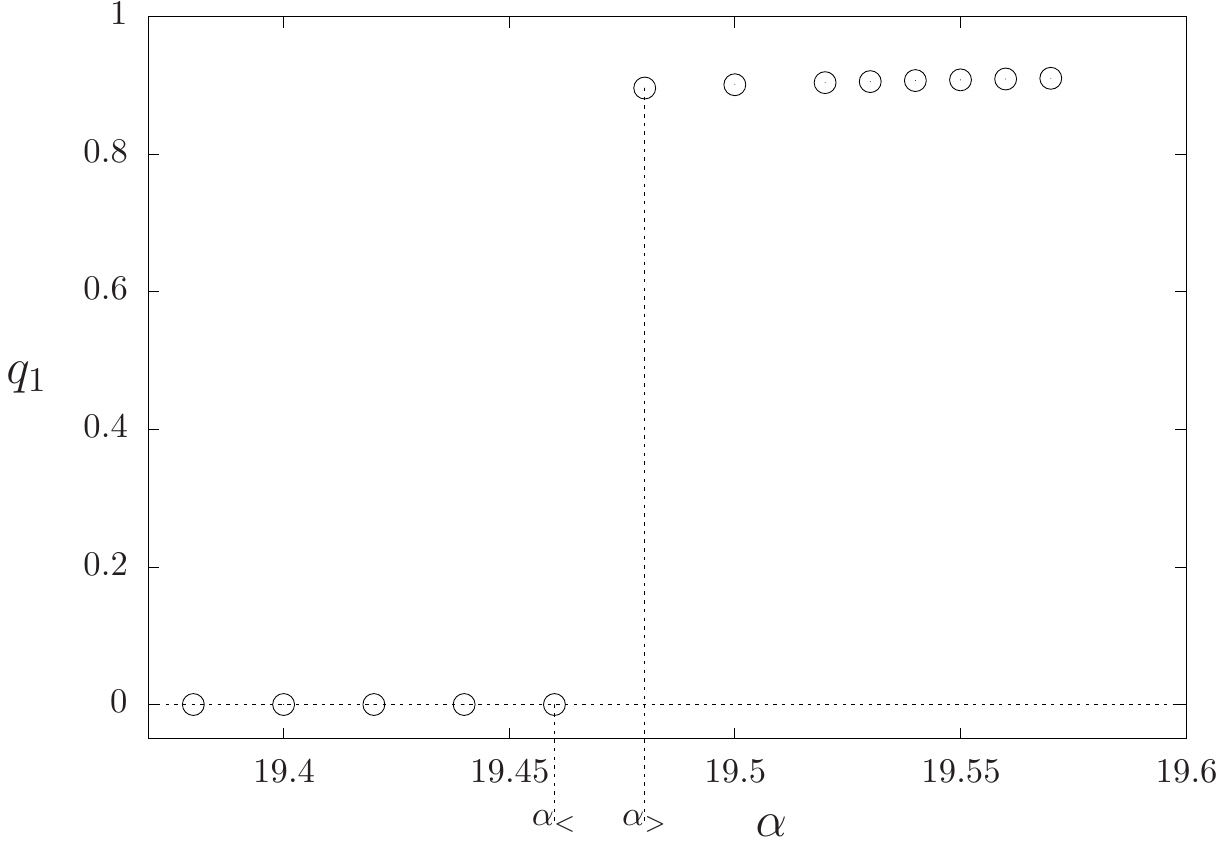}
   \end{tabular}   
   \caption{Exemple of a discontinuous dynamic transition for $k=6$, $\w_0=\w_k=0.005$, $\w_1=\w_{k-1}=0.92$, and $\w_2=\dots=\w_{k-2}=1$. Left: $q_1(\a,t)$ versus iteration time $t$ for different values of $\a$. Right: $q_1(\a)$ averaged over $t$ after equilibration. The size of the population used is $10^6$.}
   \label{plateaux}
\end{figure}

The 1RSB equations (\ref{1RSBeqnSimplifs}) can be written abstractly as a functional fixed point equation $Q=F(Q,\a)$; at variance with the scalar toy model discussed above they can only be solved approximately, for instance by the population dynamics numerical algorithm explained in Sec.~\ref{sec_simplifications_m1}. Some examples of typical numerical results that can be obtained in this way are presented in Fig.~\ref{plateaux}; we use as an observable to condense the distribution $Q_+$ into a single scalar the overlap $q_1 = \int Q_+(h) h \dd h$, which is equal to 0 for the trivial solution. On the left panel we plot the value of $q_1$ as a function of the number of iterations, for a few values of $\a$. One sees on this plot, reminiscent of the left panel of Fig.~\ref{fig_bifurcation_scalaire_discontinue}, the discontinuous birth of a non-trivial fixed point at $\ad$, with a longer and longer plateau in the low $\a$ phase as a precursor of the transition. On the right panel of Fig.~\ref{plateaux} we present the asymptotic value of $q_1$ reached for large $t$, for different values of $\a$ around the dynamic transition (corresponding to the right panel of Fig.~\ref{fig_bifurcation_scalaire_discontinue}), that jumps discontinuously from 0 when $\a$ crosses $\ad$.

It is not completely obvious how to extract a precise estimate of $\ad$ from this kind of data. The simplest approach amounts to determine the curves $q_1^{(t)}$ for several closely spaced values of $\a$, and assess that $\ad \in [\a_<,\a_>]$, where $\a_<$ is the largest value for which $q_1^{(t)}$ drops to 0 at large $t$, $\a_>$ the smallest value for which a stable plateau is encountered. This determination suffers however from inaccuracies due to the finite number of $\a$ values one can investigate, the finite number of iterations one can perform (leading to an underestimation of $\a_>$) and to the finite size of the population that approximate the distribution $Q_+$ ($\a_<$ can thus be overestimated, finite size fluctuations having a destabilizing effect).

One can try to circumvent these difficulties by getting some inspiration from the much simpler scalar bifurcation studied above. We recall that the criticality at $\ad$ showed up in three different ways: (i) $x_+(\a)$ exhibits a square root singularity when $\a \to \ad^+$, see Eq.~(\ref{x2_behavior}); (ii) the length of the plateau diverges when $\a \to \ad^-$ with a critical exponent $-1/2$, cf. Eq.~(\ref{eq_scalaire_divergencet}); (iii) the stability parameter $\lambda(\a)$ reaches 1 with a square root singularity when $\a \to \ad^+$, as written in Eq.~(\ref{eq_scalaire_lambda}). 

\begin{figure}[hbtp]
\begin{center}
\includegraphics[scale=0.6]{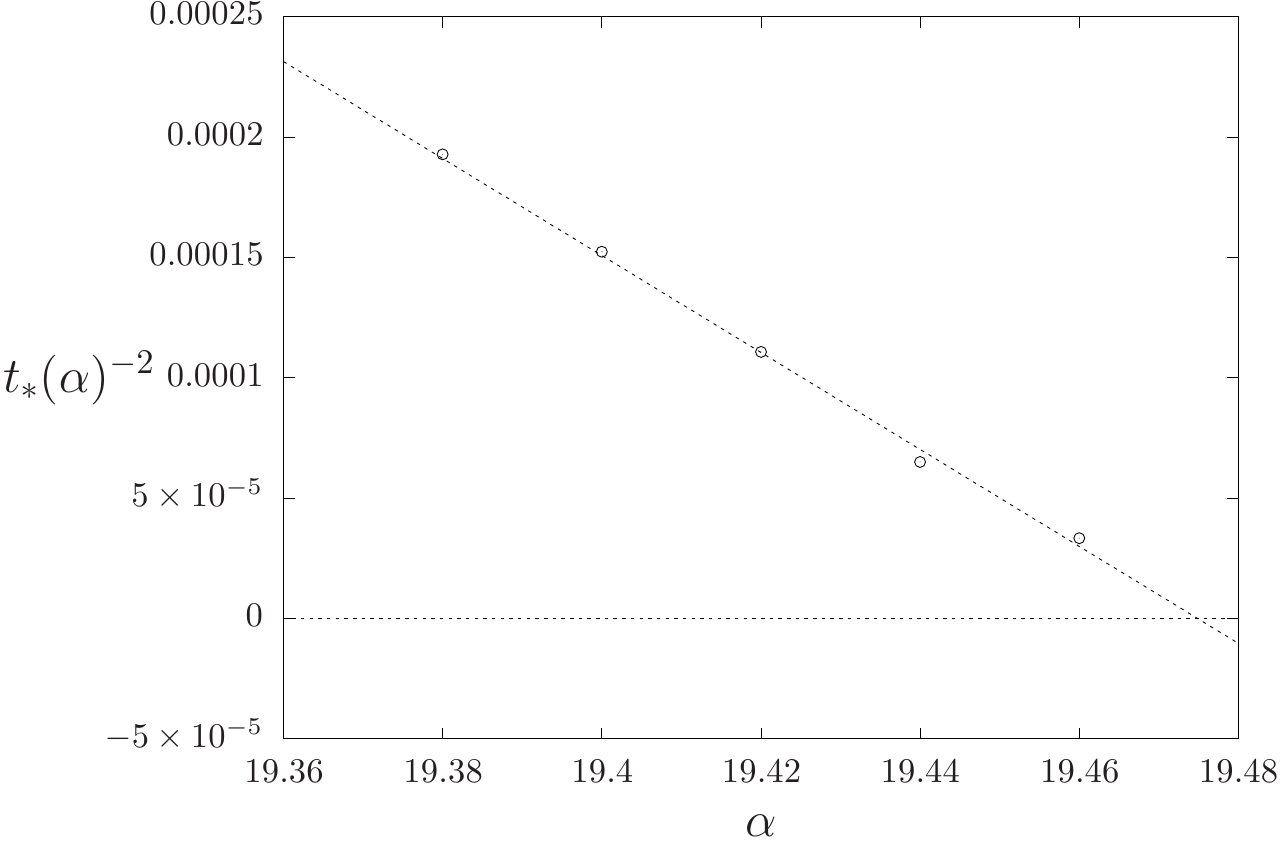}
\hspace{1cm}
\includegraphics[scale=0.6]{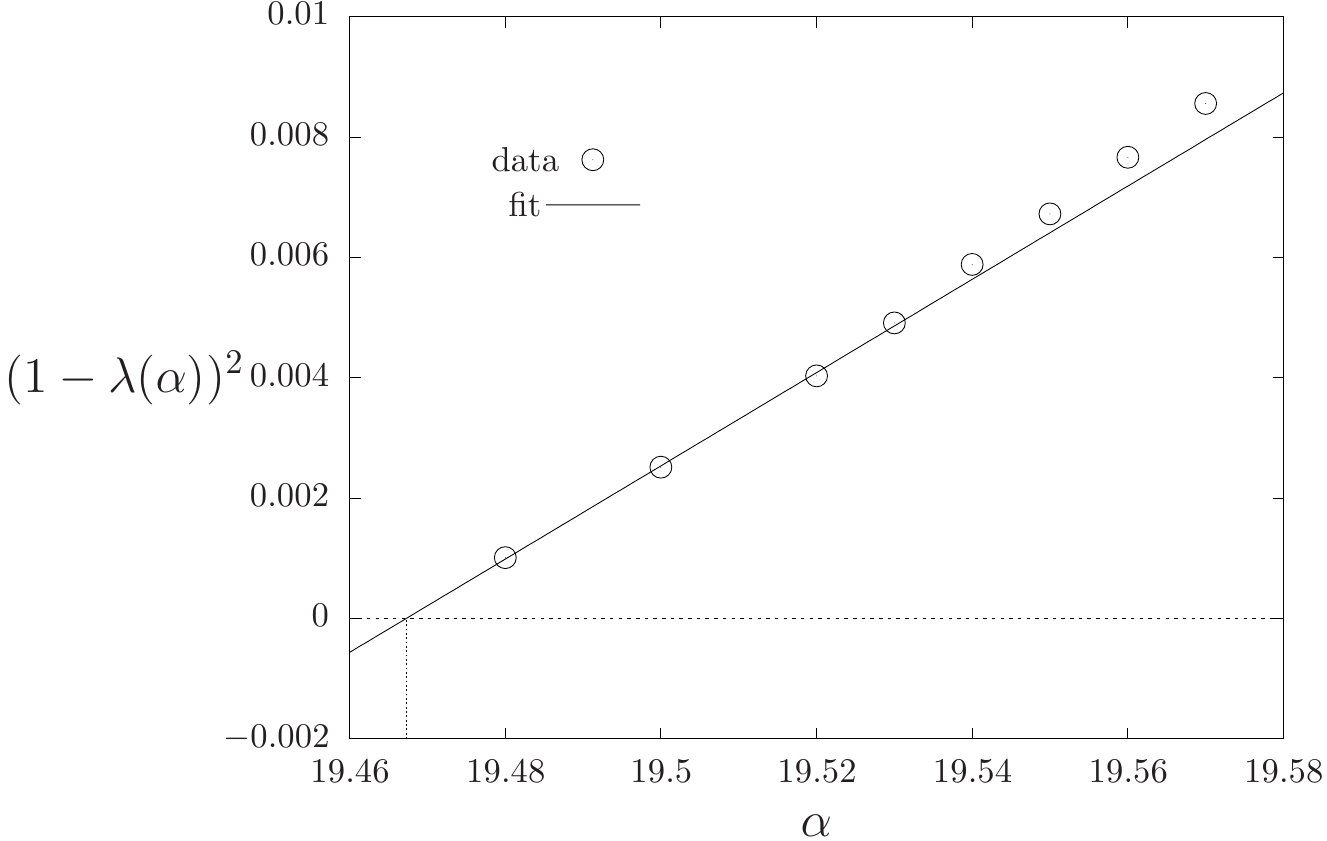}
\caption{Study of the discontinuous dynamic transition encountered as a function of $\a$ for the choice of parameters $k=6$, $\w_0=\w_k=0.005$, $\w_1=\w_{k-1}=0.92$, $\w_2=\dots=\w_{k-2}=1$. Left panel: determination of $\ad$ from the study of the decorrelation time $t_*(\a)$ for $\a<\ad$. The plot displays $t_*(\a)^{-2}$ versus $\a$,  where one has defined $t_*(\a)$ as the first time for which the overlap drops below the value $q_1=0.4$. The line is a fit of the data of the form $t_*(\a)^{-2} = A \, (\ad-\a)$, with fitting parameters $A$ and $\ad$. The linear behavior confirms the divergence of $t_*$ with a scaling exponent $-1/2$, as in the scalar case (\ref{eq_scalaire_divergencet}), the fit gives the estimation $\ad=19.47$.
Right panel: determination of $\ad$ from the study of the stability parameter $\lambda(\a)$ for $\a > \ad$. The plot displays $(1-\l(\a))^2$ versus $\a$, the linear fit reproduces the scaling behavior (\ref{eq_scalaire_lambda}) of the scalar case, and yields $\ad=19.467$. Only the first points are used for the fit, in order to avoid the higher order contributions in powers of $\a-\ad$ that are clearly visible at the largest values of $\alpha$.}
\label{tempsPlateaux}
\end{center}
\end{figure}

Assuming the same critical behavior to occur in the discontinuous functional bifurcation case (more complicated behaviors could occur in infinite dimensional spaces, but in absence of accidental degeneracies there should be a single critical direction at a bifurcation driven by a single parameter) one can try to exploit these scaling laws in order to obtain more precise estimates of $\ad$. Point (i) translates into a square root singularity of the large $t$ limit of $q_1$ in the limit $\a \to \ad^+$; this does not seem very useful to us, as it would involve a fit of $q_1(\a)$ in which both $\ad$ and $q_1(\ad)$ are unknowns. On the contrary points (ii) and (iii) yield simpler fits for the determination of $\ad$. The aspect (ii) is very easy to exploit: from the curves of the left panel of Fig.~\ref{plateaux} one can deduce immediately a value $t_*(\a)$ for the number of iterations necessary to fall below the plateau (as in the scalar case one can define $t_*(\a)$ with any threshold strictly between 0 and the plateau value). According to (\ref{eq_scalaire_divergencet}) $t_*(\a)^{-2}$ should vanish linearly at $\ad$; this is indeed what we obtain with a rather good accuracy, see the left panel of Fig.~\ref{tempsPlateaux}. However one cannot reach in this way a very large number of iterations, the numerical rounding errors and finite population size fluctuations having the tendency to accumulate over time; this cutoff on $t$ thus limits the accuracy of this determination of $\ad$. We have thus turned to the functional generalization of point (iii) above, namely the computation of a stability parameter $\lambda(\a)$ for the stable non-trivial branch $\a > \ad$, and the determination of $\ad$ as the parameter for which $\lambda$ reaches 1. This extrapolation is done using the scaling anticipated in the scalar case in (\ref{eq_scalaire_lambda}), and is illustrated in the right panel of Fig.~\ref{tempsPlateaux}. The functional nature of the unknown in the fixed point equation makes the definition of $\lambda$ more complicated than in the scalar case, where it was simply $\partial_x f$; we give detailed explainations on the numerical computation of $\lambda(\a)$ in the functional case in the next section. Before that let us emphasize that the square root behavior of $\lambda$ around $\ad$, guessed from the scalar bifurcation, is in very good agreement with the numerical results obtained in the functional case when the dynamic transition is discontinuous (see the right panel of Fig.~\ref{tempsPlateaux}). We believe the determination of $\ad$ reached by the extrapolation of $\lambda(\a)$ is more reliable and accurate than the one based on $t_*(\a)$. Indeed the former quantity is defined from a stable fixed-point of the equations, averages can be performed in a steady-state (plateau) regime to reduce the statistical error on its computation, while $t_*(\a)$ is a measure of a transient regime more sensitive to numerical inaccuracies.

\begin{figure}[hbtp]
\begin{center}
\includegraphics[scale=0.6]{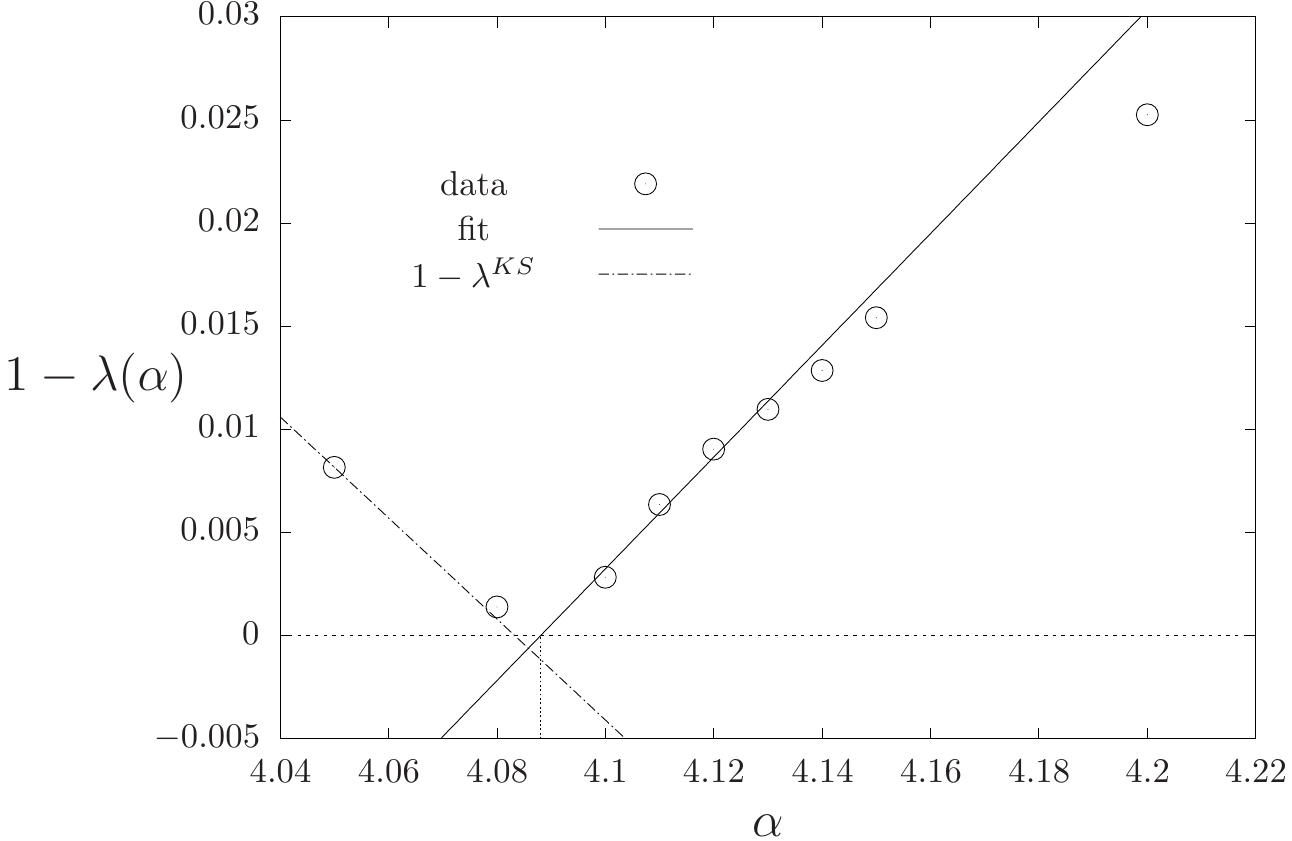}
\caption{The stability parameter $\lambda(\a)$ in the case of a continuous transition (here for $k=4$, $\omega_0=\omega_k=0$, $\omega_1=\dots=\omega_{k-1}=1$). From the fit one obtains the result $\a = 4.088$, wich is in good agreement with the analytical one $\a_{KS} = 4.083$.}
\label{fit-continu}
\end{center}
\end{figure}

As a consistency check we also present in Fig.~\ref{fit-continu} a similar study in the case of a continuous transition. We see that the stability parameter $\lambda$ computed on the non-trivial solution, i.e. for $\a > \ad$, reaches 1 with a linear behavior (as in the scalar case, see Eq.~(\ref{eq_lambda_scalaire_continue})), and that its extrapolation is in good agreement with the analytically computed value of $\aKS$ from Eq.~(\ref{eq_KS_generic}). Moreover the numerical computation of the stability parameter of the trivial fixed-point coincides for $\a < \ad$ with the analytical one, $\lambda^{KS}=\a k(k-1) \theta^2$.

\subsection{The stability parameter $\lambda$ in the functional case}

As an intermediate step in the generalization from the scalar to the functional case let us consider a fixed point equation of the form $\vec{x}=f(\vec{x},\a)$, where the unknown $\vec{x}$ is a finite-dimensional real vector. The stability of a branch of solutions $\vec{x}(\a)$ can be determined by considering the Jacobian matrix $J$ of the first derivatives of $f$ computed at the fixed point, that can be defined through the linearization
\beq
f(\vec{x}(\a) + \vec{\ve},\a) =  \vec{x}(\a) + J \, \vec{\ve} + o(\|\vec{\ve} \|) \ ,
\eeq
where $\vec{\ve}$ is a small perturbation around the fixed point. The stability parameter $\l(\a)$ can then be defined as the spectral radius of $J$, i.e. the largest absolute value of the elements of its spectrum. This spectral radius can be expressed in terms of successive applications of $J$ on a perturbation $\vec{\ve}$ as
\beq
\l(\a) = \lim_{n \to \infty} \left( \frac{\| J^n \vec{\ve} \|}{\|\vec{\ve}\| } \right)^\frac{1}{n} \ ,
\eeq
where we assume that $\vec{\ve}$ has a non-vanishing projection on the eigenspace associated to the relevant eigenvalue, and where $\| \bullet \|$ can be any norm. For future use let us define $\vec{\ve}_n = J^n \vec{\ve}$ and rewrite this expression as
\beq
\l(\a)= \lim_{n \to \infty} \left( 
\frac{\|\vec{\ve}_n \|}{\| \vec{\ve}_{n-1}\| } 
\frac{\|\vec{\ve}_{n-1} \|}{\| \vec{\ve}_{n-2}\| } \dots
\frac{\|\vec{\ve}_1 \|}{\| \vec{\ve}\| }
\right)^\frac{1}{n}
\label{eq_spectral_radius}
\eeq

We would like now to extend the computation of a stability parameter to the 1RSB equations (\ref{1RSBeqnSimplifs}) that can be rewritten as $Q_+=F(Q_+,\a)$ by grouping the two lines together. $Q_+$ being a probability distribution the Jacobian of $F$ is now an infinite-dimensional operator, which makes the study of its spectrum rather difficult. Even worse, we do not have at our disposal an exact description of the fixed point $Q_+$ around which we would like to expand $F$: we only have a sequence of approximations of $Q_+^{(t)}$ by the population representation written in Eq.~(\ref{eq_def_popu_Qp}). The individual elements of these representations still evolve at each iteration step, even when the observables computed as averages of $Q_+^{(t)}$ have reached convergence (within the numerical accuracy fixed by the population size ${\cal N}$). To circumvent these difficulties we have followed a strategy inspired by the expression (\ref{eq_spectral_radius}): we consider $Q_+^{(t)}$ and a slight perturbation of it, $Q_+^{(t)} + \delta Q_+^{(t)}$, and assess the rate of growth of the perturbation along the iterations by the functional $F$. In order to implement this idea in practice one needs to choose a specific form for the perturbation; given that $Q_+^{(t)}$ is represented as a sum of Dirac deltas we perturb it by giving an infinitesimal width to each of the peaks, that we replace by Gaussian distributions with a small variance. We thus define
\bea
&& Q_+^{(t)}(h) \approx \frac{1}{{\cal N}} \sum_{i=1}^{{\cal N}} \delta(h - h_i^{(t)}) \ ,
\qquad 
(Q_+^{(t)} + \delta Q_+^{(t)})(h) \approx \frac{1}{{\cal N}} 
\sum_{i=1}^{{\cal N}} {\cal G}(h; h_i^{(t)}, \ve_i^{(t)}) \ , \\
&& \hQ_+^{(t)}(u) \approx \frac{1}{{\cal N}} \sum_{i=1}^{{\cal N}} \delta(u - u_i^{(t)}) \ ,
\qquad 
(\hQ_+^{(t)} + \delta \hQ_+^{(t)})(u) \approx \frac{1}{{\cal N}} 
\sum_{i=1}^{{\cal N}} {\cal G}(u; u_i^{(t)}, \hve_i^{(t)}) \ ,
\label{eq_hQ_perturbed}
\eea
where ${\cal G}( \cdot ; a, b)$ denotes the density of a Gaussian random variable of average $a$ and variance $b$. Consider now the insertion of the form (\ref{eq_hQ_perturbed}) in the right hand side of (\ref{1RSBeqnSimplifs}); the choice of the $d$ peaks indexed by $i_1,\dots,i_d$ produces a random variable $h$ equal in distribution to $f(u_{i_1} + \sqrt{\hve_{i_1}} z_1,\dots, u_{i_d} + \sqrt{\hve_{i_d}} z_d)$, where $z_1,\dots,z_d$ are independent standard Gaussians (of zero mean and unit variance). As the $\hve$ are infinitesimally small one can linearize $f$ to compute the mean and variance of this random variable.

In summary, the determination of $\l(\a)$ is done by tracking the evolution of $Q_+$, $\hQ_+$ and their perturbed versions with populations of couples of real numbers, $(h_i,\ve_i)$ and $(u_i,\hve_i)$, that evolve in time according to the following generalization of the update rules given in Sec.~\ref{sec_simplifications_m1}. To obtain $(h_i^{(t+1)},\ve_i^{(t+1)})$ one repeats, independently for $i=1,\dots,\N$, these steps:
\begin{itemize}
\item draw an integer $d$ from the law $p_d$
\item draw $d$ indices $i_1,\dots, i_d$ uniformly at random in $\{1,\dots,\N\}$
\item set $h_i^{(t+1)}=f(u_{i_1}^{(t)},\dots,u_{i_d}^{(t)})$ and $\ve_i^{(t+1)} = \sum_{j=1}^d (\partial_j f)^2 \hve_{i_j}^{(t)}$, where $\partial_j f$ denotes the derivative of $f$ with respect to its $j$-th argument, computed in $(u_{i_1}^{(t)},\dots,u_{i_d}^{(t)})$
\end{itemize}
Similarly the population $(u_i^{(t)},\hve_i^{(t)})$ is generated according to, again independently for $i=1,\dots,\N$:
\begin{itemize}
\item draw $\s_1,\dots,\s_{k-1}$ from the probability law $\tp(\s_1,\dots,\s_{k-1}|+)$
\item draw $k-1$ indices $i_1,\dots, i_{k-1}$ uniformly at random in $\{1,\dots,\N\}$
\item set $u_i^{(t)}=g(\s_1 h_{i_1}^{(t)},\dots, \s_{k-1} h_{i_{k-1}}^{(t)} )$ and $\hve_i^{(t)}=\sum_{j=1}^{k-1} (\partial_j g)^2 \ve_{i_j}^{(t)}$
\end{itemize}

The rate of growth of the perturbation during the iteration $t \to t+1$ is estimated as the ratio of the $L_1$ norms of the perturbation parameters,
\beq
\l_t = \frac{\sum_{i=1}^\N \ve_i^{(t+1)}}{\sum_{i=1}^\N \ve_i^{(t)}} \ ,
\eeq
and the stability parameter is finally computed as
\beq
\l(\a) = (\l_{t_0} \l_{t_0+1} \dots \l_{t_0+n-1} )^\frac{1}{n} \ .
\eeq
Indeed the first $t_0$ iterations are done with the usual population dynamics algorithm, evolving only the $h_i$'s and $u_i$'s, in order to reach an approximate convergence in distribution of the populations to their fixed points, and the perturbation is then initialized with $\ve_i^{(t_0)}=1$. A large number $n$ of additional iterations during which the growth rates are recorded are then performed, and averaged geometrically as in (\ref{eq_spectral_radius}); in the large $n$ limit the value of $\l(\a)$ should be independent of the norm used to define $\l_t$. In practice we divide the $\ve_i^{(t+1)}$ by $\l_t$ after each iteration in order to keep the norm constant and avoid numerical underflows.

This method is similar to the one presented in~\cite{PaRiRi14,parisi2015erratum} to determine the location of a continuous RSB transition from a non-trivial RS solution.

\section{Results of the cavity method}
\label{sec_results_cavity}

\subsection{The existence of a RS phase for $\a > \adu$}
\label{sec_optimalRS}

We shall address now the main question raised in the introduction, namely the evolution of the dynamic phase transition when the measure $\mu(\us)$ over the proper bicolorings of a typical Erd\H os-R\'enyi random hypergraph is not uniform anymore. In the setting considered in this paper this corresponds to take the parameters $\{\w_p\}$ of the interaction function (\ref{generalConstraint}) different from the uniform choice $\w_0=\w_k=0$, $\w_1=\dots=\w_{k-1}=1$. 

We will concentrate first on the ``zero-temperature'' case, i.e. on the measures that give a non-zero weight to proper bicolorings only, which implies $\w_0=\w_k=0$. The choice of the other parameters is constrained by the global spin-flip symmetry that we want to preserve, hence $\w_p=\w_{k-p}$; as it is obvious from (\ref{measure}), multiplying all the $\w_p$ by a common constant does not change the properties of the model. One realizes that for $k=3$ there is no free parameter left, we will thus concentrate on the cases $k \ge 4$ from now on. For arbitrary large values of $k$ there will be of the order of $k/2$ free parameters in the $\w_p$; we will however make the following choice for the zero-temperature measure:
\beq
\w_0=\w_k=0 \ , \qquad  \w_1=\w_{k-1}=1-\e \ , \qquad \w_2=\dots=\w_{k-2}=1 \ ,
\label{eq_parameters_epsilon}
\eeq
where $\e$ is the sole parameter that quantifies the deviation from the uniform measure (that is recovered for $\e=0$). This slight loss in generality is made for the sake of simplicity, and motivated by considerations on the large $k$ limit presented in Sec.~\ref{sec_largek}. The parameter $\e$ controls the relative weight given to the ``almost monochromatic'' constraints that contain a single vertex of one of the possible colors (positive values of $\e$ disfavoring them); as discussed in Sec.~\ref{sec_1RSB_hardfields} these are precisely those responsible for the existence of frozen variables, one of the mechanism of RSB. 

\begin{figure}[ht]
\begin{center}
  \includegraphics[scale=0.43]{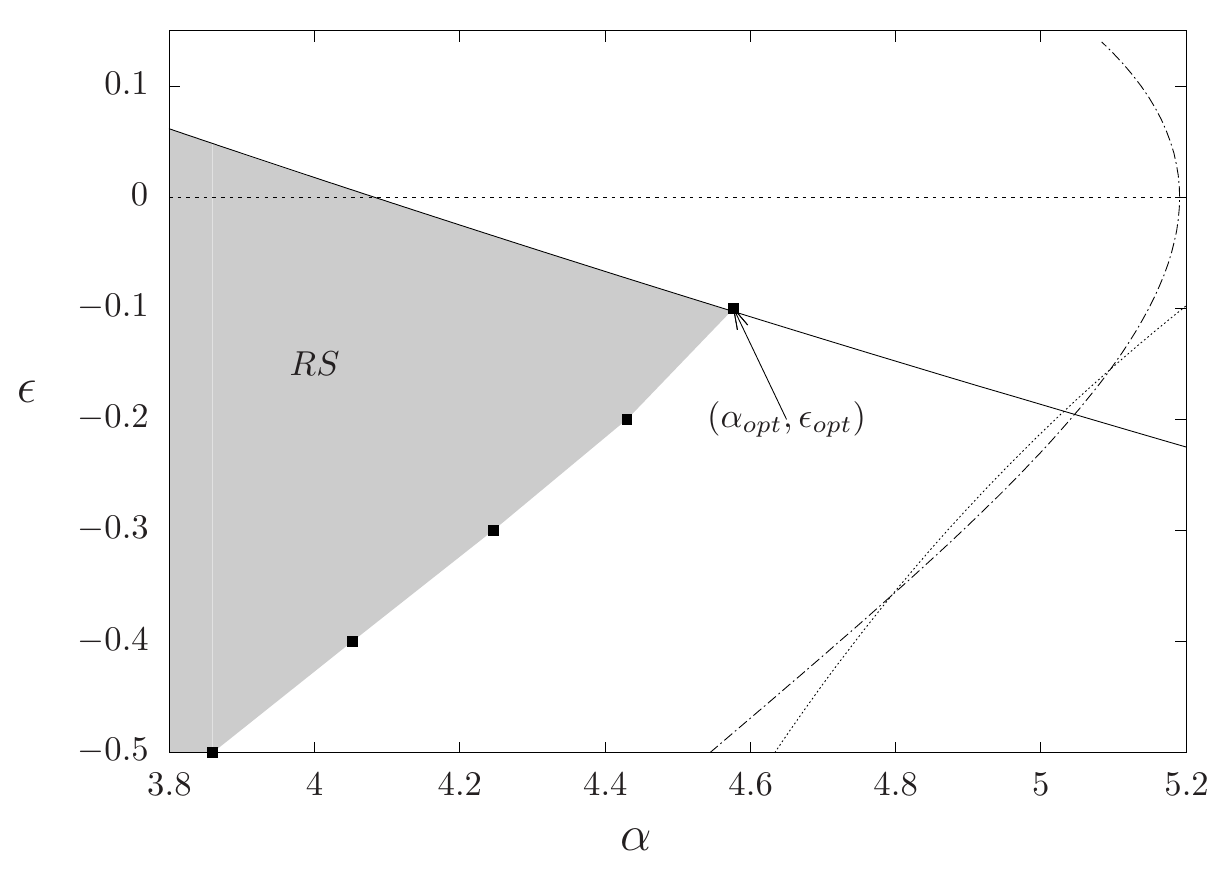}\hspace{3mm}
  \includegraphics[scale=0.43]{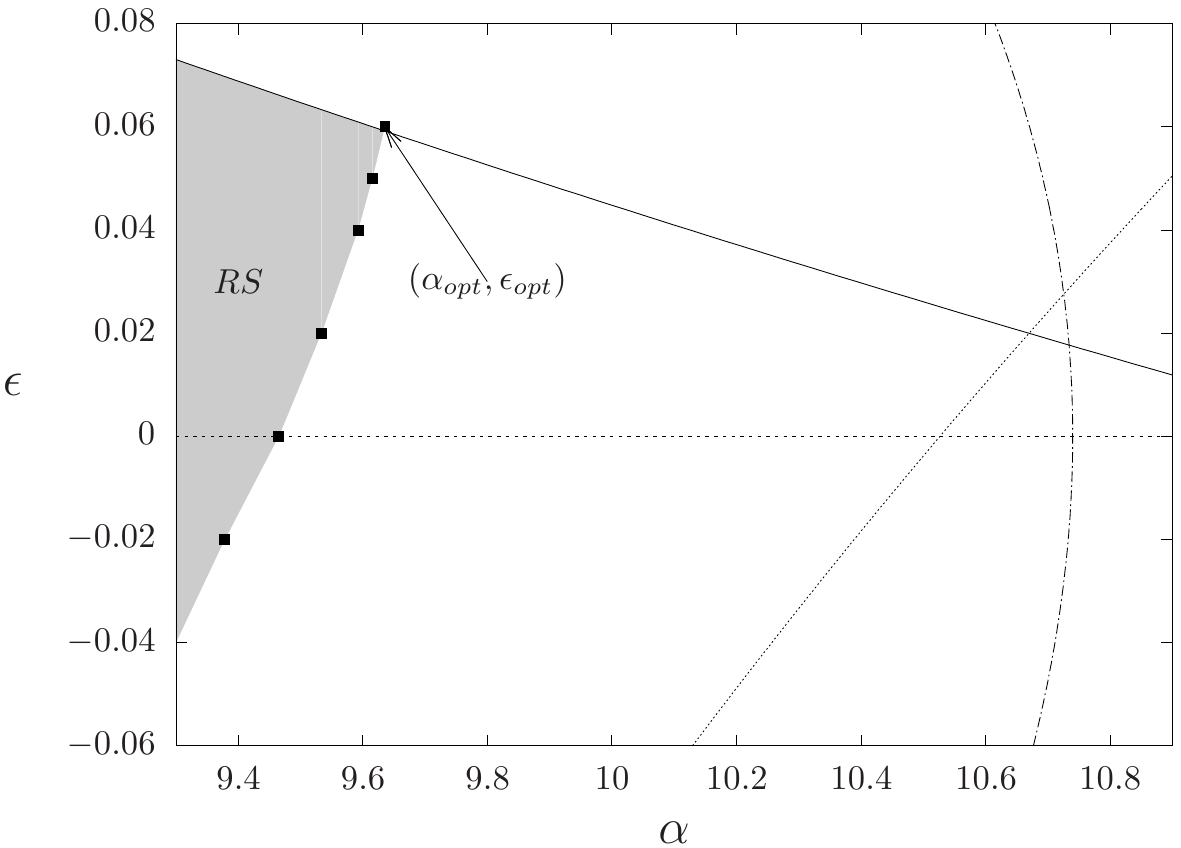}\hspace{3mm}
  \includegraphics[scale=0.43]{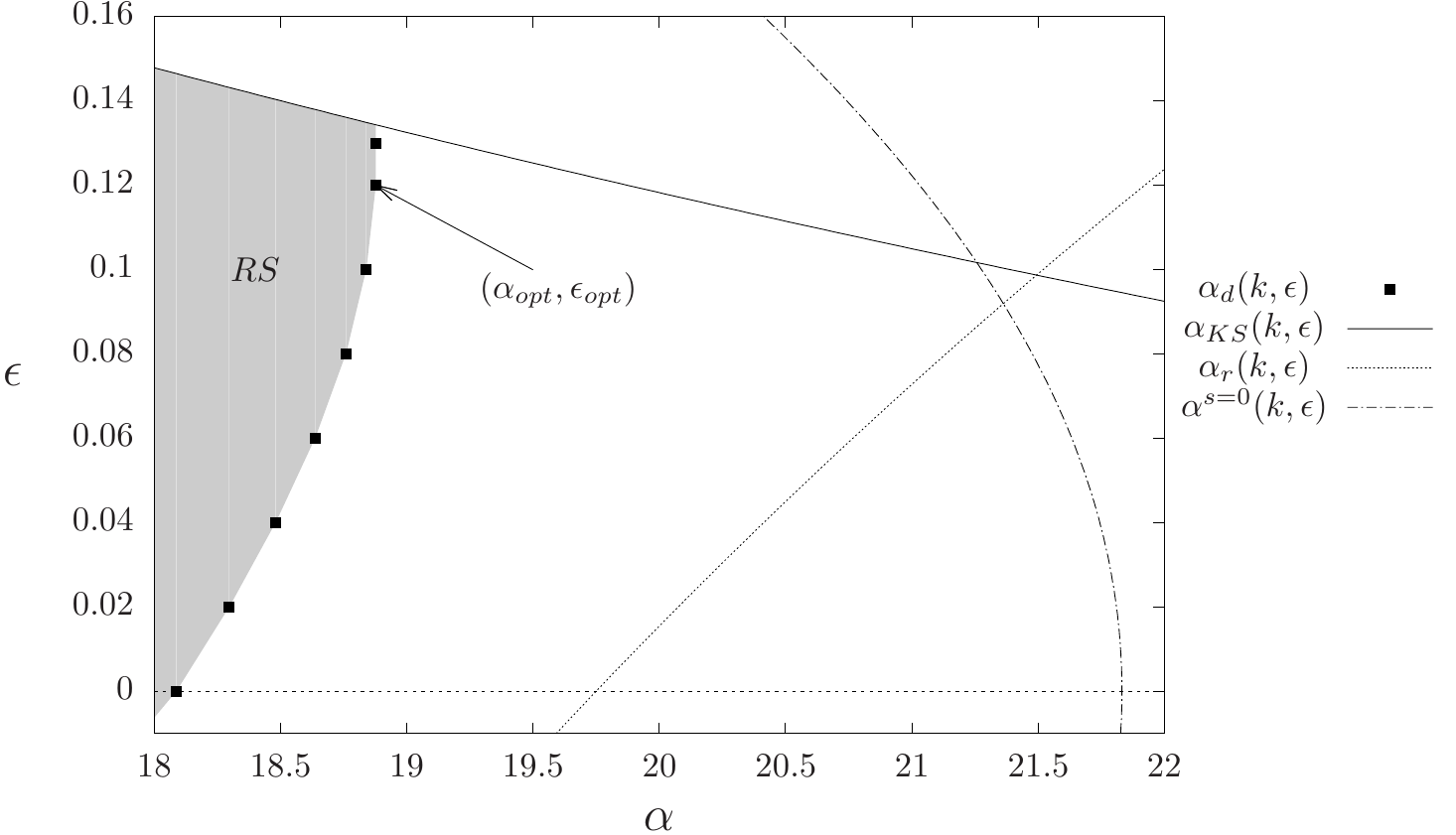}
\end{center}
\caption{Phase diagram for $k=4$, $k=5$ and $k=6$ (from left to right), in the plane $(\e, \a)$, at zero temperature $\w_0=0$. The RS phase, painted in gray, is on the left of $\ad(\e)$, the latter corresponds either to a continuous transition with $\ad(\e) = \aKS(\e)$ (solid line, see (\ref{eq_KS_epsilon})) or to a discontinous transition (black squares). The dashed horizontal line corresponds to $\e=0$, the uniform measure, which intersects $\ad$ at $\adu$. The arrow points to the optimal point that maximizes $\ad$. The dotted line is the rigidity threshold $\ar$ from (\ref{eq_rigidity_epsilon}), the dot-dashed line marks the vanishing of the RS entropy (see Eq.~(\ref{eq_s0_epsilon})).}
\label{fig_pd_T0_simplified}
\end{figure}

We present in Fig.~\ref{fig_pd_T0_simplified} phase diagrams in the $(\a,\e)$ plane for $k=4$, $k=5$ and $k=6$. The three lines in these plots correspond to the thresholds defined in (\ref{eq_s0_generic}) from the vanishing of the RS entropy, in (\ref{eq_KS_generic}) from the instability of the RS solution (Kesten-Stigum threshold), and in (\ref{eq_rigidity_generic}) from the appearance of hard fields in the solution of the 1RSB equations at $m=1$ (rigidity threshold); specializing these three expression with the choice of parameters (\ref{eq_parameters_epsilon}) yields 
\bea
\a^{s=0}(k,\e) &=& \frac{\ln 2}{\frac{k (1-\e) \ln (1-\e)}{2^{k-1} - 1 - k \e}-
\ln\left(1-\frac{1+k\e}{2^{k-1}} \right)} \label{eq_s0_epsilon} \ , \\
\aKS(k,\e) &=& \frac{1}{k(k-1)} \left(\frac{2^{k-1} -1 - k \epsilon}{1+(k-4) \epsilon} \right)^2 \ , \label{eq_KS_epsilon} \\
\ar(k,\e) &=& \frac{1}{k} \Gr(k) \frac{2^{k-1}-1-k \e}{1-\e} \ .
\label{eq_rigidity_epsilon}
\eea
In addition the black squares in Fig.~\ref{fig_pd_T0_simplified} signal a discontinuous appearance of a non-trivial solution of the 1RSB equations at $m=1$ upon increasing $\a$, that we located by a numerical resolution of these equations following the methods explained in Sec.~\ref{sec_methodology}. One can see on these plots that for all values of $\e$ there is a critical density of constraints, $\ad(\e)$, such that a non-trivial solution of the 1RSB equations at $m=1$ exist if and only if $\a > \ad(\e)$. To make this separation more visible the area on the left of $\ad(\e)$, i.e. the RS phase of the model, has been painted in gray in Fig.~\ref{fig_pd_T0_simplified}. Let us call $(\aopt,\eopt)$ the coordinates of the point on the line $\ad(\e)$ which maximizes the density $\a$ of constraints, $\aopt = \max_\e \ad(\e)$, that corresponds to an optimal choice of the bias parameter. The numerical values of these optimal parameters can be found in Table~\ref{table:result} for $k=4$, $5$ and $6$. By definition $\aopt \ge \adu=\ad(\e=0)$, the dynamic transition of the usual model, with the uniform measure over the proper bicolorings; the non-trivial result here is that the inequality is strict, i.e. that a well chosen value of the biasing parameter $\e$ is able to turn the clustered uniform measure into an unclustered biased one (for $\a \in [\adu,\aopt]$). 

\begin{table}
\centering
\begin{tabular}{ | c | c | c | c |}
\hline
$k$ & $\adu$ & $\aopt$ & $\eopt$ \\
\hline
4 & 4.083 & 4.578 & -0.10 \\
\hline
5 & 9.465 & 9.636 & 0.06 \\
\hline
6 & 18.088 & 18.879 & 0.12 \\
\hline
\end{tabular}
\caption{Dynamic threshold for the uniform measure ($\adu=\ad(\e=0)$), and largest $\a$ reachable in the RS phase, this optimal point having coordinates $(\aopt,\eopt)$.}
\label{table:result}
\end{table}

A further scrutiny of the phase diagrams reveals different scenarios depending on the value of $k$. For $k=4$ the nature of the bifurcation on the line $\ad(\e)$ changes precisely at $\eopt$: for $\e > \eopt$ the transition is continuous and thus $\ad$ coincides with the Kesten-Stigum line $\aKS$, while it is discontinuous for $\e < \eopt$ and there is a cusp at the optimal point (we shall come back on this point later on). It turns out that for $k=4$, $\eopt < 0$: this is rather counterintuitive at first sight, as it means that favoring the almost violated configurations of variables actually makes the measure less frustrated. This peculiarity can be explained by noticing that for $k=4$ the dynamic transition of the uniform measure ($\e=0$) is continuous and that $\aKS$ decreases with $\e$. As the dynamic transition of the uniform measure is discontinuous for $k \ge 5$~\cite{gabrie2017phase} this peculiarity is restricted to $k=4$, and one has $\eopt(k \ge 5)>0$. Turning now to the phase diagram for $k=5$ in Fig.~\ref{fig_pd_T0_simplified} one observes similarly a cusp in $\ad(\e)$ at $\eopt$, that separates a continuous and discontinuous branch of the dynamic transition line, but with now $\eopt >0$. Finally for $k=6$ the optimal point is on the discontinuous branch of $\ad(\e)$; increasing further $\e$ one encounters a cusp at some value of $\e > \eopt$ and then a continuous branch $\ad(\e)=\aKS(\e)$. The large $k$ behavior of the model will be further discussed in Sec.~\ref{sec_largek}; we can nevertheless anticipate that for large enough $k$ the Kesten-Stigum threshold becomes irrelevant, as it happens in the negative RS entropy region (compare the leading orders of Eqs.~(\ref{eq_s0_epsilon},\ref{eq_KS_epsilon})). In this case the whole line $\ad(\e)$ corresponds to a discontinuous bifurcation. As a last remark on the phase diagrams of Fig.~\ref{fig_pd_T0_simplified} let us emphasize that for all $\e$ one has necessarily $\ad(\e) \le \min(\aKS(\e),\ar(\e),\a^{s=0}(\e))$, these three thresholds implying a mechanism of failure for the hypotheses underlying a purely RS phase. This should easily convince the reader of the necessity of discontinuous branch of $\ad(\e)$ in some parts of the phase diagrams. For instance when $k=4$ and $\e \le -0.3$ the rigidity and negative entropy bounds imply $\ad(\e) < \aKS(\e)$, in other words the dynamic transition must be discontinuous.

\begin{figure}[ht]
\begin{center}
  \includegraphics[scale=0.45]{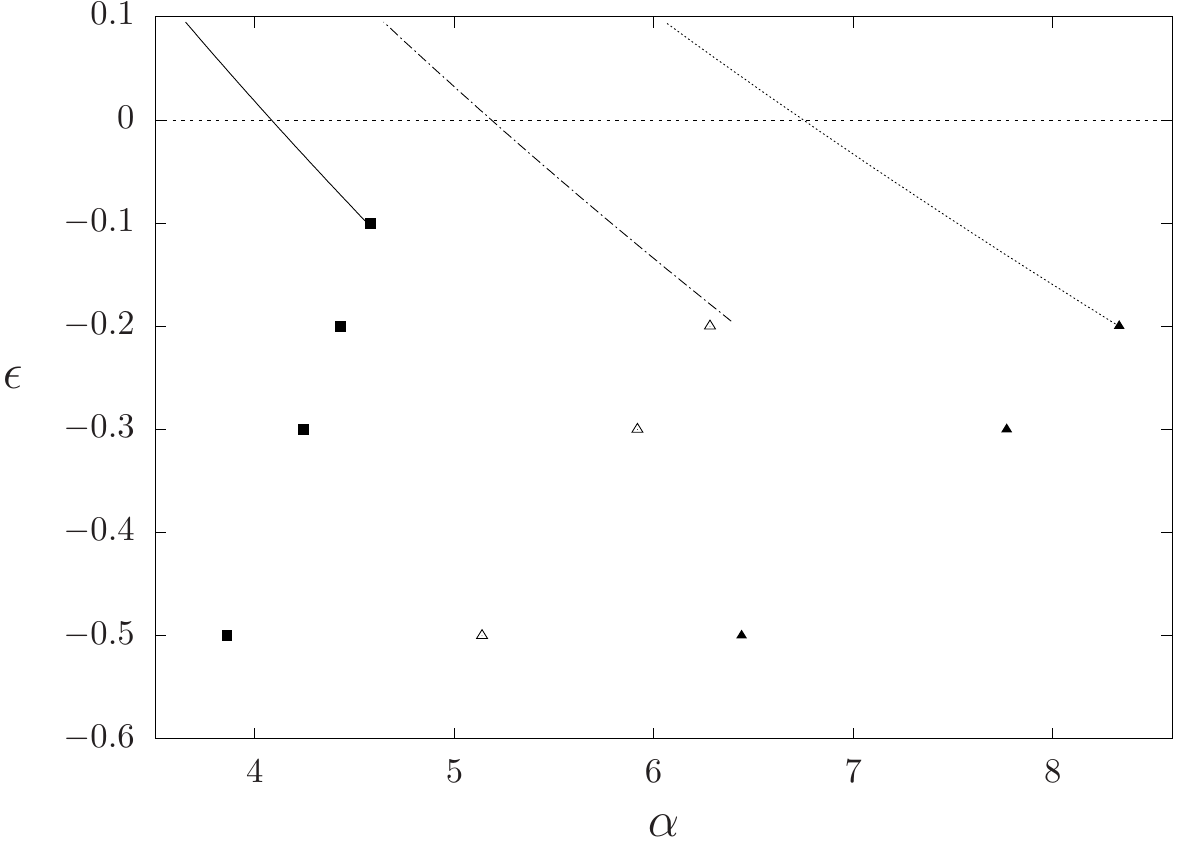}\hfill
  \includegraphics[scale=0.45]{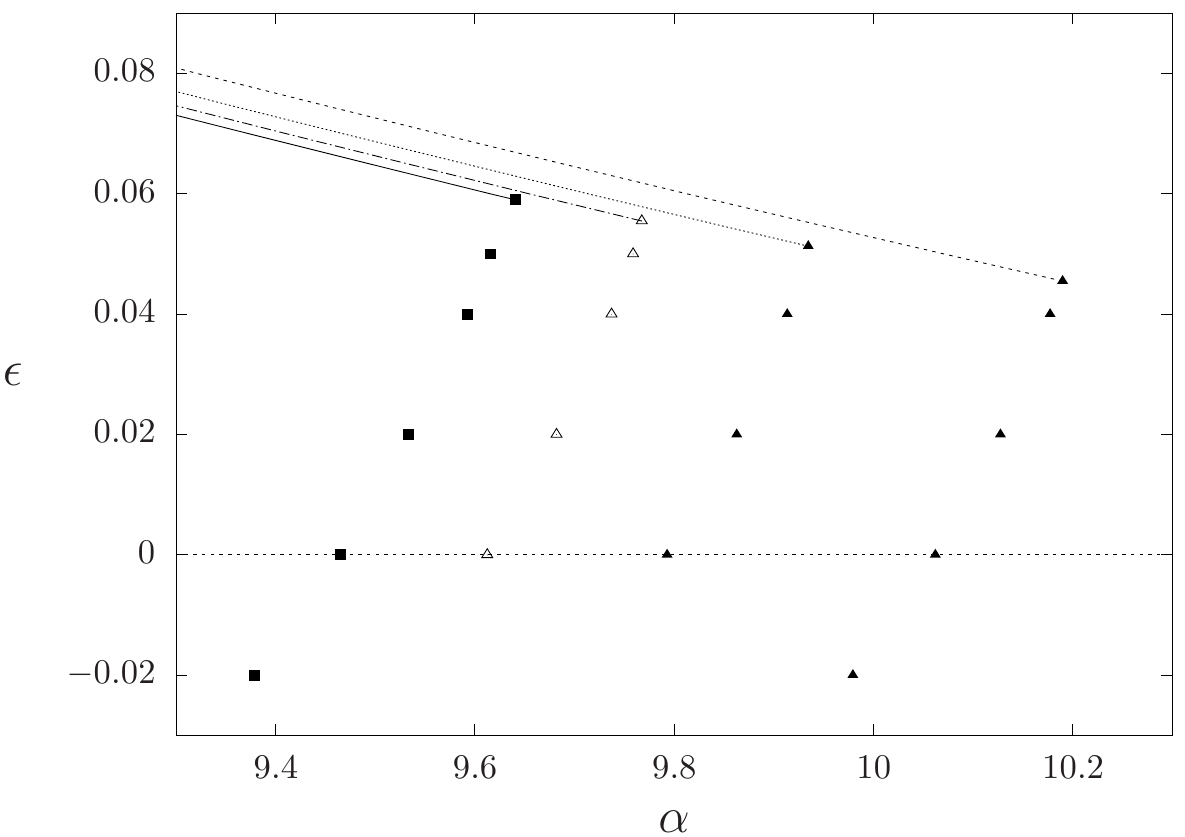}\hfill
  \includegraphics[scale=0.45]{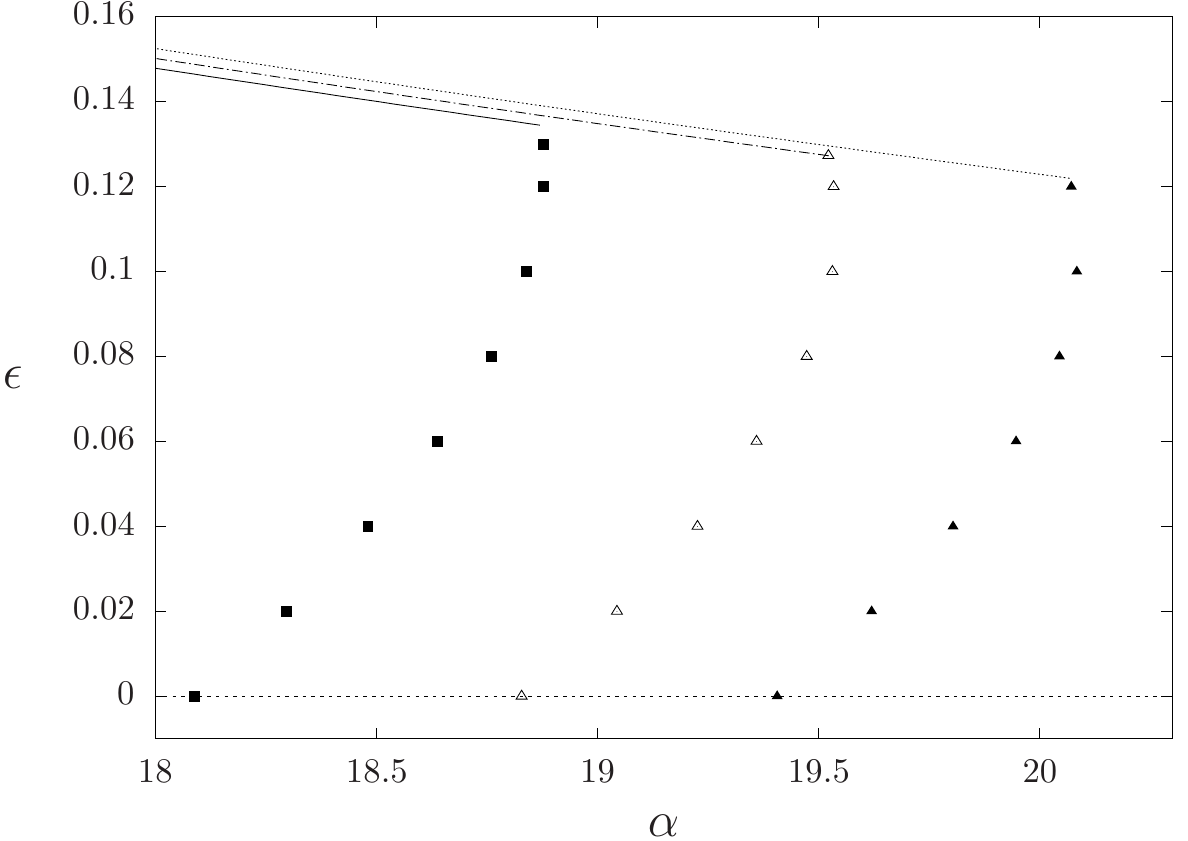}
\end{center}
\caption{Phase diagrams for $k=4$, $k=5$ and $k=6$ (from left to right) in the plane $(\a, \e)$, giving the RS phase delimitated by the KS bound and a dynamic line where the transition toward a non-trivial solution is discontinuous, for different temperatures. Left $(k=4)$:  the dynamic line is given from left to right for $\w=0$ (filled square and solid line), $\w=0.1$ (empty triangle and dashed line), $\w=0.2$ (filled triangle and dotted line). Middle $(k=5)$: from left to right $\w=0$, $\w=0.002$, $\w=0.005$, $\w=0.01$. Right $(k=6)$: from left to right $\w=0$, $\w=0.005$, $\w=0.01$.}
\label{diagram_withT}
\end{figure}

We have motivated earlier our study of the boundaries of the RS phase by algorithmic considerations, Monte Carlo Markov Chains being expected to equilibrate rapidly inside such a phase. However in a practical simulation one cannot assume that the initial configuration belongs to the support of a zero-temperature measure (otherwise the problem of finding a solution of the CSP would be already solved), it is thus necessary to make an annealing in temperature for a random initial condition to be allowed. For this reason we have also studied the evolution of the phase diagrams at positive temperature, modifying the parameters (\ref{eq_parameters_epsilon}) with $\w_0=\w_k>0$, see the results in Fig.~\ref{diagram_withT}. These plots show the absence of ``reentrance'' in temperature, in the sense that the lines $\ad(\e)$ move towards higher density of constraints when $\w_0$ is increased. Hence in principle a simulated annealing procedure with parameters $(\a,\e)$ in the zero temperature RS domain, progressively decreasing $\w_0$, should be able to remain equilibrated on polynomial time scales, hence finding solutions for $\a < \aopt$ if the appropriate bias is used. A numerical test of this conjecture is presented in Sec.~\ref{sec_results_sa}.

\subsection{More detailed zero temperature phase diagrams}

\begin{figure}[ht]
\begin{center}
  \includegraphics[scale=0.6]{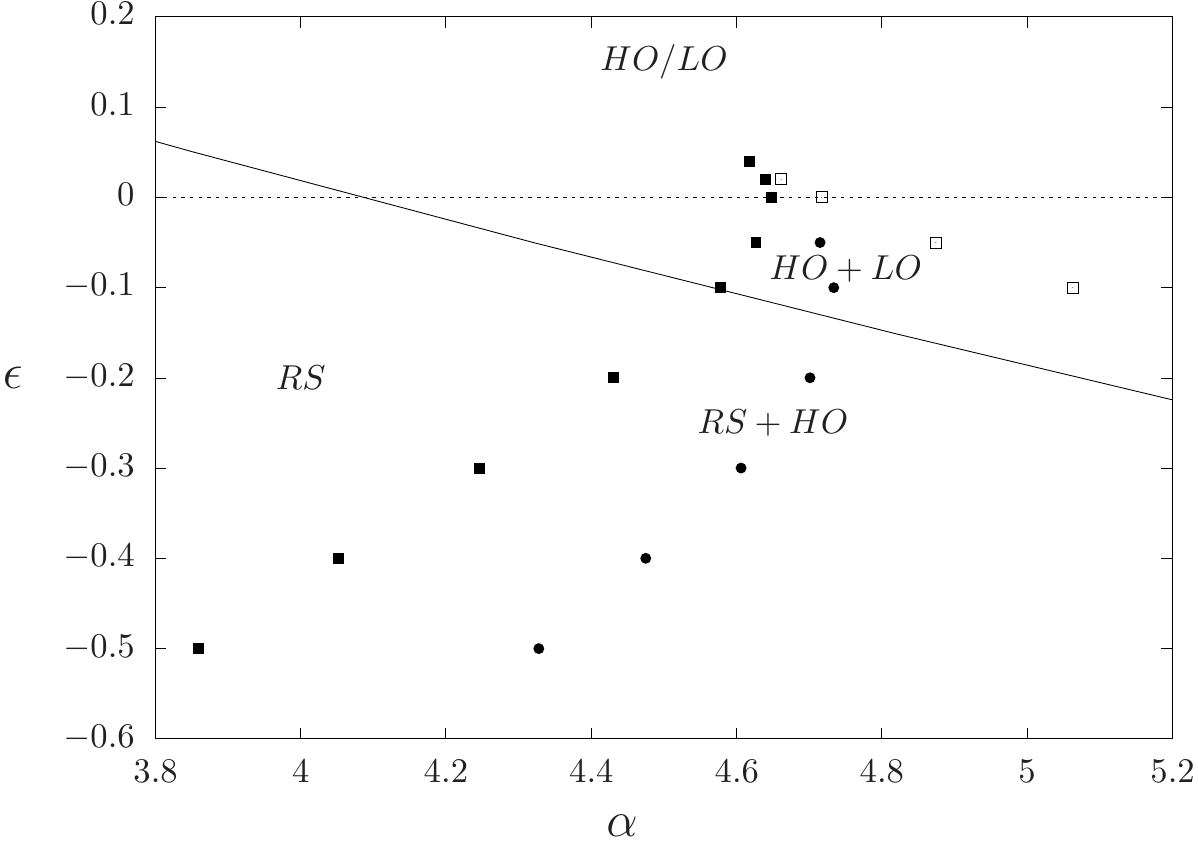}\hspace{3mm}
  \includegraphics[scale=0.6]{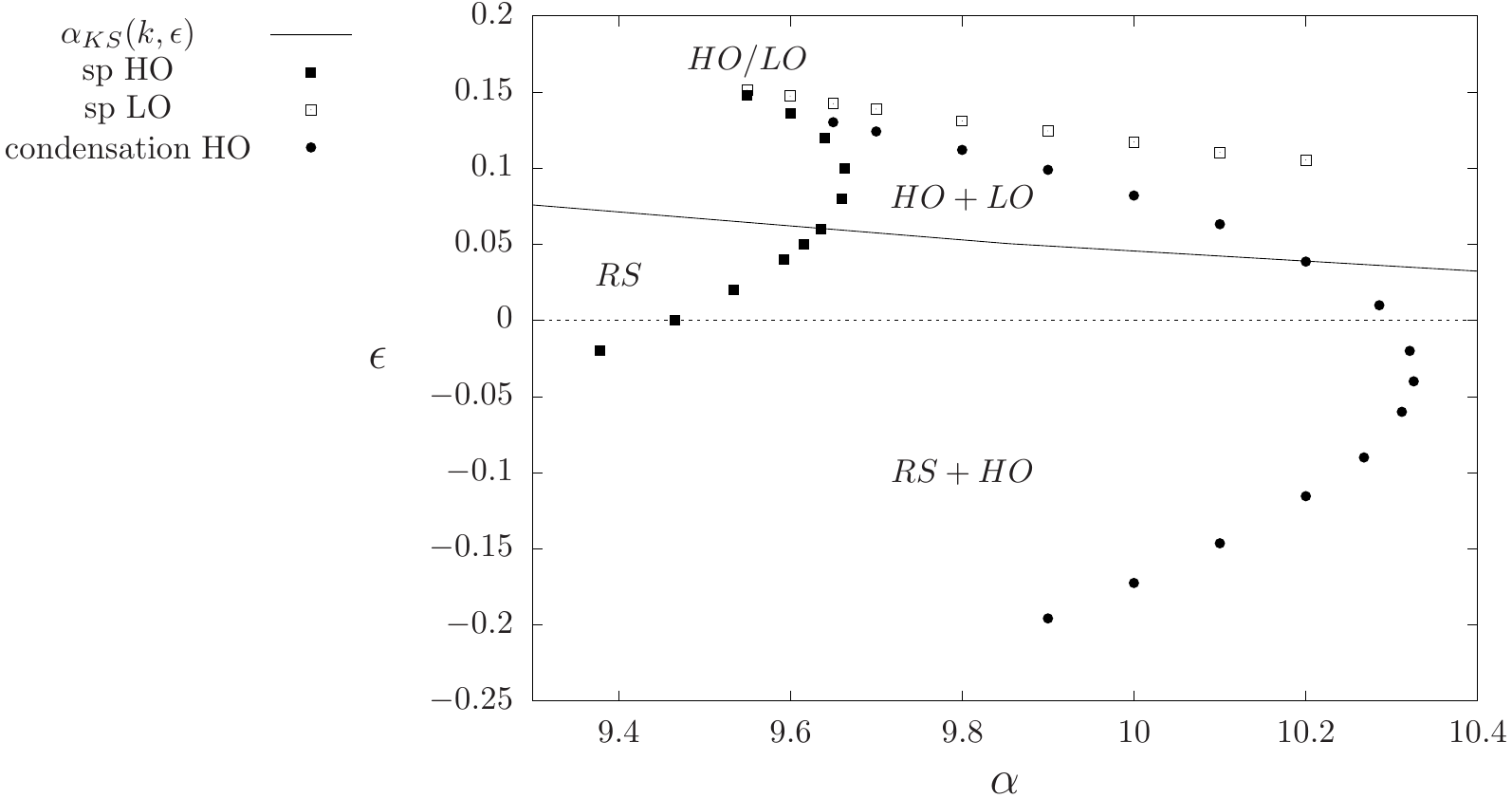}
\end{center}
\caption{Phase diagram for $k=4$ (left), and $k=5$ (right), in the plane $(\a, \e)$, at zero temperature ($\w_0=0$): the solid line is the Kesten-Stigum (KS) transition where a non-trivial solution of the 1RSB equations emerge continuously from the trivial one upon increasing $\alpha$, the filled (resp. empty) squares corresponds to the spinodal (sp) of the HO (resp. LO) branch that appears discontinuously when $\alpha$ is increased (resp. decreased). The filled circles are defined by the vanishing of the complexity of the HO branch.}
\label{diagram_epsilonVSalpha_T0}
\end{figure}

The extent of the RS domain in the $(\a,\e)$ phase diagram presented in Fig.~\ref{fig_pd_T0_simplified} was the most interesting information to extract from the cavity formalism in the perspective of this paper. For the sake of completeness we shall nevertheless discuss with slightly more details some properties of the RSB phase, and present another version of the phase diagrams for $k=4$ and $k=5$ in Fig.~\ref{diagram_epsilonVSalpha_T0}.

\begin{figure}[ht]
\begin{center}
  \includegraphics[scale=0.45]{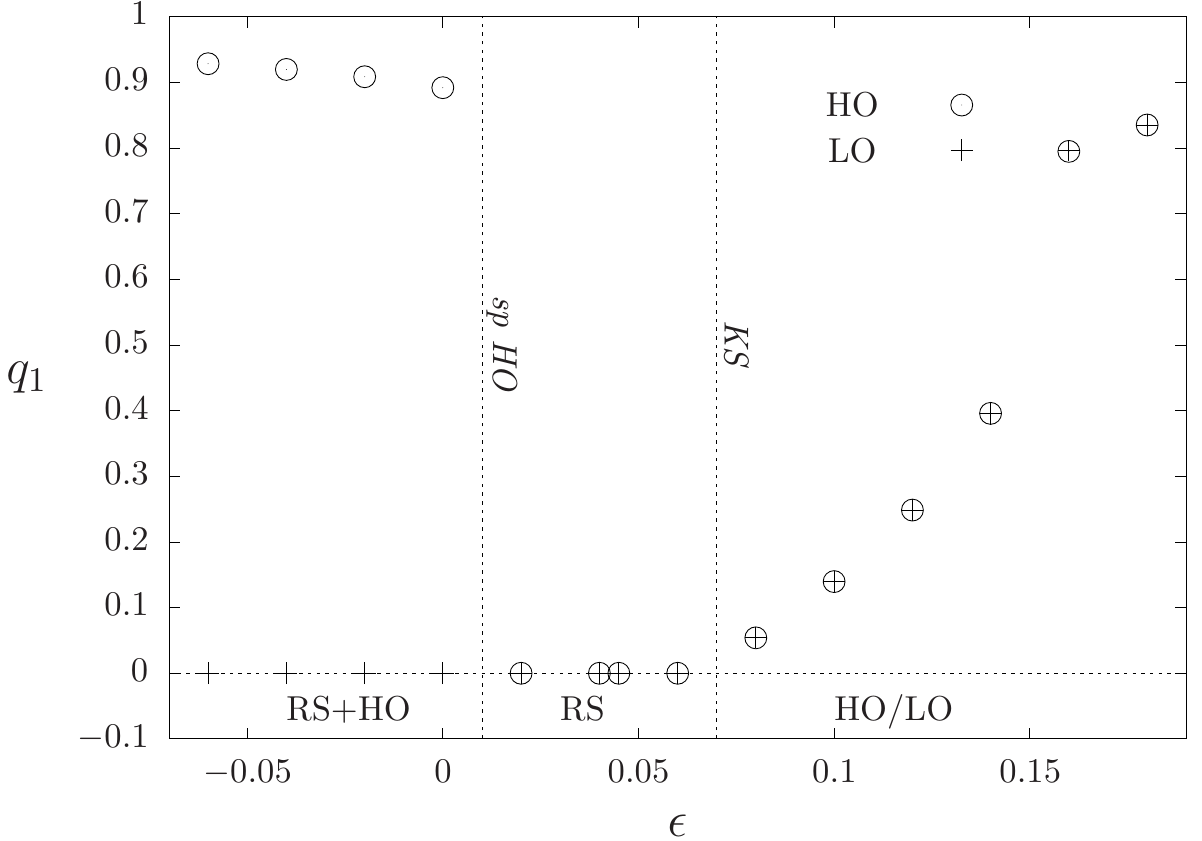}\hfill
  \includegraphics[scale=0.45]{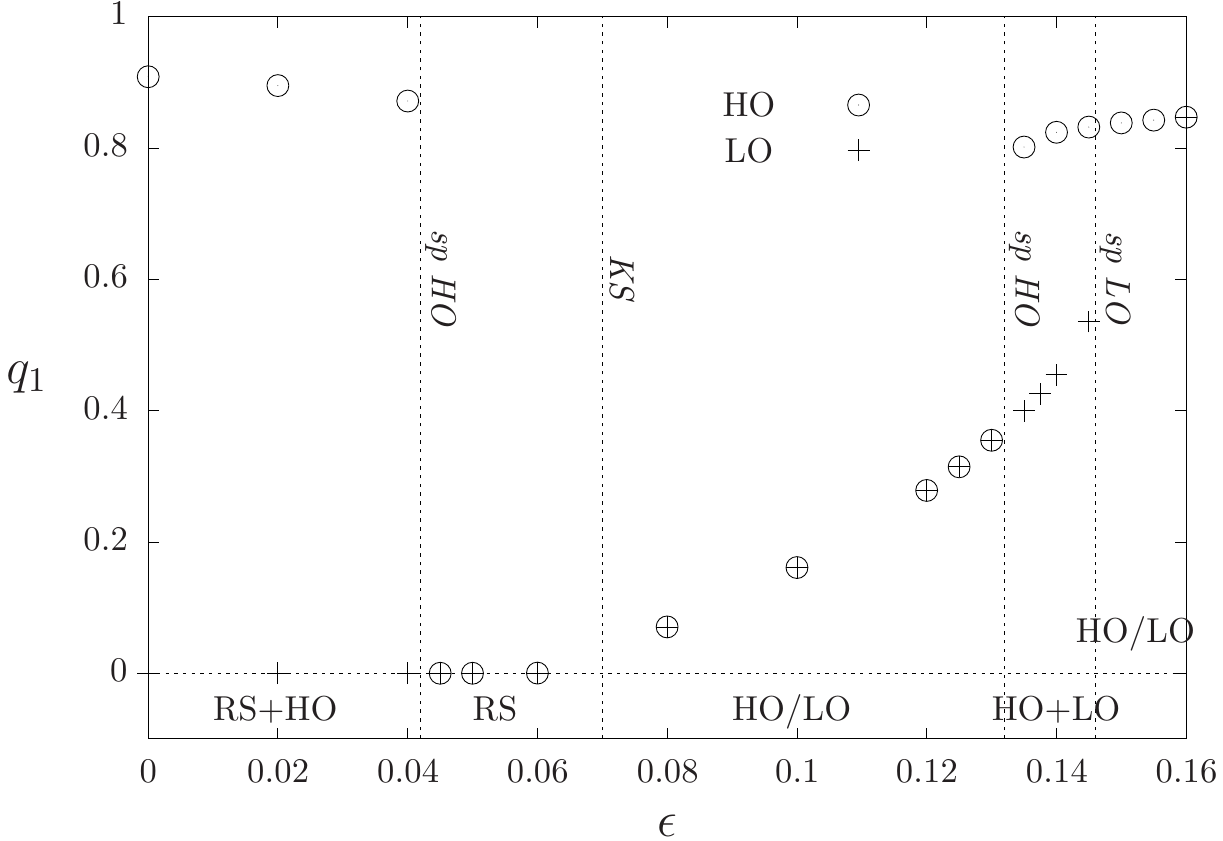}\hfill
  \includegraphics[scale=0.45]{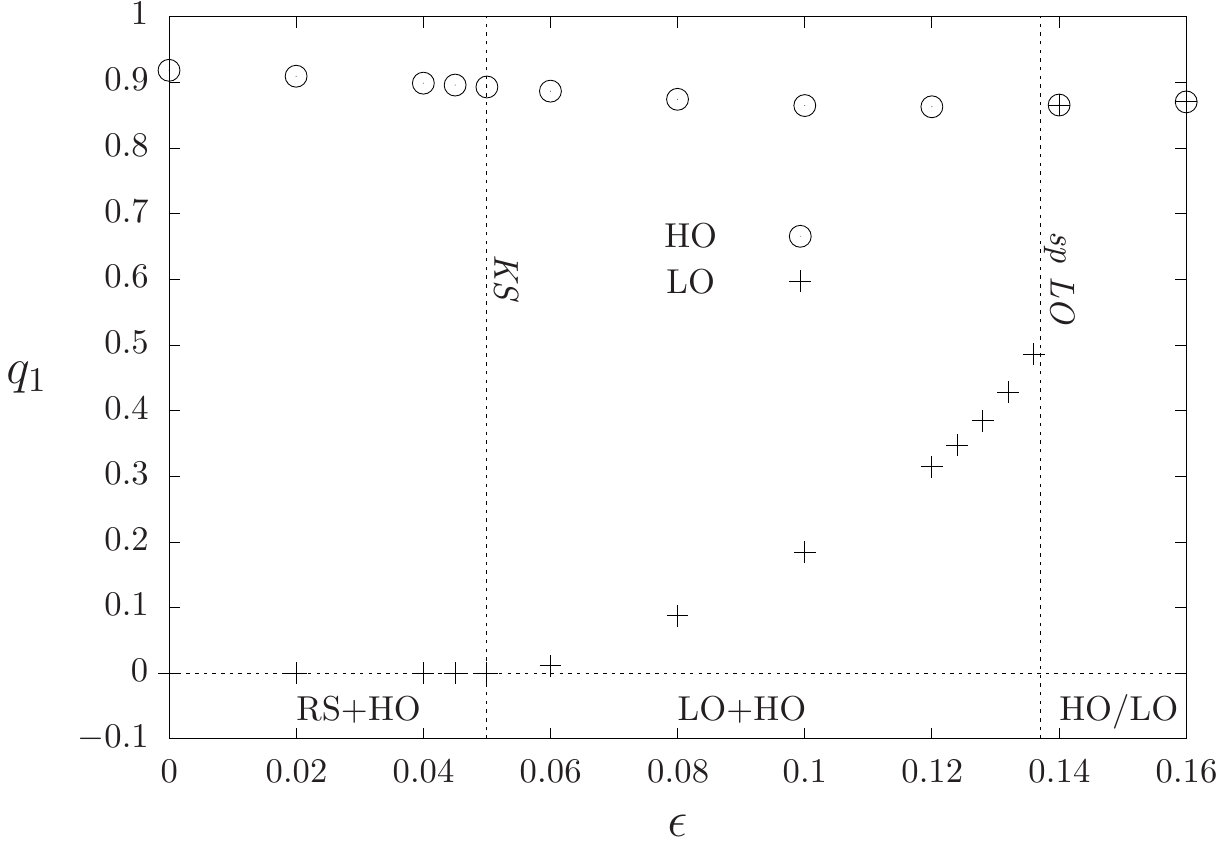}\\
  \includegraphics[scale=0.45]{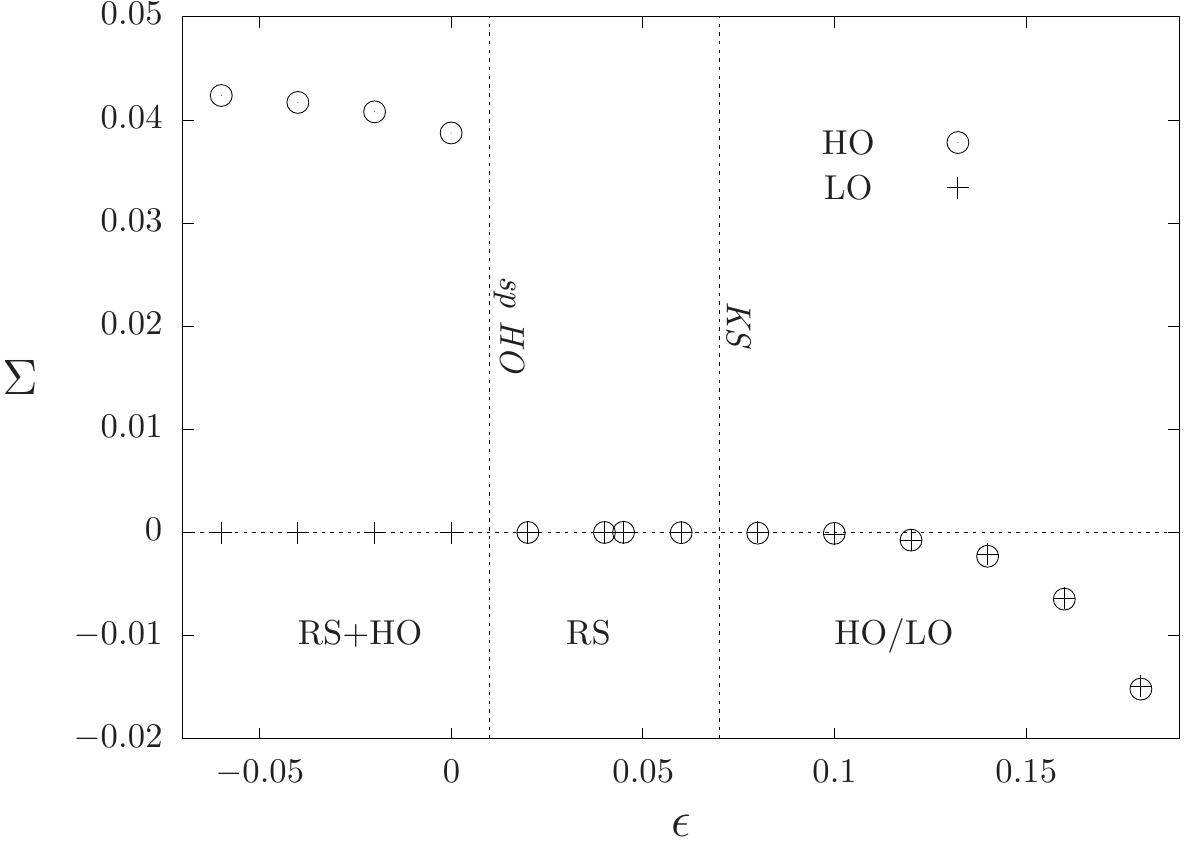}\hfill
  \includegraphics[scale=0.45]{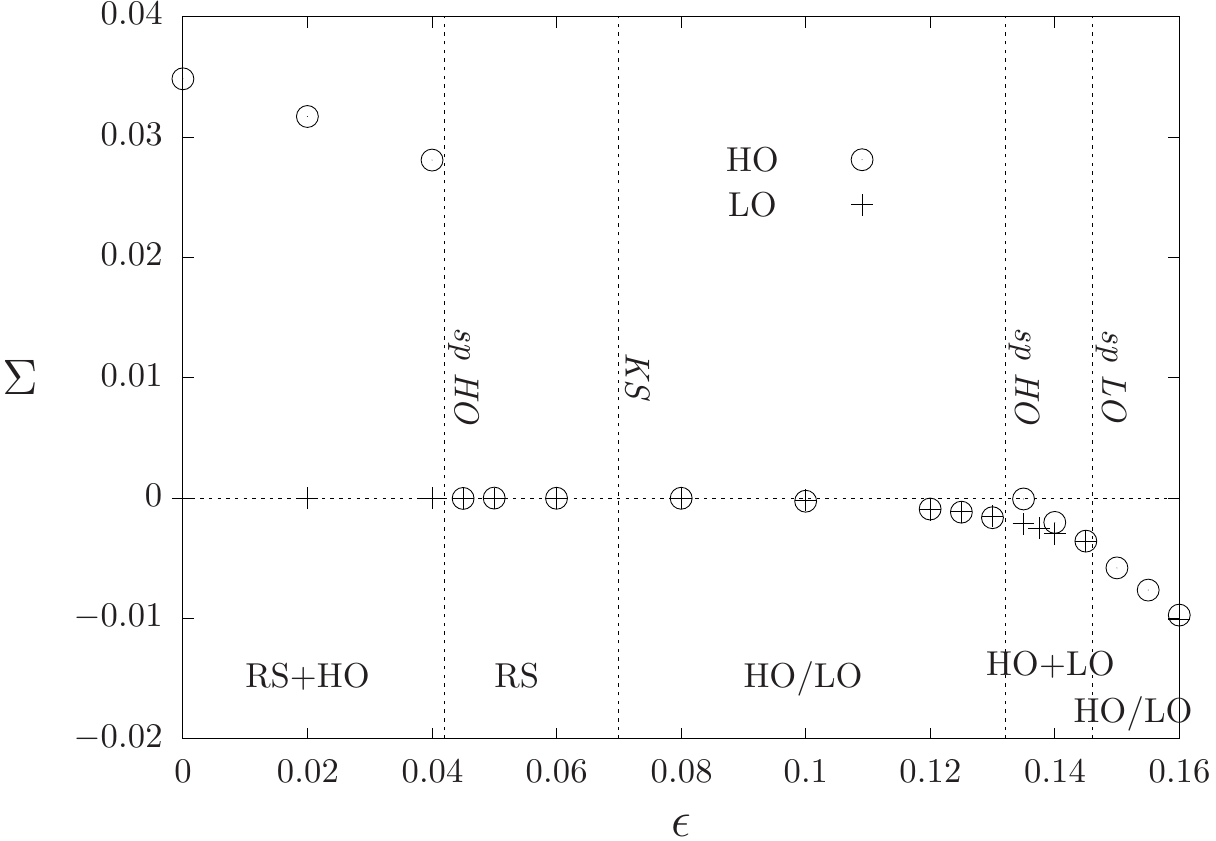}\hfill
  \includegraphics[scale=0.45]{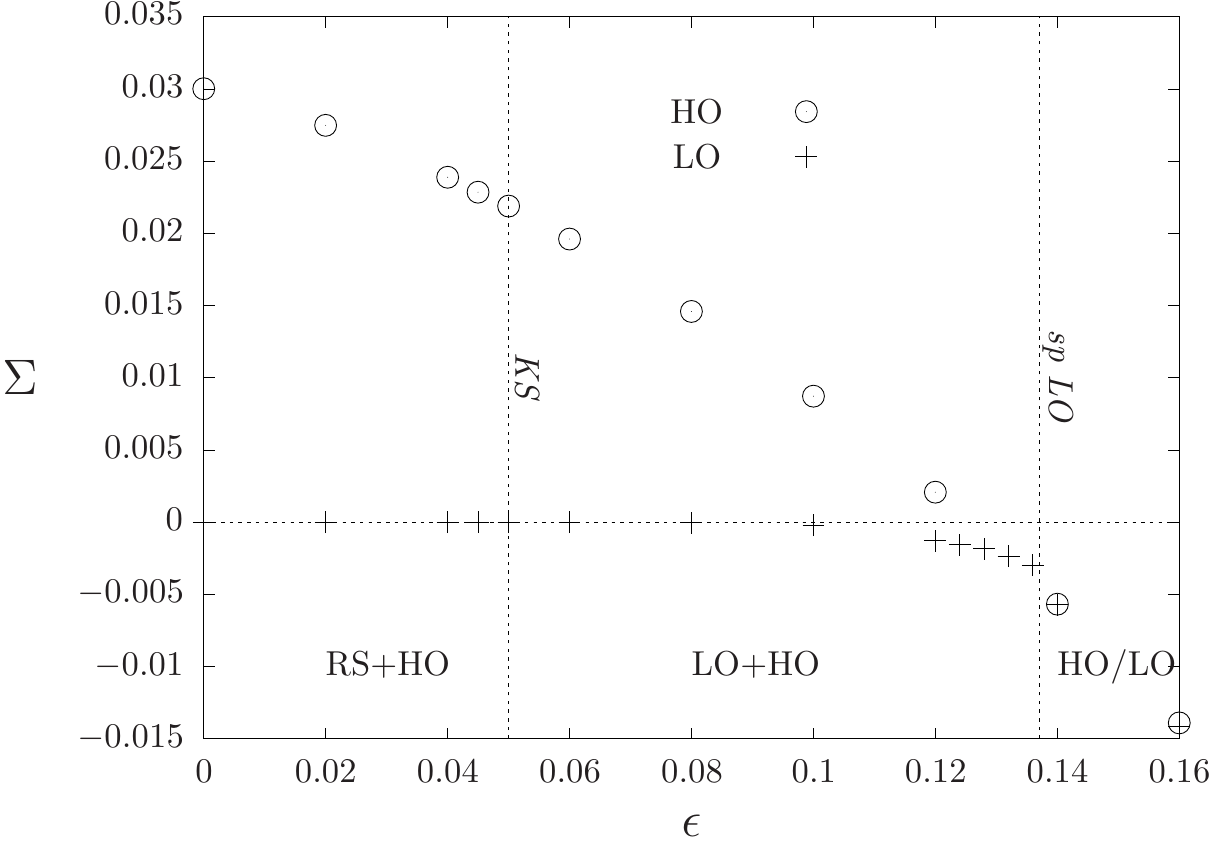}
\end{center}
\caption{Overlaps (top) and complexities (bottom) versus $\e$ at $k=5$, for $\a=9.5$ (left), $\a=9.6$ (center), and $\a=9.7$ (right).}
\label{k5-evol-HO-LO}
\end{figure}

The most important additional feature unveiled by these phase diagrams is that for some values of the parameters $k$, $\a$, $\e$, there exits (at least) two different non-trivial solutions of the 1RSB equations at $m=1$ (\ref{1RSBeqnSimplifs}). This type of behavior was described in~\cite{gabrie2017phase} for a family of random CSPs generalizing the hypergraph bicoloring, and its consequences for inference problems (or planted CSPs) have been discussed in~\cite{RiSeZd18}. In order to reach numerically these different solutions we used the population dynamics algorithm explained in Sec.~\ref{sec_simplifications_m1} with an initial condition generalizing (\ref{eq_initialcondition}) into
\beq
Q_+^{(t=0)}(h)=(1-\varepsilon) \, \delta(h) + \varepsilon \, \delta(h-1) \ .
\eeq
For each choice of the parameters we ran twice the population dynamics algorithm, once with $\varepsilon=1$ and once with a small value of $\varepsilon>0$ (in practice we used $\varepsilon=0.01$); in the tree reconstruction interpretation the latter correspond to a variant known as robust tree reconstruction~\cite{JansonMossel04}, in which only a fraction $\varepsilon$ of the variables at large distance from the root are revealed to the observer. We will call HO, for high overlap, the initialization with $\varepsilon=1$, and LO (low overlap) the small $\varepsilon$ one. Depending on the parameters these two procedures can produce different solutions of the 1RSB equations, or not. More precisely, the different phases located in Fig.~\ref{diagram_epsilonVSalpha_T0} are defined as follows:
\begin{itemize}
\item RS: both HO and LO initial conditions lead to the trivial solution.
\item RS+HO: LO initial condition leads to the trivial solution, whereas HO initial condition leads to a non trivial solution. 
\item LO+HO: LO initial condition leads to a non trivial solution, HO initial condition leads to a non trivial solution with a higher overlap. 
\item HO/LO: both HO and LO initial conditions lead to the same non-trivial solution.
\end{itemize}
The frontiers between these different phases are:
\begin{itemize}
\item $\aKS$, the limit of stability of the trivial fixed point, that undergoes a bifurcation at the Kesten-Stigum transition.
\item two spinodal lines (denoted sp HO and sp LO) that correspond to the limit of existence of the two non-trivial branches of solution of the 1RSB equations.
\end{itemize}
We invite the reader to consult also the top panels of Fig.~\ref{k5-evol-HO-LO} where the evolution of the overlap is plotted as a function of $\e$ for different fixed $\a$ at $k=5$, which should help to grasp the meaning and succession of the different phases. The frontiers between the various phases are indicated with the same names in Fig.~\ref{diagram_epsilonVSalpha_T0} and Fig.~\ref{k5-evol-HO-LO}.

This more complete study of the number and domain of existence of solutions of the 1RSB equations should clarify the cusp at $\eopt$ of the line $\ad(\e)$ found for $k=4$ and $k=5$ in Fig.~\ref{fig_pd_T0_simplified}: a first look at these figures could suggest that the two parts of the $\ad(\e)$ line join at a tri-critical point, in the sense that the discontinous transition becomes less and less discontinuous before crossing over to a continuous transition. However from Fig.~\ref{diagram_epsilonVSalpha_T0} one sees that this is not the case, the discontinuous branch of $\ad(\e)$ extends to the RSB phase as a spinodal unrelated to the continuous transition, which can only make sense in the context of coexistence of two non-trivial solutions. A tri-critical point does exist in these phase diagrams, but it is located strictly inside the RSB phase, not at the cusp, and corresponds to the merging of the two spinodals.

Finally we have also indicated on Fig.~\ref{diagram_epsilonVSalpha_T0} the threshold for the cancellation of the complexity of the HO solution (see in addition the bottom panels in Fig.~\ref{k5-evol-HO-LO}); this corresponds to the condensation transition of the model in the $\a < \aKS$ part of the phase diagram (for $\a > \aKS$ the LO solution has a negative complexity hence the problem is condensed, see~\cite{gabrie2017phase} for a discussion of this point).

\section{Results of simulated annealing}
\label{sec_results_sa}

In this Section we present the results of extensive simulations, where we have used the Simulated Annealing (SA) algorithm \cite{KirkpatrickGelatt83} to find solutions of the hypergraph bicoloring problem.
Our main aim is to show that SA finds solutions more easily if the biased measure is used: although the uniform measure ($\e=0$) has a larger entropy, the biased one is more concentrated on solutions that can be reached in an easier way by SA and thus the SA algorithmic threshold improves if $\e\neq0$ is used.

As in the rest of paper, we consider Erd\H os-R\'enyi random hypergraphs with $k=4,5,6$ and sizes ranging from $N=10^4$ to $N=10^6$. The parameters $\a$ and $\e$ are taken in the relevant region where we expect an algorithmic phase transition to take place, that is around $\ad$. Let us rewrite the biased measure that we are willing to sample via the SA at a generic finite temperature $T=1/\beta$ as
\begin{equation}
\mu(\us) = \frac{1}{Z(G)} e^{-\beta U(\us)} (1-\e)^{F(\us)}
\label{eq_mu_SA}
\end{equation}
where $U(\us)$ is the number of unsatisfied constraints (i.e.\ monochromatic hyperedges) and $F(\us)$ is the number of freezing clauses (i.e.\ hyperedges with exactly $k-1$ variables of the same color). This corresponds to the measure (\ref{measure}) with parameters 
\beq
\w_0=\w_k=e^{-\beta} \ , \qquad  \w_1=\w_{k-1}=1-\e \ , \qquad \w_2=\dots=\w_{k-2}=1 \ .
\eeq
The solutions of the CSP have $U=0$, and non-uniform weights if $\e\neq0$.
Our SA implementation uses the Metropolis algorithm with single-spin flip dynamics: at each time step one considers a configuration $\us'$ that differs from the current configuration $\us$ by the reversal of an uniformly chosen spin. The move $\us \to \us'$ is accepted with the probability
\beq
\min\left(\frac{\mu(\us')}{\mu(\us)},1\right) = \min\left(e^{-\beta \Delta U} (1-\e)^{\Delta F},1\right) \ ,
\eeq
where $\Delta U=U(\us')-U(\us)$ and $\Delta F=F(\us')-F(\us)$, in such a way that the detailed balance (reversibility) condition with respect to the measure (\ref{eq_mu_SA}) is ensured. We store the quantity $\sum_{i=1}^k \sigma_i$ for each clause, which allows a fast computation of the changes $\Delta U$ and $\Delta F$ when a spin is flipped.

We run SA with a very simple piecewise constant, uniformly spaced, temperature scheduling: the first Monte Carlo Sweep (MCS, i.e. $N$ elementary steps described above) is performed with $T$ fixed to $T_{\rm max}$ (we used $T_{\rm max}=0.5$ in all our simulations), then $T$ is reduced by $\Delta T = T_{\rm max} / \tau$ and a new MCS is performed, $T$ is again reduced by $\Delta T$, and so on and so forth. We perform in this way $\tau+1$ MCS, the last one being at zero temperature, the running time of the algorithm thus scales as $N \tau$ elementary steps.

The lowest value of $U(\us)$ is always reached at the end of each run, when the annealing has reached zero temperature. So we present results only for the quantity $U_0=U(T=0)$, that is the smallest number of violated clauses that the SA is able to reach in a running time of $\tau$ MCS. SA is successful as a solver if and only if $U_0=0$, but we will be interested in estimating the lowest energy reachable by SA even in the regime where it is not successful. In particular we are going to study the lowest intensive energy reached by SA, $u_0 = U_0/N$, in the large size limit where it becomes independent on the problem size $N$.

\subsection{Estimating the algorithmic threshold for Simulated Annealing}
\label{sec:epsZero}

We shall first discuss the problem of the estimation of the algorithmic threshold for a stochastic algorithm like SA, concentrating for simplicity on the unbiased ($\e=0$) case, the extension to $\e \neq 0$ will be considered later on.

The behavior of the algorithm depends on the density of constraints $\a$, the annealing time $\tau$, and the size of the problem $N$; it can be described in terms of the average energy density $u_0(\a,\tau,N)$ reached at the end of the run, or in terms of the probability (with respect to the random instance generation and the stochasticity of the algorithm) $p_{\rm succ}(\a,\tau,N)$ that the algorithm discovers a solution of the instance. It is clear that the energy (resp. success probability) reached by SA is a decreasing (resp. increasing) function of the running time $\tau$. We are interested in the limit of large times but sub-exponential with respect to the problem size $N$ (on exponentially large timescales any Monte Carlo simulation of a finite size system is ergodic and $u_0=0$ as long as $\a<\asat$, but this is not the regime we are interested in). An idealized definition of the algorithmic threshold $\a_{\rm algo}$ would be the smallest density of constraints such that 
\beq
\lim_{N \to \infty} u_0(\a,\tau=N^c,N) >0 \qquad \text{or} \ \ \ 
\lim_{N \to \infty} p_{\rm succ}(\a,\tau=N^c,N) = 0 \ ,
\eeq
for any fixed exponent $c$, corresponding to polynomial time algorithms. Of course time and space requirements impose strong constraints on the values of $\tau$ and $N$ that can be used in practice. The limit above must thus be performed by an extrapolation from finite $N$ results, and if $c$ is free any running time could be considered as ``polynomial'' as long as $N$ is finite. To resolve this ambiguity we shall restrict our study to linear times (this time scale is the only one practically accessible on very large problems), i.e. consider $\tau$ fixed (but arbitrary large) in the thermodynamic limit $N \to \infty$.

\begin{figure}[t]
\begin{center}
  \includegraphics[width=0.48\textwidth]{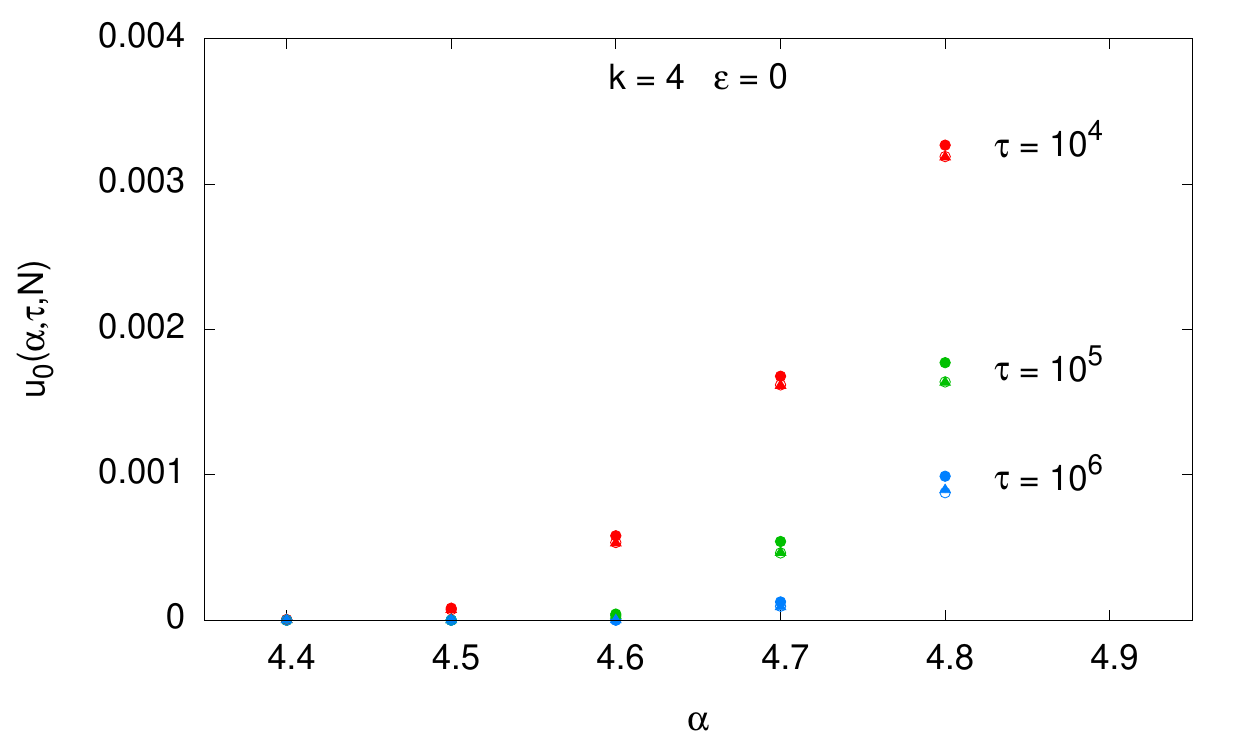}\hspace{5mm}
  \includegraphics[width=0.48\textwidth]{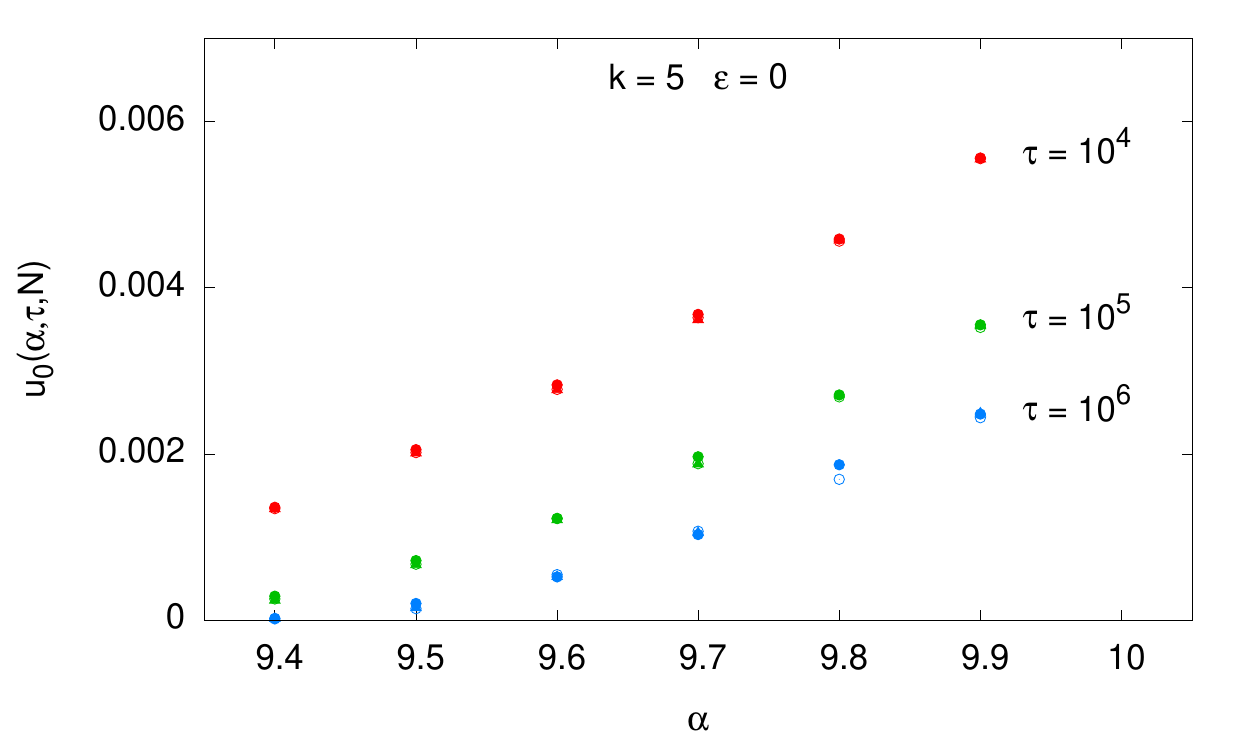}
\end{center}
\caption{Lowest intensive energy reached by SA for $k=4$ (left) and $k=5$ (right) as a function of $\a$ for different cooling times $\tau$. For each cooling time $\tau$ we show results for 3 problem sizes: $N=10^4$ with filled circles, $N=10^5$ with empty circles and $N=10^6$ with triangles. The latter two values do always coincide (except for $k=5$ and $\a=9.8$, where the $N=10^6$ datapoint is missing). Estimating the algorithmic threshold from these plots is very difficult due to the strong $\tau$ dependency.}
\label{fig:epsZeroVsAlpha}
\end{figure}

Even with this restriction the numerical extrapolation necessary to estimate $\a_{\rm algo}$ is far from being an easy task. The definition given above relies on the behavior of the asymptotic intensive energy $u_0$ as a function of $\a$. We plot the corresponding data in Figure~\ref{fig:epsZeroVsAlpha} for $k=4,5$: these data have been obtained for the unbiased measure ($\e=0$) and different problem sizes ($10^4\le N\le 10^6$); note that data points with different $N$ values are very close, i.e.\ the size dependence is very weak, and the values for $N=10^5$ and $N=10^6$ always coincide within errorebars. Unfortunately the asymptotic energy $u_0$ is strongly dependent on the running time $\tau$ and it is thus very difficult to extract from this figure the algorithmic threshold, i.e.\ the value of $\a$ where $u_0$ becomes positive in the large $\tau$ limit.

\begin{figure}[t]
\begin{center}
  \includegraphics[width=0.48\textwidth]{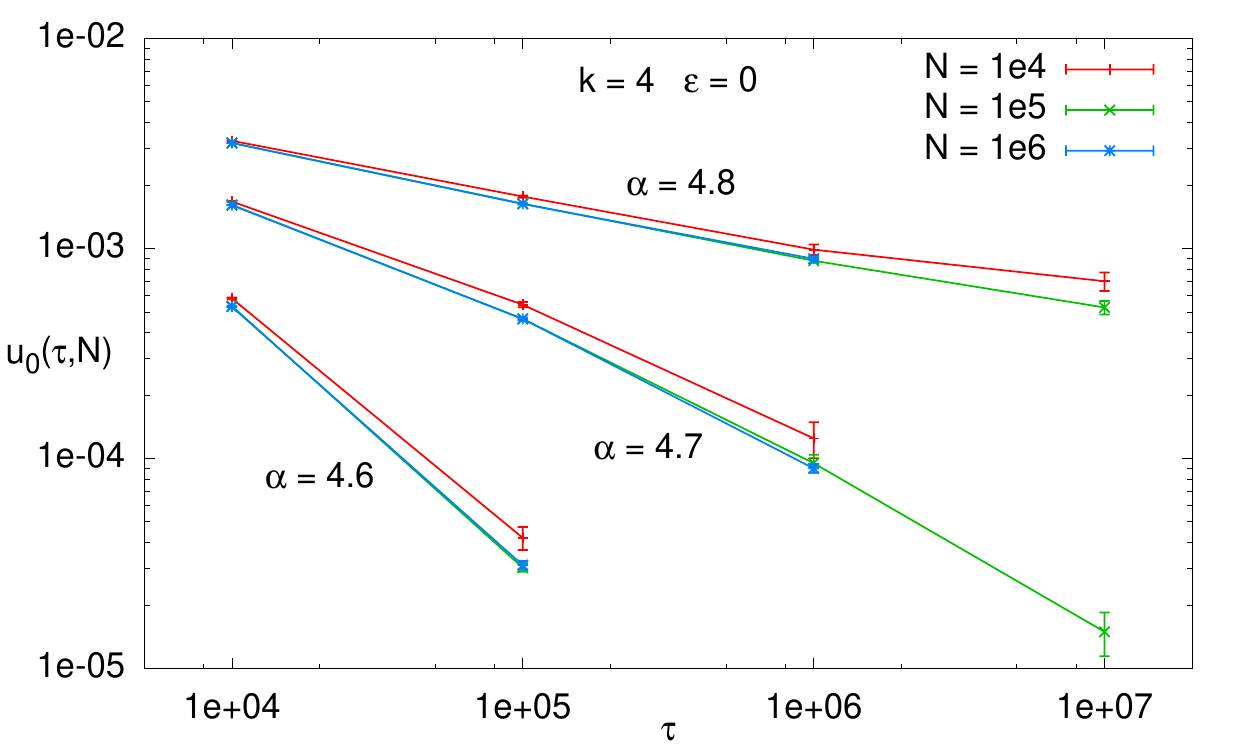}\hspace{5mm}
  \includegraphics[width=0.48\textwidth]{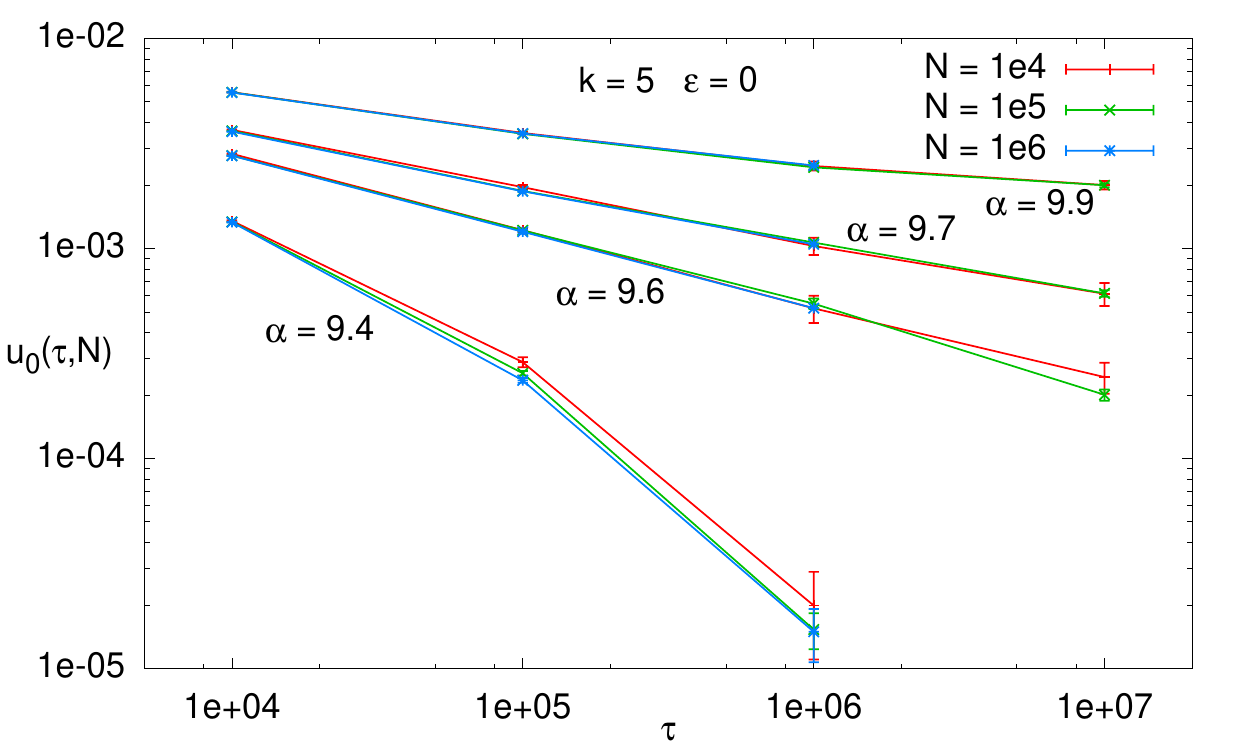}\\
  \includegraphics[width=0.48\textwidth]{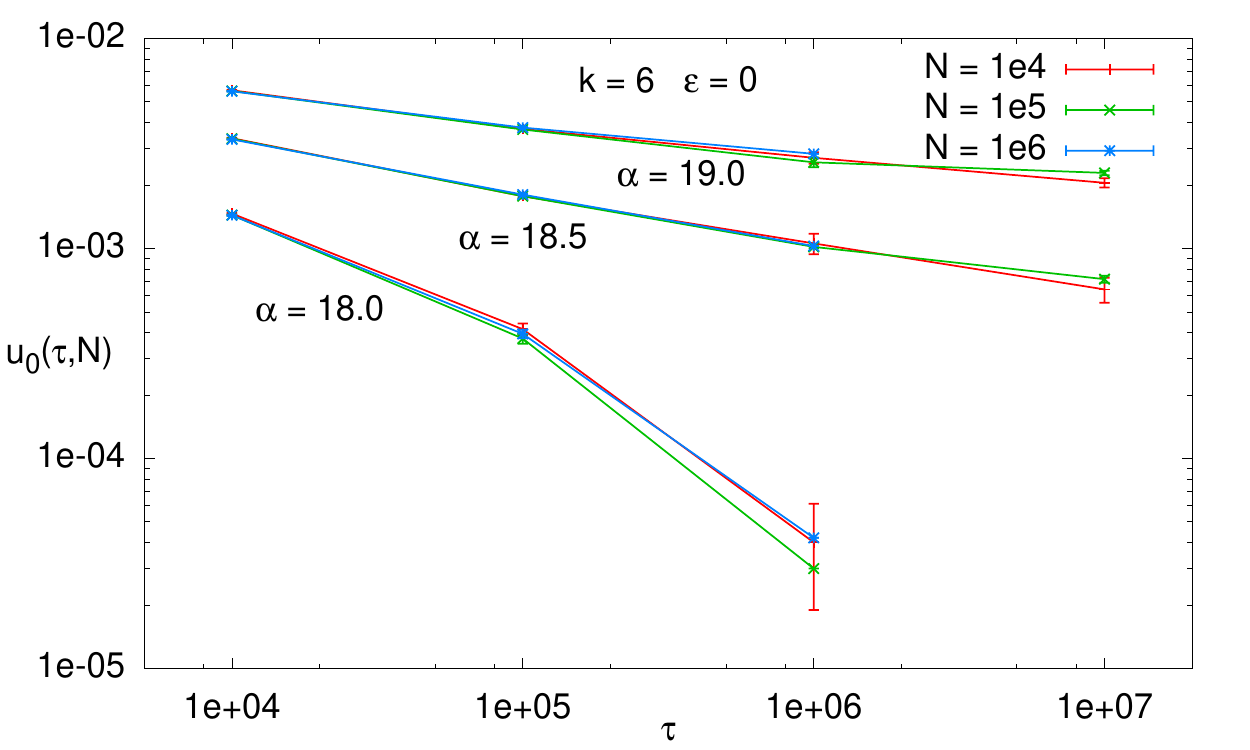}
\end{center}
\caption{Plotting the lowest intensive energy reached by SA with no bias ($\e=0$) as a function of $\tau$ provides a better way to estimate the algorithmic threshold $\aalgo$: the decay is faster (resp.\ slower) than a power law if $\a<\aalgo$ (resp.\ $\a>\aalgo$).}
\label{fig:epsZero}
\end{figure}

A much more convenient way of analyzing the same data is presented in Figure~\ref{fig:epsZero}, where for each value of $\a$ we study the dependence of $u_0$ on $\tau$. The $\a$ values shown are such that the relative difference between the smallest and the largest $\a$ values is around 5\%. Again the size dependence is weak and we can mostly ignore it. Error bars have been computed only from sample to sample fluctuations. The main observation now --- note the log-log scale in the plots --- is that for the smallest $\a$ values shown in the plots the asymptotic energy is decreasing very fast with $\tau$, faster than a power law (data not shown have $u_0\simeq 0$); on the contrary, for the largest $\a$ values, $u_0$ decreases slower than a power law. In the latter case we even observe an upwards curvature, suggesting a non-zero value for $u_0$ in the $\tau\to\infty$ limit.

In practice, our best estimate for the SA algorithmic threshold is given by the $\a$ value such that $u_0$ decays as an inverse power law of $\tau$, thus separating the regimes where $u_0$ decays faster and slower than a power law in $\tau$. For $\e=0$, we find the following approximate values $\aalgo(k=4) \approx 4.7$, $\aalgo(k=5) \approx 9.6$ and $\aalgo(k=6) \approx 18.5$.
We notice that all these algorithmic thresholds are larger than the threshold $\adu$ listed in Table~\ref{table:result}, where the ``dynamic'' phase transition, defined as the appearance of a solution of the 1RSB equations, takes place. This observation is consistent with the idea that sampling solutions uniformly is more difficult than just finding one or few solutions. Indeed, while a MCMC is expected to sample uniformly the solutions efficiently only for $\a<\adu$, SA can find a solution in linear time until $\aalgo$, which is greater than $\adu$. In other words, in the range $[\adu,\aalgo]$ the SA algorithm does not thermalize at the lowest temperatures explored during the annealing, but it can be seen as an efficient out of equilibrium process converging in linear time to a solution, as discussed for instance in~\cite{ZdeborovaKrzakala10}.

In particular for $k=4$ the model has a continuous phase transition and the SA algorithm seems to be very efficient in this case: the algorithmic threshold $\aalgo(k=4) \approx 4.7$ is well beyond the dynamic threshold $\adu=4.083$ and not far from the 1RSB estimate of the satisfiability threshold $\asat(k=4) \approx 4.9$ \cite{gabrie2017phase} (this is only expected to be an upperbound on the true satisfiability threshold due to an instability towards higher levels of RSB). On the contrary for $k\ge5$ the phase transition taking place at $\adu$ is of the random first order type (discontinuous) and this seems to have a dramatic effect on the performance of SA, which is able to find solutions only slightly beyond $\adu$, stopping far from the $\asat$ threshold. For example, for $k=5$ we have $\adu=9.465$, $\aalgo \approx 9.6$ and $\asat=10.46$ \cite{gabrie2017phase}.

\subsection{Performances of Simulated Annealing with optimal RS parameters}

\begin{figure}[t]
\begin{center}
  \includegraphics[width=0.6\textwidth]{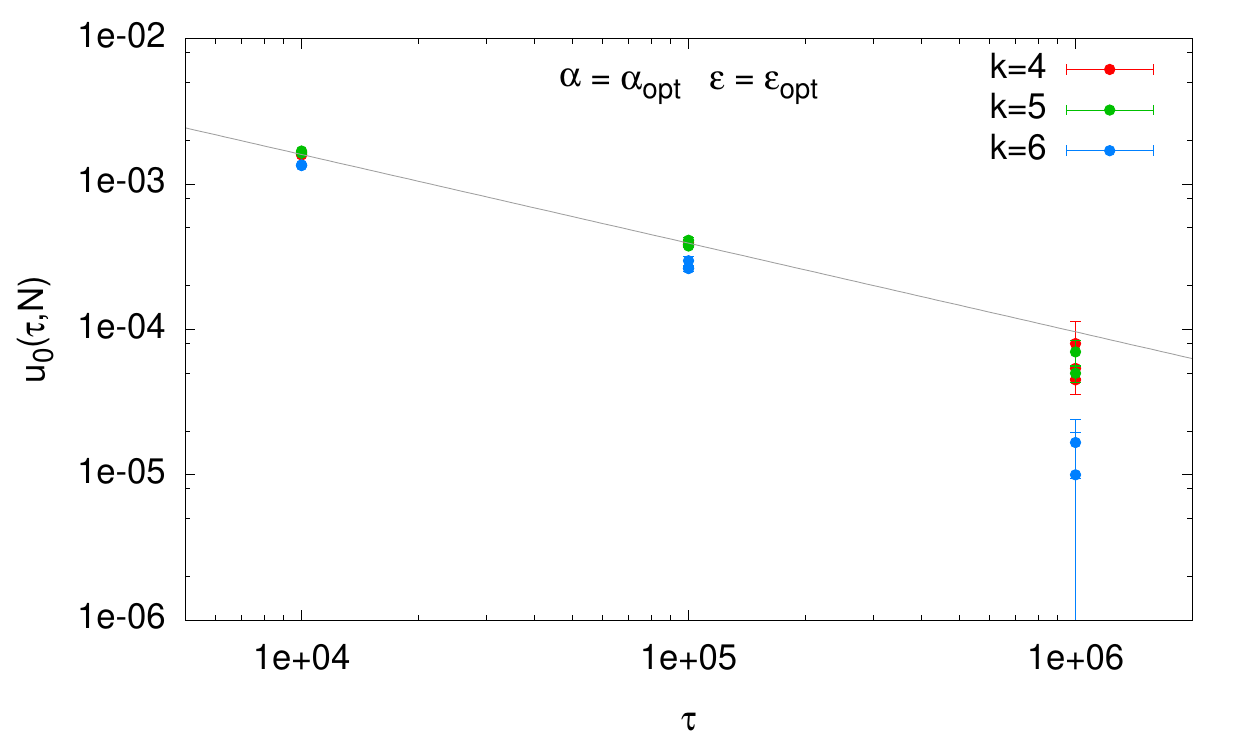}
\end{center}
\caption{The fast decays of $u_0$ as a function of $\tau$ for $\a=\aopt$ and $\e=\eopt$ confirms that for these ``optimal'' parameters SA is effective in finding the ground state in linear time.}
\label{fig:aOpt}
\end{figure}

As shown in Sec.~\ref{sec_optimalRS} we can extend the RS phase in the region $\a>\adu$ by tuning appropriately the bias $\e$. In Figure~\ref{fig:aOpt} we show the asymptotic energy as a function of the running time $\tau$ for the parameters $\a=\aopt$ and $\e=\eopt$ given in Table~\ref{table:result}, that are optimal from the point of view of extending the RS phase to the largest $\a$ possible. For each value of $\tau$ we report the results obtained with 3 sizes $N=10^4,10^5,10^6$ although the different data points are hardly visible due to their strong overlap (for $\tau=10^6$ fewer sizes are shown). Errors are computed from sample to sample fluctuations.

For all the values of $k=4,5,6$ the behavior of the asymptotic energy is compatible with a power law decay or even faster than that (the straight line is just a guide to the eye with a slope $-0.61$). So, as expected, SA seems to be an efficient algorithm to find solutions in the RS phase, even when this phase extends beyond $\adu$ via the optimization of the bias $\e$.

\subsection{Performances of Simulated Annealing with the biased measure}

In Section~\ref{sec_results_cavity} we have shown how the phase diagram and the corresponding thresholds change in presence of a non-zero bias ($\e\neq0$).
The suggestion we get from this analytical study is that a non-zero bias should make easier for the SA algorithm to find solutions at large $\a$ values.
However the connection between the phase diagram and the behavior of the SA is not obvious, as already shown in Section~\ref{sec:epsZero} for the $\e=0$ case.

The aim of the present section is to show the results of extensive numerical simulations running SA with the biased measure in order to gather evidence that a non-zero bias is in general beneficial for the performances of SA in finding a solution to the random hypergraph bicoloring problem.

We have already shown that finite size effects are very small and slightly visible only for $N=10^4$. So in the following we present uniquely data obtained with size $N=10^5$. We have checked these are practically indistinguishable from the results with $N=10^6$ on the time scales reachable in the latter case.

\begin{figure}[t]
\begin{center}
  \includegraphics[width=\textwidth]{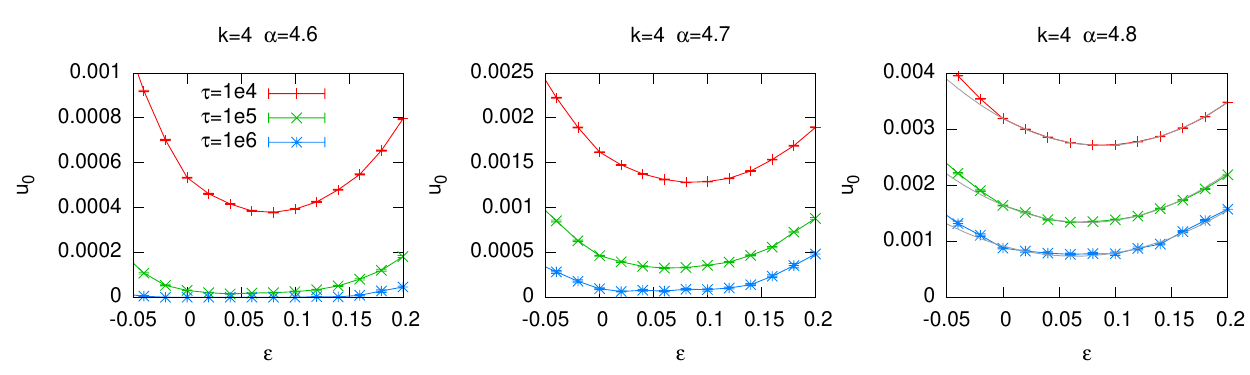}\\
  \vspace{3mm}
  \includegraphics[width=\textwidth]{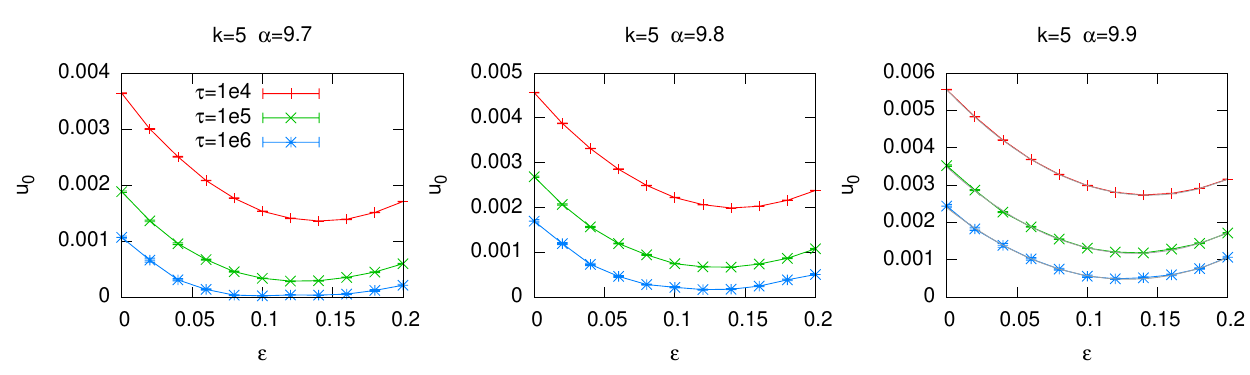}\\
  \vspace{3mm}
  \includegraphics[width=\textwidth]{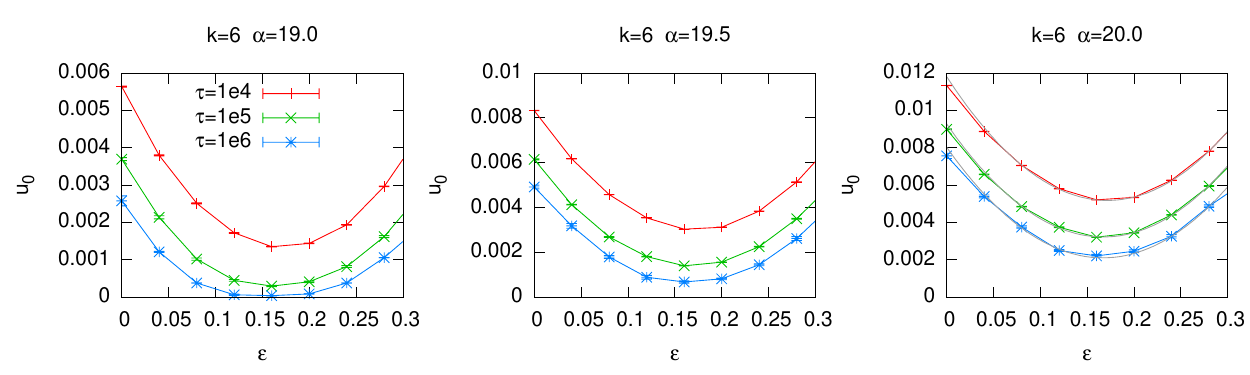}
\end{center}
\caption{The lowest intensive energy reached by SA reaches its minimum for a positive bias parameter $\e$.}
\label{fig:mainPlot}
\end{figure}

In Figure~\ref{fig:mainPlot} we show the data collected at the three largest $\a$ values for $k=4$ (upper row), $k=5$ (middle row) and $k=6$ (lower row).
In each panel we plot $u_0$ as a function of $\e$ for three different cooling rates $\tau=10^4,10^5,10^6$ (from top to bottom in each panel).
The plots provides a clear evidence that reweighting solutions with a bias $\e>0$ enhances the probability that SA reaches lower energies.

Already a simple qualitative analysis reveals the advantage of using $\e>0$.
In every panel we see that $u_0$ reaches a minimum for a strictly positive value of $\e$. The value $\e_{\rm SA}$ that minimizes $u_0$ is only weakly dependent on the SA cooling time $\tau$, so it is likely to assume that $\lim_{\tau\to\infty} \e_{\rm SA}>0$ and the bias is effective even in the limit of large times.

The data in Figure~\ref{fig:mainPlot} suggest that the SA algorithmic threshold may grow for moderately small values of $\e$ with respect to its $\e=0$ value.
For example for $k=6$ the SA algorithmic threshold for $\e=0$ was estimated around $\aalgo\approx 18.5$, but looking at the plots in the lower row it is evident that at least for $\a=19$ and $\e\simeq 0.15$ SA reaches the ground state $u_0=0$ and for $\a=19.5$ and $\e\simeq 0.15$ the convergence to $u_0=0$ is very fast in $\tau$.

\begin{figure}[t]
\begin{center}
  \includegraphics[width=0.6\textwidth]{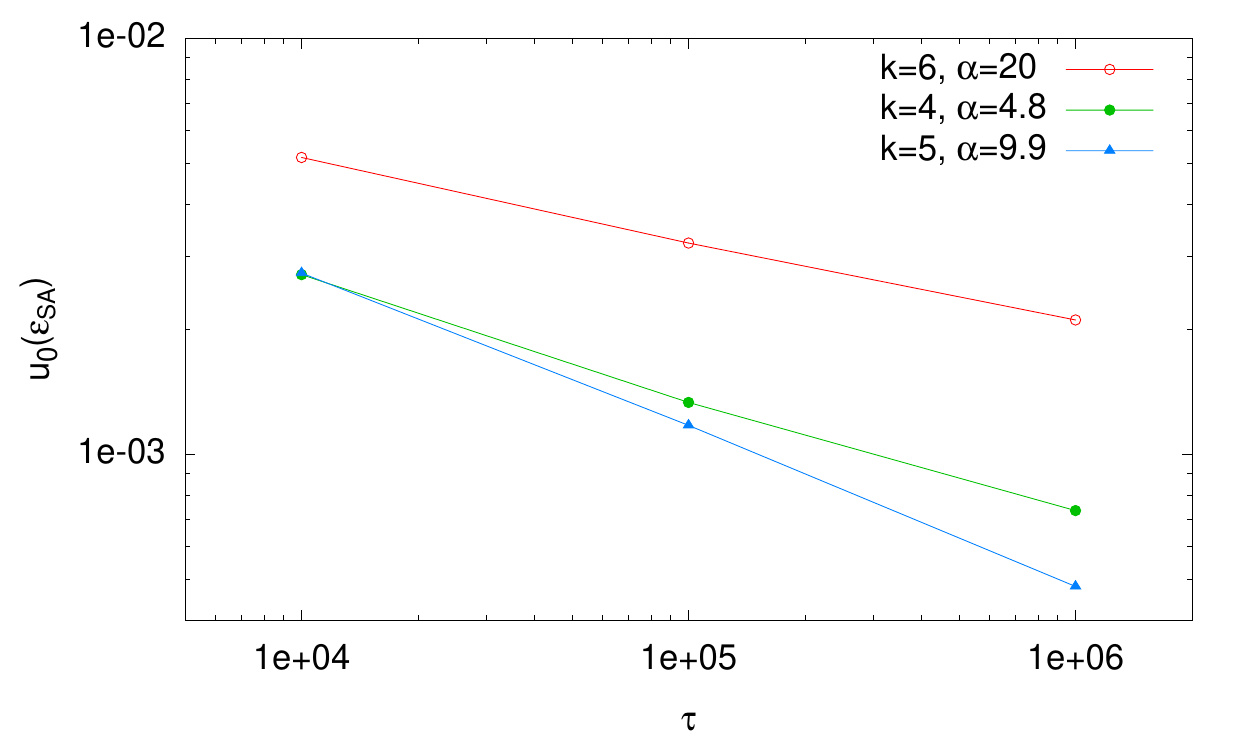}
\end{center}
\caption{The lowest energy reached with the optimal bias $\e_{\rm SA}$ for the largest $\a$ value simulated as a function of $\tau$.}
\label{fig:paramFit}
\end{figure}

We have done a more quantitative analysis for the largest $\a$ values, shown in the three panels on the right. We have interpolated the data of $u_0$ with a quadratic function of $\e$, the interpolating parabolas are shown in the right panels in Figure~\ref{fig:mainPlot}. Fitting the minimum of the parabola $u_0(\e_{\rm SA})$ as a power law in $\tau$ we find the results shown in Figure~\ref{fig:paramFit}. For $(k=5,\a=9.9)$ the behavior is faster than a power law and thus we expect the SA algorithmic threshold with the biased measure to be slightly greater than $\a=9.9$. On the contrary for both $(k=4,\a=4.8)$ and $(k=6,\a=20)$ the behavior looks slightly slower than a power law and thus we are tempted to believe $\lim_{\tau\to\infty}u_0(\e_{\rm SA})>0$ in these cases and the SA algorithmic threshold is slightly below.

Unfortunately the quantitative analysis cannot be made more robust, due to the strong $\tau$ dependence observed. Nevertheless we believe that the evidence that $\e>0$ makes ground states more accessible to Simulated Annealing is strong enough, both for finite $\tau$ values and in the large $\tau$ limit.

Let us finally compare the optimal value $\e_{\rm SA}$ of the bias that improves most the performances of SA with the optimal value $\eopt$ found in Section~\ref{sec_optimalRS} that increases most the extent of the RS phase. We notice that $\e_{\rm SA}$ is always larger than $\eopt$. Considering that, for the small values of $k$ studied in the simulations, it is approximately true that for $\e>\eopt$ the phase transition increasing $\a$ is continuous, while for $\e<\eopt$ the model undergoes a random first order transition, we believe the most natural explanation for the observation $\e_{\rm SA}>\eopt$ is the following. 
The ergodicity breaking taking place at a discontinuous (i.e.\ random first order) transition is much more severe than the one taking place at a continuous phase transition. In the case of a discontinuous phase transition, the SA algorithm can find solutions only slightly above the dynamic threshold $\ad$, while in the continuous case SA remain efficient in finding solutions even well above $\aKS$. The analysis supporting this scenario has been presented in Section~\ref{sec:epsZero}. Thus it is natural to expect that SA presents its best performances for $\e>\eopt$ where the phase transition is continuous and the ergodicity breaking not too severe. The finding that $\e_{\rm SA}(k=4)>0$ also resolves the rather counterintuitive result $\eopt(k=4)<0$. So even for $k=4$ the SA algorithm finds the ground state more easily if frozen variables are avoided.

\section{Large $k$ asymptotics}
\label{sec_largek}

The numerical resolution of the cavity equations presented in Sec.~\ref{sec_results_cavity} shows that for small values of $k$ one has $\aopt(k) > \adu(k)$, in other words that distorting the measure over solutions can make it RS for larger densities of constraints than the uniform one. We want now to investigate the large $k$ limit, for which the gap between the satisfiability threshold and the algorithmic ones is most clearly demonstrated. As $\aopt(k)$ is defined from a discontinuous bifurcation of a functional equation (in other words a reconstruction problem for which the Kesten-Stigum bound is not tight) we do not have an analytical expression for it, we will in consequence aim at a more modest objective, namely deriving asymptotic bounds on its large $k$ behavior.

\begin{figure}[hbtp]
\begin{center}
\includegraphics[scale=0.6]{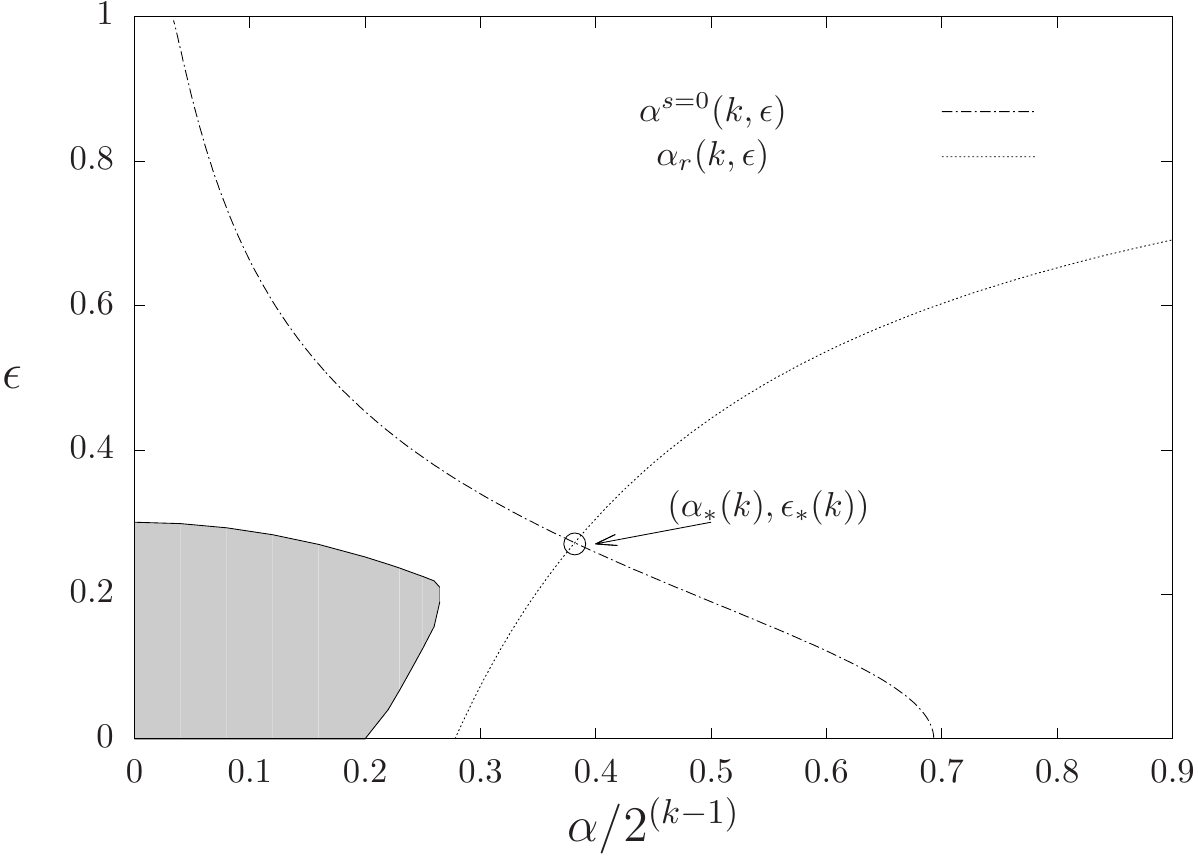}
\caption{For $k=20$, plot in the plane $(\a, \e)$ of the rigidity line and zero RS entropy line, that intersect at the point $(\a_*(20), \e_*(20))$. The gray zone is a qualitative guess of the RS zone delimitated by $\a_d(k,\e)$.}
\label{analyticalbounds}
\end{center}
\end{figure}

As explained before for any $\e$ one has $\ad(\e) \le \min(\aKS(\e),\ar(\e),\a^{s=0}(\e))$, these three upperbounds having simple expressions given in Eqs.~(\ref{eq_s0_epsilon}-\ref{eq_rigidity_epsilon}). At large enough $k$ it is easy to convince oneself that the Kesten-Stigum transition occurs after $\a^{s=0}$, hence is completely irrelevant (the dominant term in the asymptotic expansion of $\aKS$ is of the order $2^{2(k-1)}$ instead of $2^{k-1}$ for $\a^{s=0}$ and $\ar$). We show in Fig.~\ref{analyticalbounds} the lines $\ar(\e),\a^{s=0}(\e)$ for a large value of $k$, as well as a guess on the qualitative behavior of $\ad(\e)$. We define $(\a_*(k),\e_*(k))$ as the coordinates of the intersection of the rigidity and the zero-entropy line, in such a way that $\aopt(k) \le \a_*(k)$: for larger densities either the RS entropy is negative, or there exists a 1RSB solution with hard-fields (or both), in any case no RS phase can exist for $\a \ge \a_*$. We will now derive an asymptotic expansion at large $k$ of this upperbound $\a_*(k)$.

In the expression (\ref{eq_rigidity_epsilon}) of $\ar$ the coefficient $\Gr(k)$ is a series depending solely on $k$, that was defined in Eq.~(\ref{eq_qrGr}); in order to obtain more easily its asymptotic equation we define a series $w_k$ by $\qr(k) = 1-e^{-w_k}$, in such a way that $w_k$ is solution of the implicit equation $e^w = 1 + (k-1) w$. Taking the logarithm of this equation and iterating once yields 
\beq
w_k = \ln k + \ln \ln k + O\left( \frac{\ln \ln k}{\ln k} \right) \ .
\eeq
One can then compute
\beq
\Gr(k) = w_k \left(1+\frac{1}{(k-1) w_k}\right)^{k-1} = \ln k + \ln\ln k+ O(1)
\ . 
\label{eq_dev_Gr}
\eeq
This gives immediately the expansion of the rigidity threshold for the uniform measure ($\e=0$),
\beq
\aru(k) = 2^{k-1} \frac{1}{k} \left( \ln k + \ln\ln k+ O(1) \right) \ .
\label{eq_dev_aru}
\eeq

We come back to the determination of $(\a_*(k),\e_*(k))$; this intersection of the rigidity and zero-entropy line is solution of the two following equations, immediately obtained from (\ref{eq_s0_epsilon},\ref{eq_rigidity_epsilon}):
\beq
\a = \frac{\ln 2}{\frac{k (1-\e) \ln (1-\e)}{2^{k-1} - 1 - k \e}-
\ln\left(1-\frac{1+k\e}{2^{k-1}} \right)} =
\frac{1}{k} \Gr(k) \frac{2^{k-1}-1-k \e}{1-\e} \ .
\label{eq_astar}
\eeq
At large $k$, as can be seen for instance on (\ref{eq_dev_aru}), the asymptotic expansions of constraint densities are organized in different scaling behaviors, namely exponential, polynomial and logarithmic in $k$. Neglecting only exponentially small corrections we can simplify (\ref{eq_astar}) into
\beq
\frac{\a}{2^{k-1}} = \frac{\ln 2}{k (1-\e) \ln (1-\e) + 1+k\e} = \frac{1}{k} \Gr(k) \frac{1}{1-\e} \ .
\label{eq_astar_simplified}
\eeq

Without making additional approximations we see that $\e_*(k)$ is solution of
\beq
\e + (1-\e) \ln(1-\e) = \frac{\ln 2}{\Gr(k)} - \frac{1}{k} - \e \frac{\ln 2}{\Gr(k)} \ .
\label{eq_estar}
\eeq
Given the asymptotic behavior of $\Gr(k)$ stated in (\ref{eq_dev_Gr}) it is easy to see that $\e_*(k)$ must vanish in the limit; one can thus expand the l.h.s. of (\ref{eq_estar}) and obtain
\beq
\e_*(k) = \sqrt{\frac{2 \ln 2}{\Gr(k)}} + O \left(\frac{1}{\ln k} \right) \ .
\label{eq_estar_dev}
\eeq
Reinserting in (\ref{eq_astar_simplified}) we obtain
\beq
\a_*(k) = 2^{k-1} \frac{1}{k} \Gr(k) \left(1 + \sqrt{\frac{2 \ln 2}{\Gr(k)}} + O \left(\frac{1}{\ln k} \right) \right) \ .
\eeq
Using the expansion (\ref{eq_dev_Gr}) of $\Gr(k)$ we have finally
\beq
\a_*(k) = 2^{k-1} \frac{1}{k} \left( \ln k + \sqrt{2 \ln 2} \sqrt{\ln k} + O( \ln \ln k) \right) \ .
\eeq
The comparison with the expansion of $\aru(k)$ given in (\ref{eq_dev_aru}) shows that the leading order is not modified, the term $\ln \ln k$ in the correction being replaced by a (larger) term of order $\sqrt{\ln k}$.

The leading order expansion of $\adu(k)$ has been rigorously derived in~\cite{MoReTe11_recclus} for a family of model encompassing the hypergraph bicoloring one, yielding in this case $2^{k-1} \ln k/k$. By analogy with other rigorous results~\cite{Sly08,SlyZhang16} (obtained for the $q$-coloring problem) we shall assume that $\adu(k)$ has the same asymptotic expansion (\ref{eq_dev_aru}) as $\aru$ (with a strictly smaller constant hidden in the $O(1)$ term). We thus conclude that
\beq
\adu(k) =  2^{k-1} \frac{1}{k} \left[ \ln k + \ln\ln k+ O(1) \right] \le
\aopt(k) \le \a_*(k) = 2^{k-1} \frac{1}{k} \left( \ln k + \sqrt{2 \ln 2} \sqrt{\ln k} + O( \ln \ln k) \right) \ ,
\label{eq_bounds_opt}
\eeq
hence that the best improvement of $\aopt(k)$ with respect to $\adu(k)$ that can be hoped for with the bias considered in this paper is a replacement of $\ln \ln k$ by $\sqrt{\ln k}$ in the second order term of their asymptotic expansions.

We shall finally come back briefly on the choice of parameters we made in (\ref{eq_parameters_epsilon}), where we used a single parameter $\epsilon$ for the bias instead of trying to exploit all the $k/2$ free values of $\omega_p$. Let us define $\a'_{\rm opt}(k)=\sup \ad(k,\{\w_p\})$, where the maximization is now over all possible values of the $\w_p$, under the conditions $\w_0=\w_k=0$ and $\w_p=\w_{k-p}$. We have certainly $\a'_{\rm opt}(k) \ge \aopt(k)$, and the inequality is probably strict at least for small enough values of $k$; however we shall now show that $\a'_{\rm opt}(k) \le \a_*(k)$, hence that the larger freedom of choice of generic parameters does not allow to beat the upperbound derived and discussed in the special case (\ref{eq_parameters_epsilon}). To see this more easily let us exploit the invariance of the measure (\ref{measure}) under a multiplication of all $\w_p$ by a common constant, and fix their normalizations in such a way that
\beq
\sum_{p=1}^{k-1} \binom{k}{p} \w_p = 1 \ .
\label{eq_w_normalization}
\eeq
With this choice the expressions of the RS entropy (\ref{eq_sRS}) and rigidity threshold (\ref{eq_rigidity_generic}) become
\beq
s^{RS}(\a,\{\w_p\}) = \ln 2 + \a \left( \sum_{p=1}^{k-1} \binom{k}{p} \w_p \ln \w_p - k \ln 2  \right) 
\ , \qquad
\ar(k,\{\w_p\}) =\frac{1}{k} \Gr(k) \frac{1}{2 \w_1} \ .
\label{eq_s_ar_generalized}
\eeq
Consider now a given choice of the parameter $\w_1$, and a value of $\a \le \ar$; for these to allow a RS phase the corresponding entropy should be positive. Maximizing the entropy in (\ref{eq_s_ar_generalized}) with respect to $\{\w_2,\dots,\w_{k-2}\}$, under the normalization condition (\ref{eq_w_normalization}) and for a fixed value of $\w_1$ is easily seen to yield $\w_2=\dots=\w_{k-2}$, i.e. precisely the choice of parameters (\ref{eq_parameters_epsilon}). In other words this bias is the one that allows to tune the fraction of frozen variables while keeping the measure as uniform as possible, in order to minimize the entropy cost it induces.

\section{Discussion}
\label{sec_conclu}

We have studied the problem of bicoloring random $k$-regular hypergraphs, where every hyper-edge joins exactly $k$ variables, also known as NAE-$k$-SAT.
Having in mind the observation that algorithms usually reach solutions with no frozen variables and the conjecture that the computational complexity of the problem is directly related to the presence of frozen variables, we have modified the uniform measure over solutions to a biased measure where solutions with frozen variables are disfavoured (the works~\cite{BaInLuSaZe15_long,BaBo16} are based on a similar idea, with a bias favouring the regions of configuration space with a high local entropy of solutions). We have studied this biased measure both analytically via the cavity method, and numerically, running extensive Simulated Annealing processes in the search for solutions.

As a byproduct of our study we have presented two technical tools of interest by themselves: (i) how to properly determine the dynamic transition threshold, that corresponds to a first order transition in a functional space, by studying the stability parameter close to a random first order transition; (ii) how to estimate the algorithmic threshold for Simulated Annealing in random CSP by studying the dependence of the asymptotic energy on the cooling rate.

The results we got are somehow different depending on whether $k$ is small or very large.
In the range of small values of $k$ we have found that the use of the biased measure is rather effective in helping SA to find solutions, thus moving its algorithmic threshold to larger values of $\alpha$.
This numerical finding is supported by the analytical computation that predicts a shift of the dynamic threshold in presence of a bias. This is a clear result showing that, both in the thermodynamic limit and on problems of finite size, the use of the bias favouring unfrozen configurations is effective. As the non-uniformity of the measure necessarily implies a reduction of its entropy, the present result supports the idea that many solutions are not necessarily useful from the point of view of searching algorithms and actually removing these solutions (or reducing their weight) helps in the search for the remaining ones. The analytic result on the shift of the dynamic threshold implies that for $\alpha\in[\adu,\aopt]$ there are long range correlations for $\e=0$ that actually disappear for a range of non-zero biases. This result suggests that long range correlations are due to a subset of solutions where variables are very strongly correlated, while in the remaining subset of solutions variables are much more weakly correlated. The lack of reentrance of critical lines in temperature is a good news for thermal algorithms, that is  algorithms satisfying detailed balance with respect to a Gibbs-Boltzmann probability distribution, because it implies that this class of algorithms should not suffer a slowing down due to critical points at finite temperature below the threshold given by the zero-temperature critical point, and the latter should be the one eventually determining the asymptotic behavior of these algorithms.

In the large $k$ limit the bounds of (\ref{eq_bounds_opt}) show that in the most optimistic scenario the gain obtained by the present version of the biasing strategy is rather modest, corresponding to an increase from $\ln \ln k$ to $\sqrt{\ln k}$ in the second order of the expansion of the threshold density of constraints. We believe that a very interesting open problem would be to determine which of the two bounds on the expansion of $\aopt$ in (\ref{eq_bounds_opt}) is tight (or whether the scaling is actually intermediate between the two). If the lowerbound is not tight even this tiny increase would be an improvement of the algorithmic gap, and would beat the rigidity threshold of the uniform measure (that was sometimes thought to be the algorithmic barrier). Adapting the techniques of~\cite{Sly08,MoReTe11_recclus,SlyZhang16} to prove a lowerbound on the reconstruction threshold of the biased measure should clarify this issue.

Another possible direction we are currently investigating is the study of more generic bias on the set of solutions, induced by interactions between variables at larger distances; this was shown in~\cite{BrDaSeZd16} to have an even more dramatic effect on the location of the rigidity transition. It remains to see whether the dynamic transition can also be efficiently manipulated in this way.

An interesting question to investigate in the future is the effect of the decimation on the biased measure, which can be analyzed along the lines of~\cite{RiSe09}. This would probably prove analytically that the algorithmic threshold for the process that searches a solution fixing variables according to marginals provided by belief propagation is improved by a non zero bias.

In the present work we have focused our study of biased measures on the hypergraph bicoloring problem because it is technically simpler than $k$-SAT or $q$-COL, its degrees of freedom being binary and its replica symmetric solution being trivial, while its unbiased measure is known to exhibit all the transitions undergone by the more complicated random CSPs. We are convinced that the biasing idea can also be applied to the latter and we expect a similar increase of the clustering transition threshold; we leave an explicit verification of this expectation as an open problem for future work.

\acknowledgments

We thank Lenka Zdeborova for useful discussions. GS is part of the PAIL grant of the French Agence Nationale de la Recherche, ANR-17-CE23-0023-01.

\bibliography{biblio}

\end{document}